\documentclass[12pt,preprint]{aastex} 

\shorttitle{Broad Radio Recombination Line Objects}
\shortauthors{Sewi{\l}o et al.}

\newcommand{\al}{$\alpha$~}
\newcommand{\chcn}{CH$_{3}$CN~}

\newcommand{\kms}{km~s$^{-1}$~}

\begin{document}

\title{A VLA Study of Ultracompact and Hypercompact H\,{\sc ii} Regions from 0.7 to 3.6 cm}

\author{M. Sewi{\l}o\altaffilmark{1}, E. Churchwell\altaffilmark{2}, 
S. Kurtz\altaffilmark{3}, W. M. Goss\altaffilmark{4}, and  P. Hofner\altaffilmark{4,5}} 
\altaffiltext{1}{The Johns Hopkins University, Department of Physics \& Astronomy,
366 Bloomberg Center, 3400 N. Charles Street, Baltimore, MD 21218}
\altaffiltext{2}{University of Wisconsin - Madison, Department of Astronomy, 475 N. 
Charter St., Madison, WI 53706}
\altaffiltext{3}{Centro de Radioastronom\'\i a y Astrof\'\i sica, Universidad Nacional 
Aut\'onoma de M\'exico, Apdo. Postal 3-72, 58089, Morelia, Mich. Mexico}
\altaffiltext{4}{National Radio Astronomy Observatory, P.O. Box 0, Socorro, NM 87801}
\altaffiltext{5}{New Mexico Tech, Physics Department, 801 Leroy Place, Socorro, NM 87801}

\begin{abstract}
We report multi-frequency Very Large Array observations of three
massive star formation regions (MSFRs) containing radio continuum
components that were identified as broad radio recombination line
(RRL) sources and hypercompact (HC) H\,{\sc ii} region candidates in
our previous H92$\alpha$ and H76$\alpha$ study: G10.96$+$0.01
(component W), G28.20$-$0.04 (N), and G34.26$+$0.15 (B). An additional
HC H\,{\sc ii} region candidate, G45.07$+$0.13, known to have broad
H66$\alpha$ and H76$\alpha$ lines, small size, high electron density and emission
measure, was also included. We observed with high spatial resolution
(0$\rlap.{''}$9 to 2$\rlap.{''}$3) the H53$\alpha$, H66$\alpha$,
H76$\alpha$, and H92$\alpha$ RRLs and the radio continuum at the
corresponding wavelengths (0.7 to 3.6 cm).  The motivation for these
observations was to obtain RRLs over a range of principal quantum
states to look for signatures of pressure broadening and macroscopic
velocity structure.  We find that pressure broadening contributes
significantly to the linewidths, but it is not the sole cause of the
broad lines.  We compare radio continuum and dust emission
distributions and find a good correspondence. We also discuss maser
emission and multi-wavelength observations reported in the literature
for these MSFRs.

\end{abstract}


\section{Introduction}

Hypercompact (HC) H\,{\sc ii} regions represent an earlier
evolutionary state than compact and ultracompact (UC) H\,{\sc ii}
regions. These very small ($\lesssim$ 0.03 pc), high electron density
(n$_{e}$ $\sim$ 10$^{5-6}$ cm$^{-3}$) and high emission measure (EM
$\gtrsim$ 10$^{8}$ pc cm$^{-6}$) nebulae are ionized from within by O
or B stars and usually are associated with maser emission.  Many, but
not all, have unusually broad radio recombination lines (RRLs) with
full widths at half maximum intensity (FWHM) $\gtrsim$ 40 km s$^{-1}$
\citep{depree96, depree97, s04}.  HC H\,{\sc ii} regions have rising spectral
indices $\alpha$ (where $S_{\nu} \propto \nu^{\alpha}$) from
centimeter to millimeter wavelengths with typical values of
$\alpha\sim$1, which suggests a range of densities and optical depths
in the ionized gas. Broad RRLs (BRRLs) and intermediate sloped ($-0.1
< \alpha < +2$) power-law spectral indices may be associated with the
age of HC H\,{\sc ii} regions, appearing only during a small fraction
of the lifetime of the hypercompact phase.  There are very dense and
compact H\,{\sc ii} regions, however, that do not have intermediate
power-law spectral energy distributions or broad radio recombination lines 
(e.g., G34.26$+$0.15 A, \citealt{ava06}; \citealt{s04}).

Although possibly related to the age, the physical reason 
that some HC H\,{\sc ii} regions do not present BRRLs is poorly
understood.
The broad linewidths suggest highly dynamic internal
structures (e.g., bipolar outflows, expansion, stellar winds,
accretion, disk rotation, shocks) as well as electron impact
(pressure) broadening.  High spatial resolution observations over a
wide range of frequencies are required to distinguish the relative
contributions from pressure broadening and bulk motion of the
gas. Transitions with high principal quantum numbers, $n$, are
significantly affected by pressure broadening, while low $n$
transitions are mostly free of pressure broadening and are sensitive
to large-scale motions. With high spatial resolution at a range of
frequencies, it is possible to distinguish bulk motions from pressure
broadening.  Observations have shown that both pressure broadening and
gas dynamics contribute to line broadening \citep{sew08, k08}.

Theoretical progress has  been made towards understanding the nature of 
HC H\,{\sc ii} regions. The model of hierarchical clumping of the 
nebular gas proposed by \citet{ic04} was able to reproduce the 
intermediate values of spectral indices seen in HC H\,{\sc ii} 
regions. \citet{tm03} and \citet{ke03} proposed models that 
predict both confinement of the ionized gas in HC H\,{\sc ii}s and
the BRRL emission.  The photoevaporating disk model, originally
proposed by \citet{hj94} and \citet{lizano96} to explain the most
compact UC H{\sc ii} regions, has been recognized to be applicable to
HC H{\sc ii} regions \citep{lugo04}.

Here we report high spatial resolution, multi-frequency Very Large
Array\footnote{The National Radio Astronomy Observatory is a facility
of the National Science Foundation operated under cooperative agreement
by Associated Universities, Inc.} 
(VLA) observations of four candidate HC H{\sc ii} regions with the goal 
of determining their physical properties.
Three sources, G10.96$+$0.01, G28.20$-$0.04, and
G34.26$+$0.15, were selected based on an earlier H92$\alpha$ and H76$\alpha$ 
study \citep{s04}.  We also observed the HC H\,{\sc ii} candidate
G45.07$+$0.13 NE, which is known to have broad H66$\alpha$ and H76$\alpha$
lines \citep{g85,g86}, small size, and high electron density (i.e., high
emission measure).  Additional RRLs were observed
 to search for systematic broadening with increasing
principal quantum number. A strong correlation of line width with
principal quantum number would indicate pressure broadening as the main
mechanism for line broadening, as opposed to large-scale motions
(rotation, outflows, infall, etc.). The rms electron temperature and
the rms electron density of the sources can be determined for each
transition using the assumption of local thermodynamic equilibrium
(LTE). RRLs also provide information on the velocity structure of the
sources. 

We report detections of the H53$\alpha$, H66$\alpha$, H76$\alpha$, and
H92$\alpha$ RRLs and the 0.7, 1.3, 2.0, and 3.6 cm radio continuum
emission with resolutions of 1$\rlap.{''}$5, 0$\rlap.{''}$9,
1$\rlap.{''}$2, and 2$\rlap.{''}$3, respectively, toward the MSFRs
G10.96$+$0.01, G28.20$-$0.04, G34.26$+$0.15, and G45.07$+$0.13
(hereafter G10.96, G28.20, G34.26, and G45.07). The observations are
described in Section 2.  In Section 3, we derive physical properties of the
nebulae based on continuum observations, and investigate the change of
line width with principal quantum number. In Section 4 we investigate the dust
emission toward these sources using mid-infrared {\it Spitzer} Space 
Telescope GLIMPSE data. We compare the mid-infrared GLIMPSE and radio
continuum images with previous maser emission and multiwavelength
observations. We discuss the general results in Section 5 and summarize
the main conclusions in Secion 6.

\section{The Data}

\subsection{VLA Observations: Continuum and Radio Recombination Lines}
 
The VLA was used to observe H53$\alpha$, H66$\alpha$, H76$\alpha$, and 
H92$\alpha$ RRLs toward G10.96, G28.20, G34.26, and G45.07.
Table \ref{tabinstr} gives the instrumental parameters of the 
observations. Table \ref{trans} summarizes the transitions that we observed 
toward each source \citep[and this paper]{s04} and the adopted distances.

The first channel of VLA spectral line data (``channel 0``) contains
the average of the central 75\% of the observing bandwidth (generated
by vector averaging channels; \citealt{wilcots2009}). The broad-band
{\it uv} data from ''channel 0'' were edited and calibrated using the
Astronomical Image Processing System (AIPS) software package, and the
resulting calibration solutions were applied to the multi-channel
spectral-line data. All sources were self-calibrated to increase the 
dynamic range of the images. The continuum level was determined from a 
linear fit to the line-free channels. Continuum subtraction was done 
in the {\it uv} plane with the AIPS task UVLSF. A line-free continuum 
data set and a 31 channel cube of 
recombination line emission were constructed. Both data sets were Fourier 
inverted using weighting intermediate between natural and uniform 
(Robust=0 in AIPS task IMAGR) and CLEANed in the standard manner. The images 
were corrected for primary beam attenuation using the AIPS task PBCOR. 
Integration times and continuum noise values of the final images 
are listed in Table \ref{tabcont}.

\begin{figure}
\centering
\includegraphics[width=0.46\textwidth]{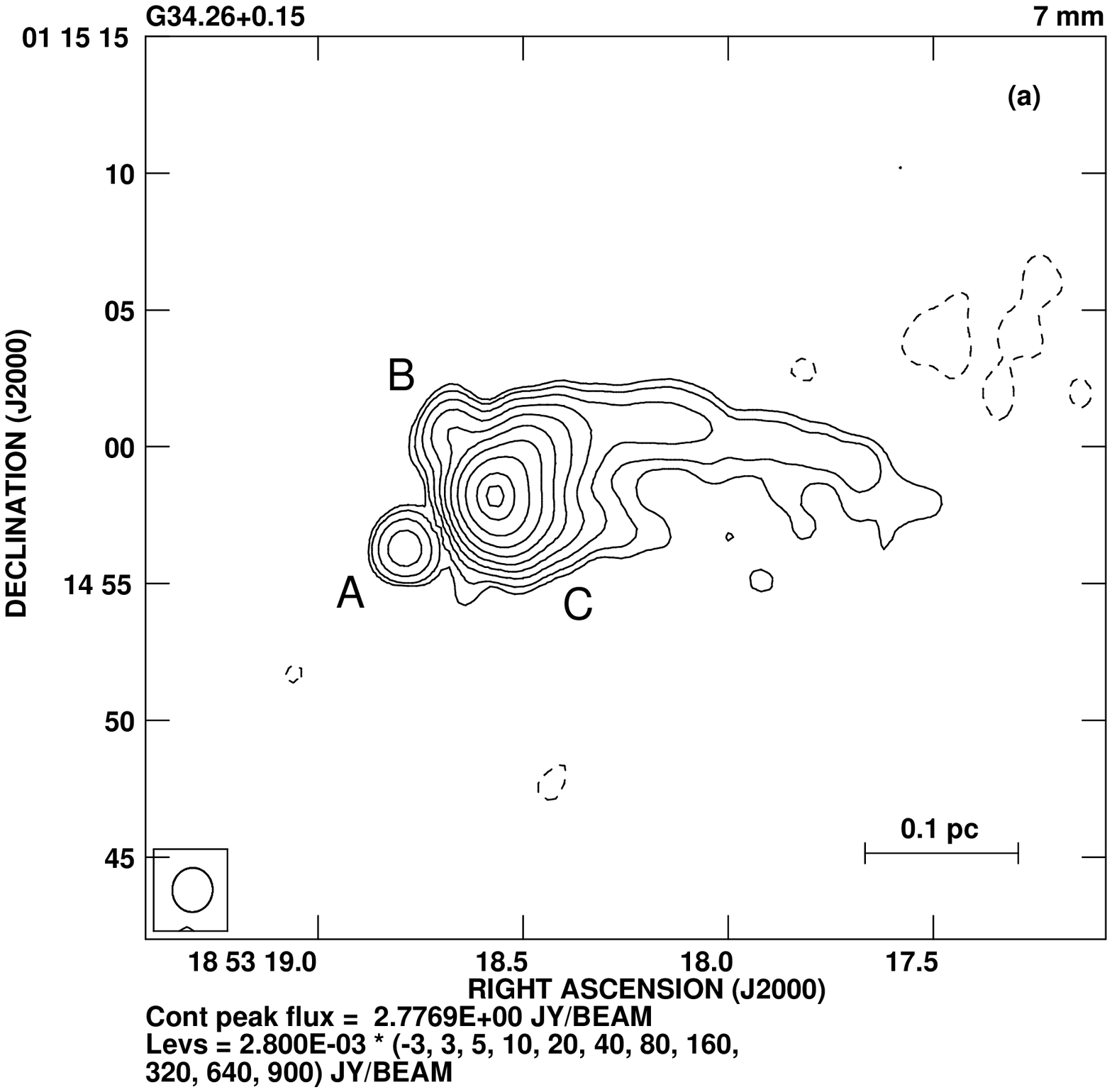}
\hfill
\hfill
\includegraphics[width=0.495\textwidth]{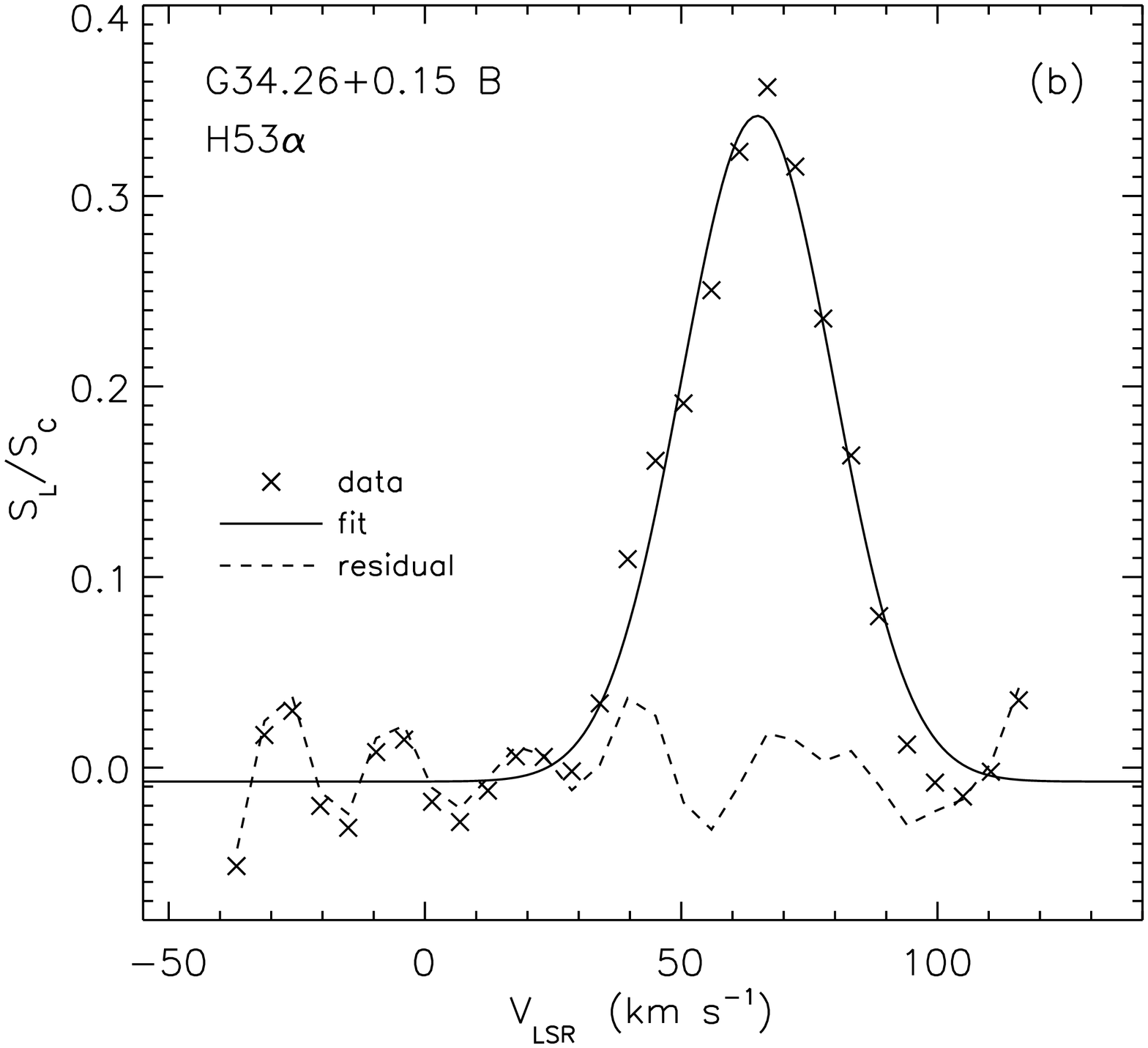}
\hfill
\vspace*{0.6cm}
\hspace*{-0.6cm}
\includegraphics[width=0.495\textwidth]{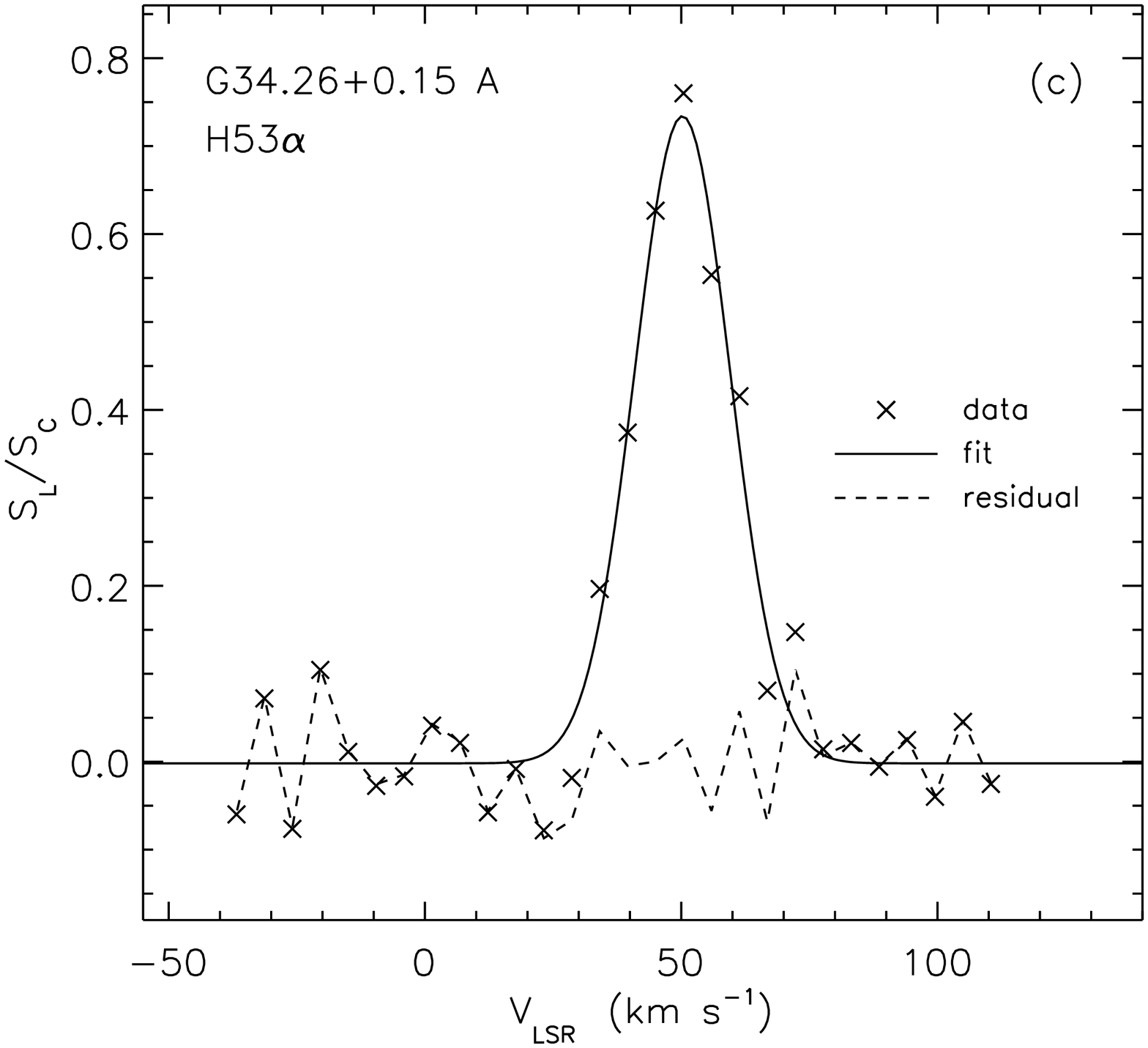}
\includegraphics[width=0.495\textwidth]{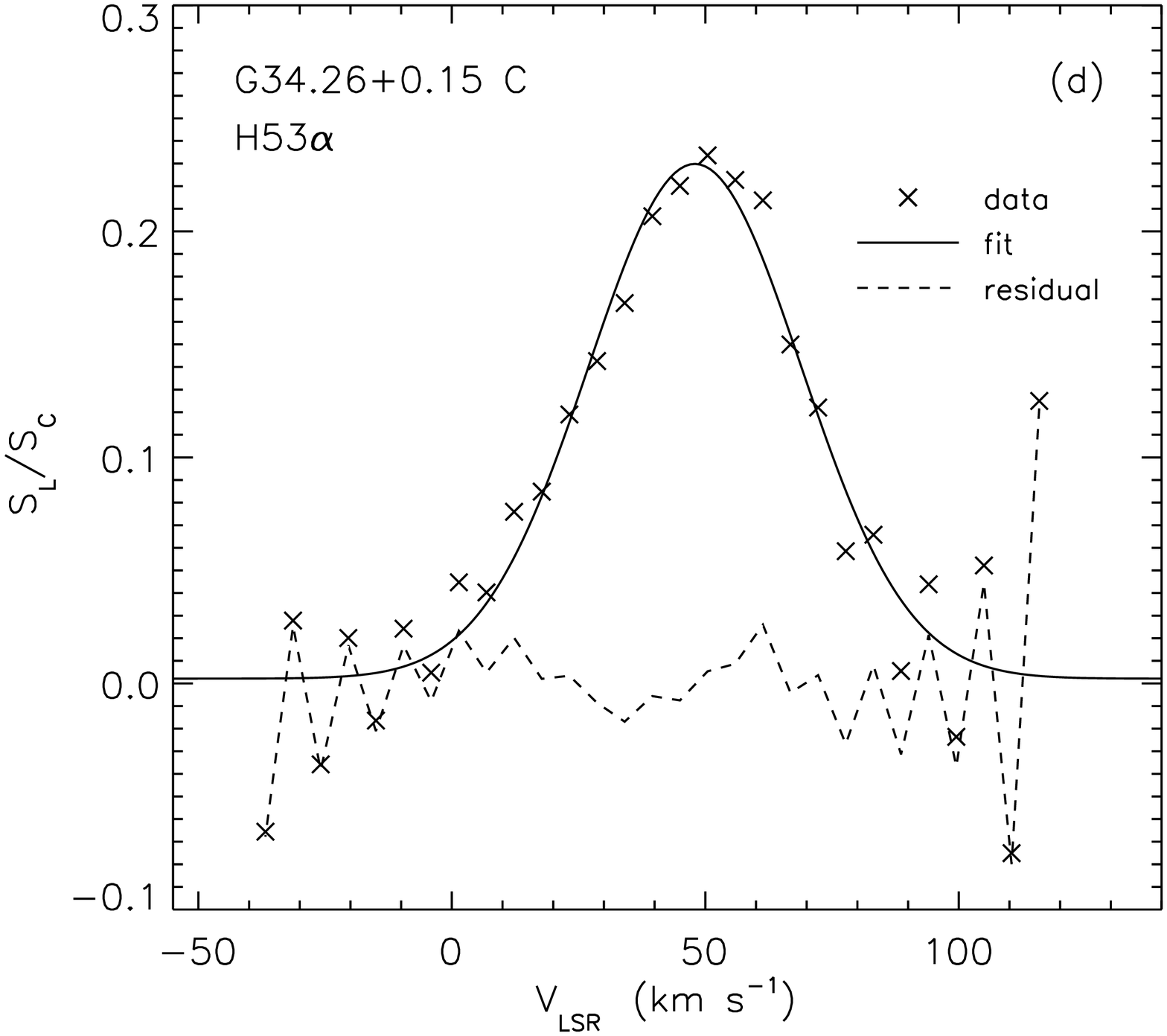}
\caption{The 7 mm VLA continuum image (a) and the integrated H53$\alpha$ 
line profiles from the A (c), B (b), and C (d) components of the MSFR 
G34.26. The beam size shown in the lower left corner in (a) is 
$\sim$1$\rlap.{''}$5 (see Table \ref{tabinstr}). 
\label{g34radio}}
\end{figure}

Source positions, deconvolved angular diameters, position angles, and
total flux densities (Table \ref{tabcont}), were determined by
elliptical Gaussian fitting using the least-squares AIPS routine JMFIT. 
Continuum-subtracted line images were used to generate the spatially
integrated H53$\alpha$, H66$\alpha$, H76$\alpha$, and H92$\alpha$ line
profiles. For each line, the integration was performed approximately 
over the area enclosed by the 50\% contour with respect to the peak value. 
The subsequent spectral analysis of the 
RRLs was done using the  IDL Programming Language
\footnote{www.ittvis.com/idl}. The spatially
integrated line profiles were fitted by a single Gaussian. The results
for individual sources are presented graphically in 
Figs.~\ref{g10wradio} -- \ref{g45radio} for the H53$\alpha$,
H66$\alpha$, H76$\alpha$, and H92$\alpha$ RRLs. The continuum images
are shown in the same figures; the projected linear scales are based
on the kinematic distances listed in Table \ref{trans}.

\subsection{Archival Data: VLA and {\it Spitzer} Space Telescope}
\label{obsarch}

In \citet{sew08}, we reported the results of high-resolution
($0\rlap.{''}$15) observations of G28.20 N that revealed the
shell-like morphology of this HC H\,{\sc ii} region. To investigate
the morphology of G28.20 N at even higher resolution, here we use VLA
K-band A-array archival data (Program ID AZ168) to create the high
resolution ($0\rlap.{''}$09) continuum image shown in
Fig.~\ref{g28masers}.  These observations were taken on 2006 March 14,
with a 3.125 MHz bandwidth of 63 channels of 48.828 kHz each. We used
line-free channels of these spectral line data to make a
self-calibrated continuum image with an rms noise of 85 $\mu$Jy~beam$^{-1}$.
The measured 1.3 cm flux density of G28.20 N is 630$\pm$63 mJy.
The synthesized beam is 0$\rlap.{''}$10$\times$0$\rlap.{''}$08 at position 
angle of $-$0$\rlap.^{\circ}$5.

We also present H$_2$O, OH, and CH$_{3}$OH maser
positions relative to the radio continuum for each HC H\,{\sc ii} region 
based on high resolution (interferometric) data from the literature.
The relative positional accuracy of 
22 GHz H$_{2}$O masers detected toward G45.07 NE is 0$\rlap.{''}$1 
\citep{hc96}. The 1665 MHz and 1667 MHz OH masers observed in 
G10.96 W and G28.20 N have an absolute positional accuracy of 
$\sim$0$\rlap.{''}$3 in right ascension and $\sim$0$\rlap.{''}$5 
in declination \citep{a00}. The absolute positions of the 6.67 GHz 
CH$_{3}$OH masers in G10.96 W and G28.20 N are accurate 
to $\sim$1$''$ \citep{w98}, while the positional uncertainty of the 
44 GHz CH$_{3}$OH masers in G45.07 NE is 0$\rlap.{''}$6 \citep{khv04}. 
The relative positional accuracy of 22 GHz H$_{2}$O masers detected 
toward G45.07 NE is 0$\rlap.{''}$1 \citep{hc96}. The distribution
of masers is shown in Figs.~\ref{g10wradio}b, \ref{g28masers}, 
and \ref{g45radio} for G10.96 W, G28.20 N, and G45.07 NE, respectively. 

We used VLA K-band spectral-line archival data for G28.20 N 
HC H\,{\sc ii} region (Program ID AP465) to supplement the single-dish 
water maser data available for this source in the literature. The 
observations were taken on 2008 April 4 with the VLA in C configuration.
A 3.125 MHz bandwidth with
255 channels of 12.2 kHz each was used. We followed standard 
VLA data reduction procedures. The synthesized beam is 
1$\rlap.{''}$19$\times$0$\rlap.{''}$94 at position angle of 
$-$12$\rlap.^{\circ}$8. The typical rms of a line-free channel is 
22 mJy beam$^{-1}$. We determined the positions and peak flux densities 
of the masers from Gaussian fits to the peak channel of each maser. 
The H$_{2}$O maser positions and observed parameters are listed in 
Table~\ref{vlawater}. The uncertainty of the H$_{2}$O maser 
positions in G28.20 N is 0\farcs1.

We also investigate dust emission for our sample using data from the
{\it Spitzer} Space Telescope GLIMPSE survey (``Galactic Legacy Infrared 
Mid-Plane Survey Extraordinaire'',  \citealt{b03}; \citealt{churchwell09}). 
The inner Galactic plane was 
imaged at 3.6, 4.5, 5.8, and 8.0 $\mu$m using the IRAC camera \citep{f04}  
on the {\it Spitzer} Space Telescope. The IRAC spatial resolution ranges from
1$\rlap.{''}$5 to 1$\rlap.{''}$9 between the 3.6 and 8.0 $\mu$m
bands. The IRAC image mosaics were created by the GLIMPSE pipeline. The
absolute accuracy of the GLIMPSE point source positions is 0$\rlap.{''}$3
\citep{meade2009}.

\section{Results}

\subsection{Physical Properties of the Sources}

Physical properties of the sources were derived from the observed
continuum parameters reported in Table \ref{tabcont} using the equations
of \citet{mh67} and \citet{pw78} for spherical, optically thin,
homogeneous, and ionization-bounded H\,{\sc ii} regions. \citet{pw78}
presented modified versions of the \citet{mh67} equations for
geometry-dependent physical parameters such as the rms electron
density (n$_{e, rms}$), the emission measure (EM), and the mass of
ionized gas (M$_{\rm H {II}}$).  The excitation parameter (U) and the
Lyman continuum photon flux required to maintain ionization of the
nebula (N$_{c}^{'}$) do not depend on the source size or its
morphology and were derived using the formulas from \citet{mh67}. The
continuum optical depth ($\tau_{c}$) was calculated from the
\citet{o61} formula in a form presented by \citet{mh67}, using EM
derived based on the results of \citet{pw78}.  The derived physical
parameters for each source are presented in Table \ref{tabphys}. 
The linear diameter of a sphere listed in Table \ref{tabphys} is 
the  diameter of the spherical model for the source, rather 
than the apparent Gaussian half-power width of the source reported in Table \ref{tabcont}.  
An optically thin homogeneous sphere has the same peak EM and total flux 
density as the observed source. The calculations take into account the 
dependence on the beam size as described in \citet{pw78}.

The N$_{c}^{'}$ values were used to
estimate a lower limit to the spectral type of the exciting star given
in the last column of Table \ref{tabphys}. The classification is based on
the non-LTE, spherically expanding (owing to the stellar wind),
line-blanketed (to account for the effects of metals on the
atmospheric structure and emerging ionizing flux) model atmospheres
derived by \citet{sm02} under the assumption that a single star 
ionizes a dust-free, ionization bounded H\,{\sc ii}
region.  The assignment of a given spectral type with N$_{c}^{'}$
depends strongly on assignment of effective temperature with spectral
type and how accurately stellar atmosphere models reflect reality. For
these reasons, spectral types listed in Table \ref{tabphys} are
approximations. The stellar spectral type corresponds to the value of 
radio flux density at the highest frequency observed.

A second set of physical properties of the nebulae was derived based
on their peak brightness temperature; these are ``peak physical properties'' 
\citep{wc89}.  These quantities
depend only on the peak brightness, not the measured diameter or morphology.
Thus, they can be derived for all the sources from our sample including the
blended components B and C in the G34.26 complex and the over-resolved
component E of G10.96. The brightness temperature $T_{B}$ is given by
the Rayleigh-Jeans approximation.

The peak optical depth was estimated from $T_{B} = T_{e} (1- e^{-\tau_c})$ 
assuming a kinetic temperature of $T_{e}$ = 10$^{4}$ K
that is constant over the synthesized beam. The peak emission measure
was calculated using the optical depth equation from \citet{mh67}. The
relation $n_{e} = <n_{e}^2>^{1/2} = (EM/\Delta s)^{1/2}$ was used to
calculate the rms electron density ($n_{e,rms}$) averaged over the beam. The
quantity $\Delta s$ is the source size along the line of sight
through the peak position. We assume the sources in our sample have
spherical geometry based on their projected circular symmetry;
for $\Delta s$ we adopt their geometrical mean angular diameters
(half-power widths, or ``source size'' in Table \ref{tabcont}) assuming that 
the extent of the source along the line of sight is the same as its extent 
in the plane of the sky. The peak physical properties are given in Table 
\ref{tabpeak}.

\subsection{Radio Recombination Line Analysis}

The Gaussian fit parameters of the line profiles are given in
Table \ref{tabfits}. $V_{LSR}$ is the central line velocity relative to
the local standard of rest, {\it FWHM} is the line full width at half
maximum, $S_{L}$ is the flux density in excess of the continuum at
line center, $S_{C}$ is the continuum flux density,
${\frac{S_{L}}{S_{C}}}$ and $\int \frac{S_{L}}{S_{C}}\,\mathrm{d}v $
are the line-to-continuum ratios at line center and integrated over
the line profile, respectively. The continuum flux densities are
determined by integrating over the same areas of the source used to
generate the line profiles; thus they may be less than the flux
densities given in Table \ref{tabcont} (integrated over the entire
source). The uncertainties given in Table \ref{tabfits} are the formal
1$\sigma$ (68.3\% confidence level) errors of the nonlinear
least-squares fit.  These uncertainties are lower limits because they
do not account for systematic errors.

The measured line widths given in Table \ref{tabfits} are 
broadened because of the finite velocity resolution of the
observations. To determine the intrinsic line widths, the instrumental
spectral resolution was deconvolved from the observed lines. The
convolution function in the spectral frequency domain was assumed to
be a Gaussian function with FWHM of 1.2 times the channel width 
(or twice the channel width if the spectrum was Hanning-smoothed off-line) and
deconvolved from the observed line widths. The deconvolved line widths
($\rm FWHM_{\it deconv}$) are given in Table \ref{tabcorr}.  

We calculated the LTE electron temperatures T$^{\star}_{e}$ for all the sources
with reliable RRL measurements using the formula from \citet{rw86}:
\begin{equation}
T_{e}^{\ast} = \bigg[ \frac{6.985 \times 10^3}{a(\nu, T_e)} \cdot \nu^{1.1} \cdot \frac{(S_L/S_C)_{peak}^{-1}}{FWHM_{deconv}} \cdot \frac{1}{1+Y^+}  \bigg]^{0.87}
\end{equation}
where $\nu$ is in GHz, {\it FWHM}$_{deconv}$ is in km s$^{-1}$, and
T$^{\star}_{e}$ in K. $Y^{+} = N(He^{+})/N(H^{+})$ is the ratio of
ionized He to H atoms and is assumed to be 0.1.  The factor $a(\nu,
T_{e})$ is approximately 1 \citep{a60}.  This equation was derived under several 
assumptions about H\,{\sc ii} regions: (1) the structure is plane-parallel, 
homogeneous and isothermal; (2) all optical depths are small: $|\tau_l + \tau_c| \ll
1$ and $\tau_c \ll 1$; and (3) the lines can be treated in the LTE approximation.

However, several of the observed sources have relatively high optical depths
($\tau_C$ from 0.02 to 0.2), causing  attenuation of the line-to-continuum
ratio. In these cases, the LTE electron temperatures have been calculated 
using the line-to-continuum ratios corrected for opacity rather than the
measured values: 
\begin{equation}
(S_L/S_C)_{corr} = (S_L/S_C)_{peak} \cdot (e^{\tau_c}-1)/\tau_c.
\end{equation}
The corrected line-to-continuum ratios and the LTE electron temperatures are listed 
in Table \ref{tabcorr}. The uncertainties in T$^{\star}_{e}$ were estimated by a 
linear propagation of errors.

\subsection{{\it Spitzer}/IRAC Mid-Infrared Emission}
\label{resirac}

High-spatial resolution mosaics from GLIMPSE
\citep{churchwell09} are a unique tool for the study of massive star formation
regions. The IRAC 3.6, 5.8, and 8.0 $\mu$m bands contain strong
polycyclic aromatic hydrocarbon (PAH) features that probe 
conditions in photodissociation regions (PDRs) that separate the
ionized gas in H\,{\sc ii} regions from the surrounding molecular gas
(e.g., \citealt{tie93}). PAHs are large organic molecules containing
$\gtrsim$50 C atoms and are very abundant in the ISM (e.g.,
\citealt{sel84}; \citealt{all89}).  PAHs are excited by sub-Lyman
ultraviolet photons but are destroyed in H\,{\sc ii} regions by
extreme ultraviolet (EUV) radiation (\citealt{voi92};
\citealt{pov07}). Nevertheless, H\,{\sc ii} regions can be traced by
the IRAC 4.5 $\mu$m band, which does not contain PAH features.  The
4.5 $\mu$m emission of H\,{\sc ii} regions is a combination of
thermal emission from small grains, scattered star light, and the Br$\alpha$ recombination
line.  The 3.6 and 4.5 $\mu$m bands are sensitive stellar tracers.
The stellar contribution at 5.8 and 8.0 $\mu$m is relatively low
because the photospheric emission is on the Rayleigh-Jeans tail of
the energy distribution and declines rapidly with increasing wavelength.

We present IRAC images in Figs.  \ref{g10irac}--\ref{g45irac}. All the images 
are presented in Galactic coordinates and the intensity is scaled logarithmically. The
positions of radio sources are marked or the radio contours are overlaid
on the images for reference.  Equatorial coordinates are
indicated to facilitate comparison with the radio images shown in Figs. 
\ref{g10eradio}-\ref{g45radio}. The GLIMPSE Point Source Catalog
(green circles) and Archive\footnote{The GLIMPSE Catalog is more
reliable but less complete than the  Archive. The GLIMPSE Data Delivery 
documents are available at http://irsa.ipac.caltech.edu/data/SPITZER/GLIMPSE/} 
sources (red circles) are overlaid on the 3.6
$\mu$m images which show the stellar distribution in each field. UC and
HC H\,{\sc ii} regions are usually bright in the IRAC bands; thus they
are often saturated in the IRAC images. G28.20 N is saturated in the 8.0
and 4.5 $\mu$m bands, G34.26 is saturated in 8.0 and 3.6 $\mu$m bands,
G45.07 is saturated in all bands. G10.96 and G28.20 S are not saturated
in any band. The fact that the sources are saturated does not affect
the conclusions we draw in this paper.  

The bright IRAC sources suffer from the ``electronic bandwidth effect''\footnote{ 
See http://ssc.spitzer.caltech.edu/irac/features.html\#2C}. This instrumental 
effect is exhibited only in the 8.0 and 5.8 $\mu$m IRAC bands for saturated
or bright sources; it produces 
artifacts appearing as a sequence of sources of decreasing intensity.
This effect is seen in the 8.0 $\mu$m image of G28.20 N, and 5.8 $\mu$m
and 8.0 $\mu$m images of G34.26 and G45.07 NE.  
As an example, the source 
appearing to be blended with G28.20 N slightly to the south in the 8.0 
\micron~ image of Figure~\ref{g28irac} is not the S component, but rather 
is an image artifact produced by this effect.

A discussion of the IR emission from individual sources is presented in Section 4.

\section{Comments on Individual Sources}
\label{indiv}

In this section, we present the results of our RRL and mid-IR study for the
individual sources.  We also summarize the previous work on radio, 
IR, maser, and molecular line (where available) emission from each source.

\subsection{G10.96$+$0.01}
\label{g10_indiv}

The G10.96 MSFR consists of two compact radio continuum components 
G10.96 E and G10.96 W surrounded by more diffuse ionized gas. They have been
detected in radio continuum between 1.4 GHz and 8.6 GHz (\citealt{becker94};
\citealt{w98}; Paper I). In Paper I, we reported the detection of a 
broad H92$\alpha$ line with a FWHM of 43.8 $\pm$ 1.5 km s$^{-1}$ toward 
the W component. The E component has a linewidth of 33.6 $\pm$ 2.8 km 
s$^{-1}$, typical for UC H\,{\sc ii} regions. Based on the 3.6 cm
and H92$\alpha$ observations, we identified G10.96 W as an HC\,{\sc ii}
region candidate (Paper I). 

\begin{figure}
\centering
\includegraphics[width=0.47\textwidth]{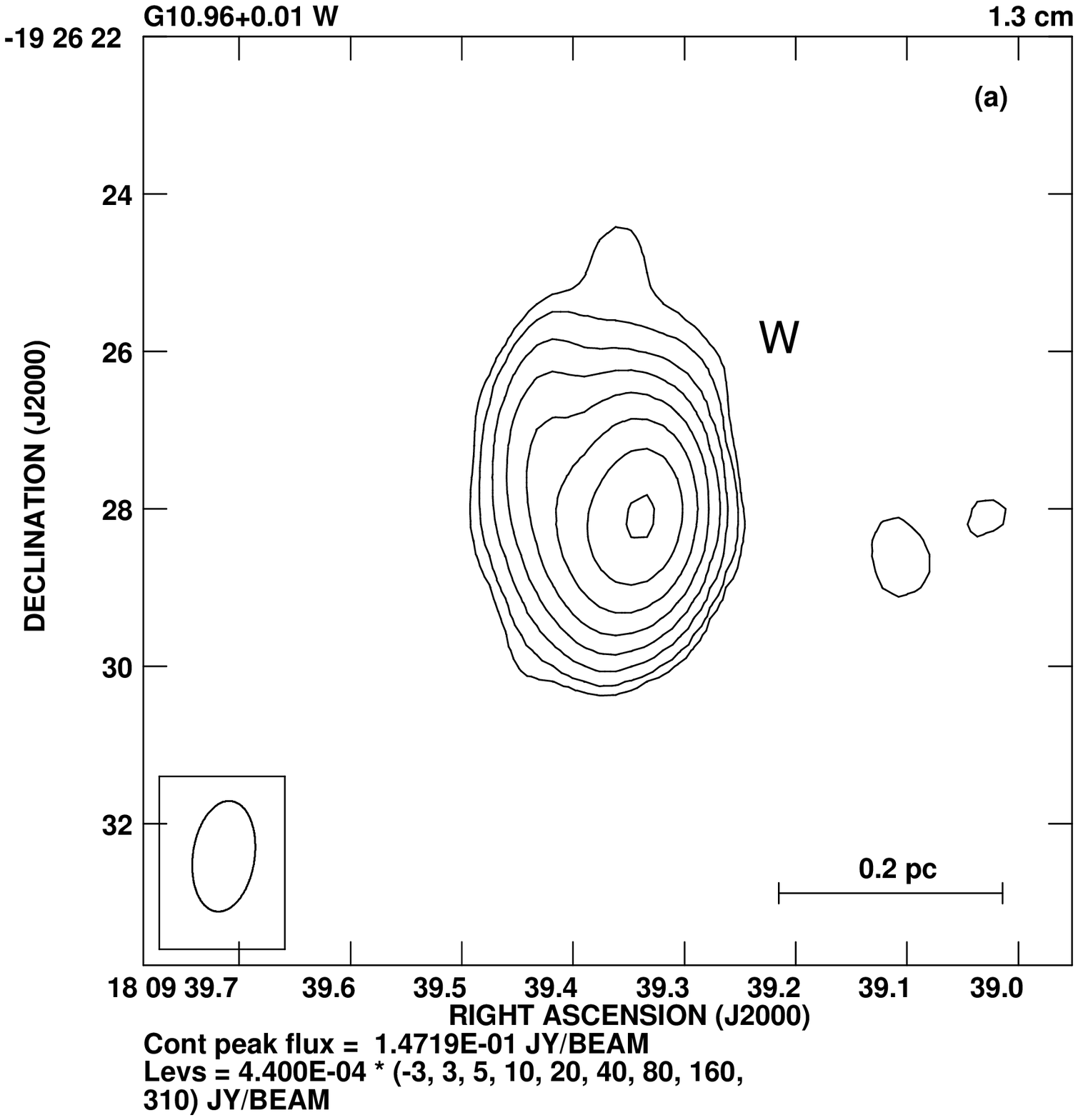} 
\hfill
\hfill
\includegraphics[width=0.47\textwidth]{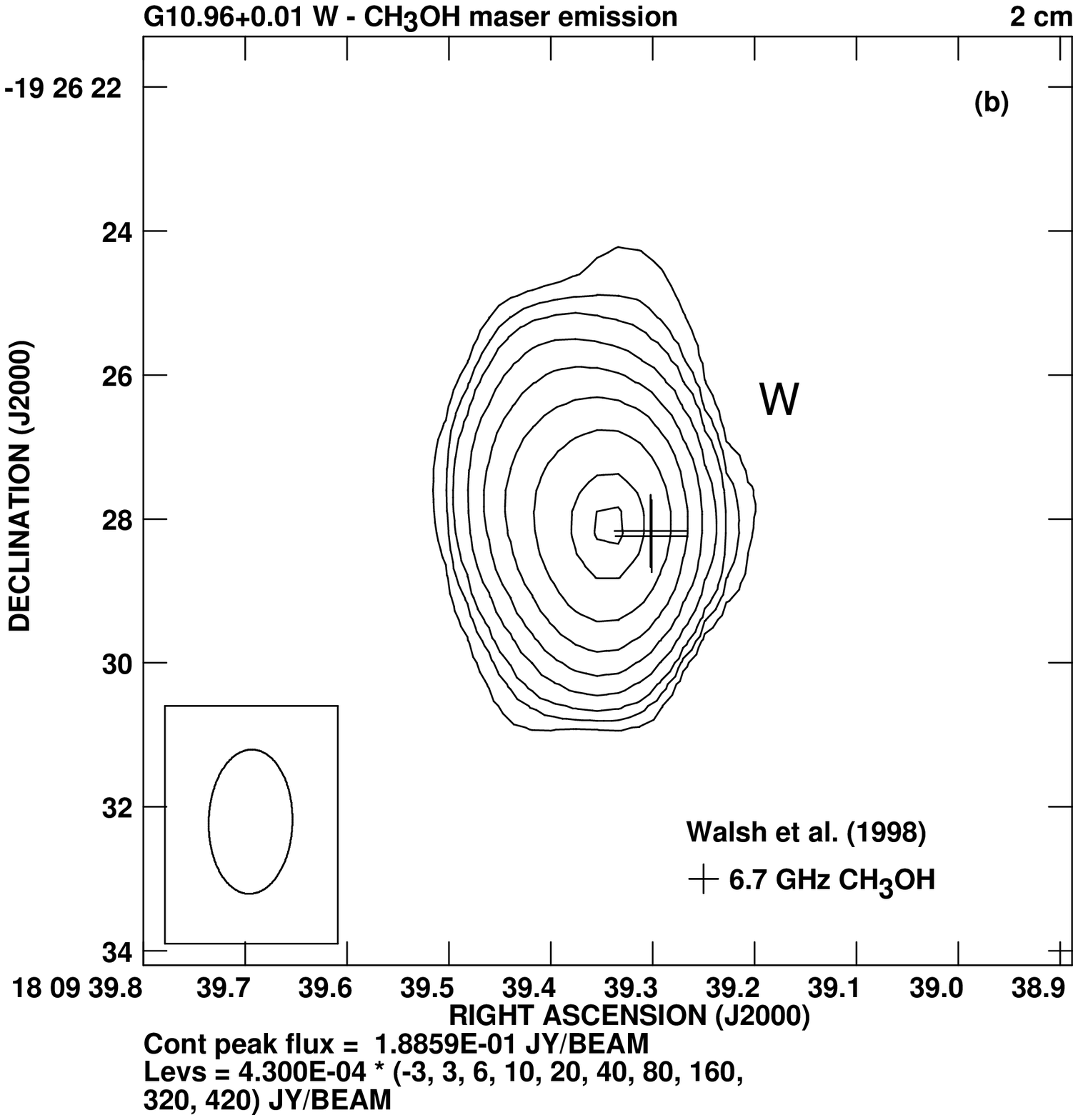}   
\hfill
\vspace*{0.6cm}
\hspace*{-0.6cm}
\includegraphics[width=0.495\textwidth]{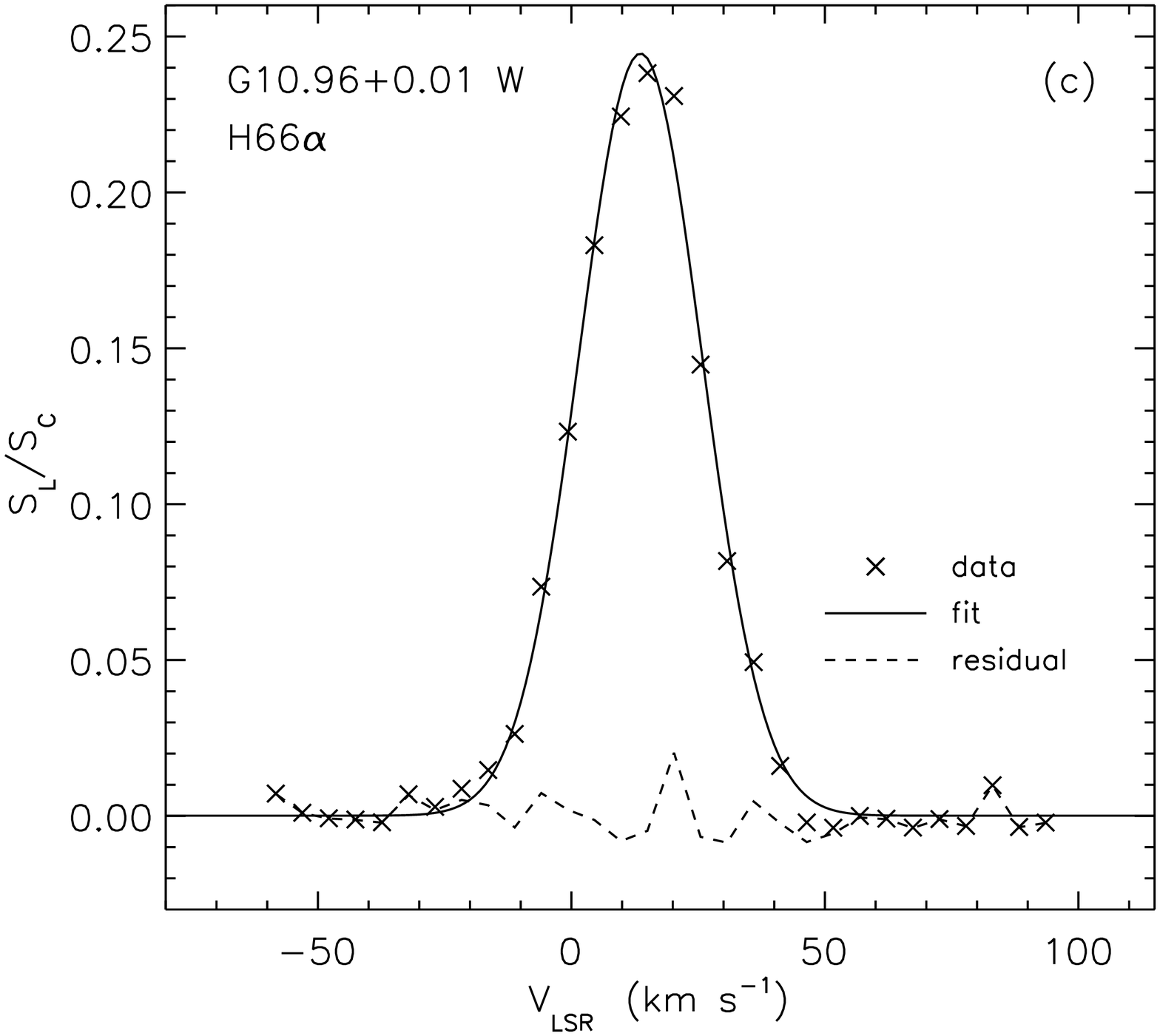}
\hfill
\includegraphics[width=0.495\textwidth]{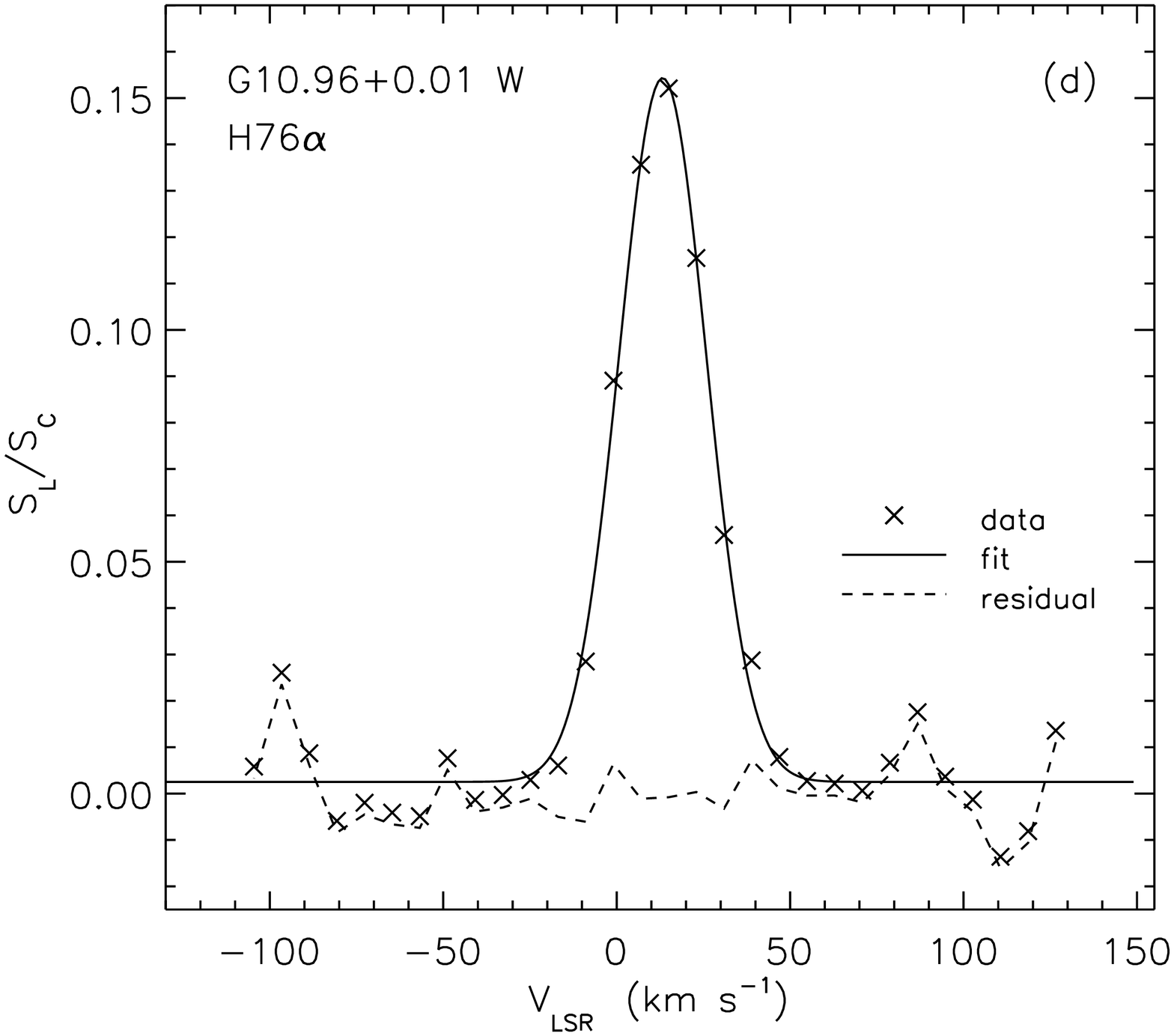}
\caption{The 1.3 (a) and 2 cm (b) VLA continuum images and the H66$\alpha$ (c) and 
H76$\alpha$ (d) line profiles integrated over the central portion of the W continuum 
component of the MSFR G10.96.  The positions of the two CH$_{3}$OH maser spots in 
G10.96 W are superimposed on the 2 cm image (b). Beam sizes are the same as in 
Figure \ref{g10eradio}.  \label{g10wradio}}
\end{figure}

\begin{figure}
\centering
\includegraphics[width=0.48\textwidth]{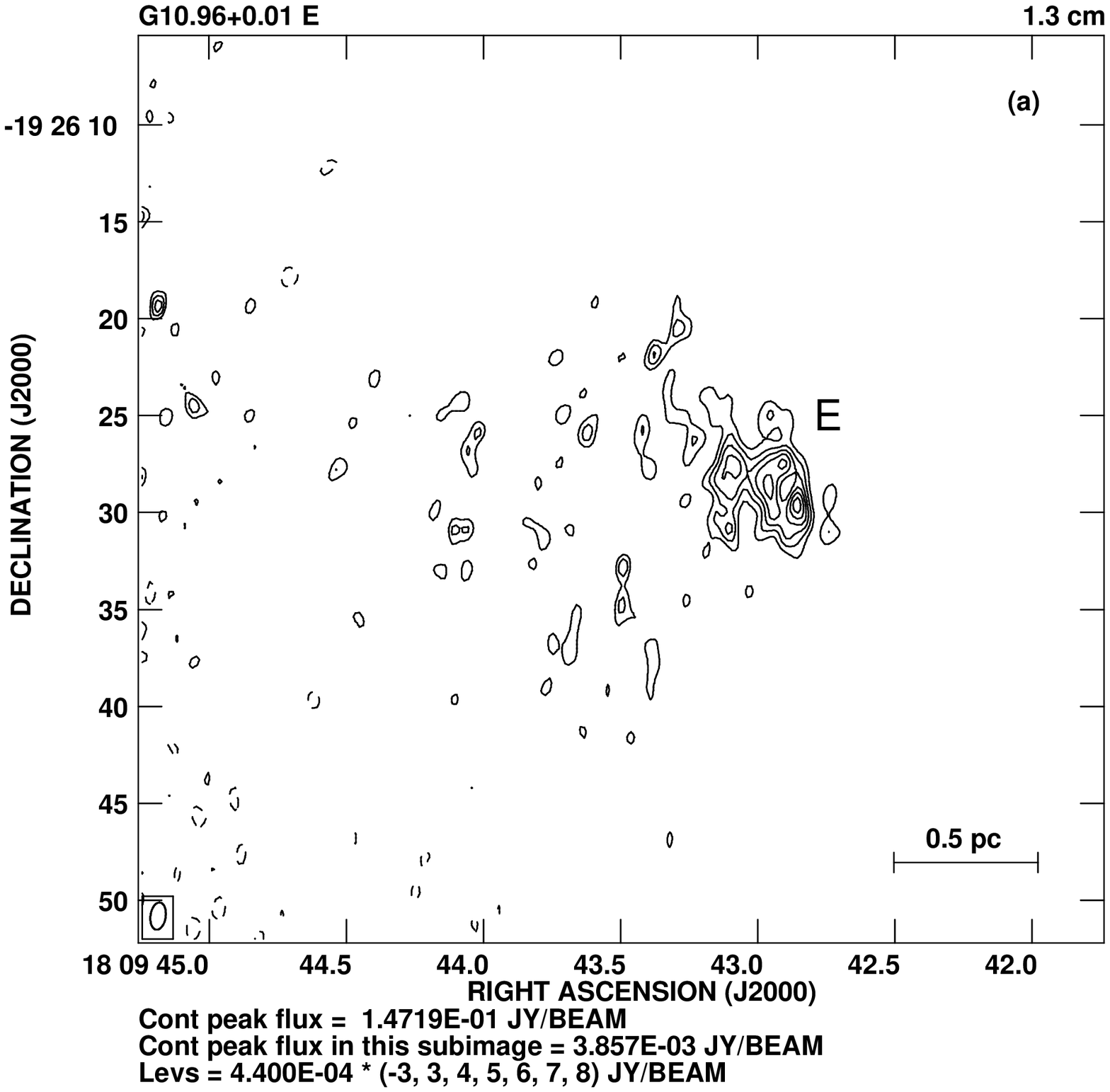}
\hfill
\hfill
\includegraphics[width=0.48\textwidth]{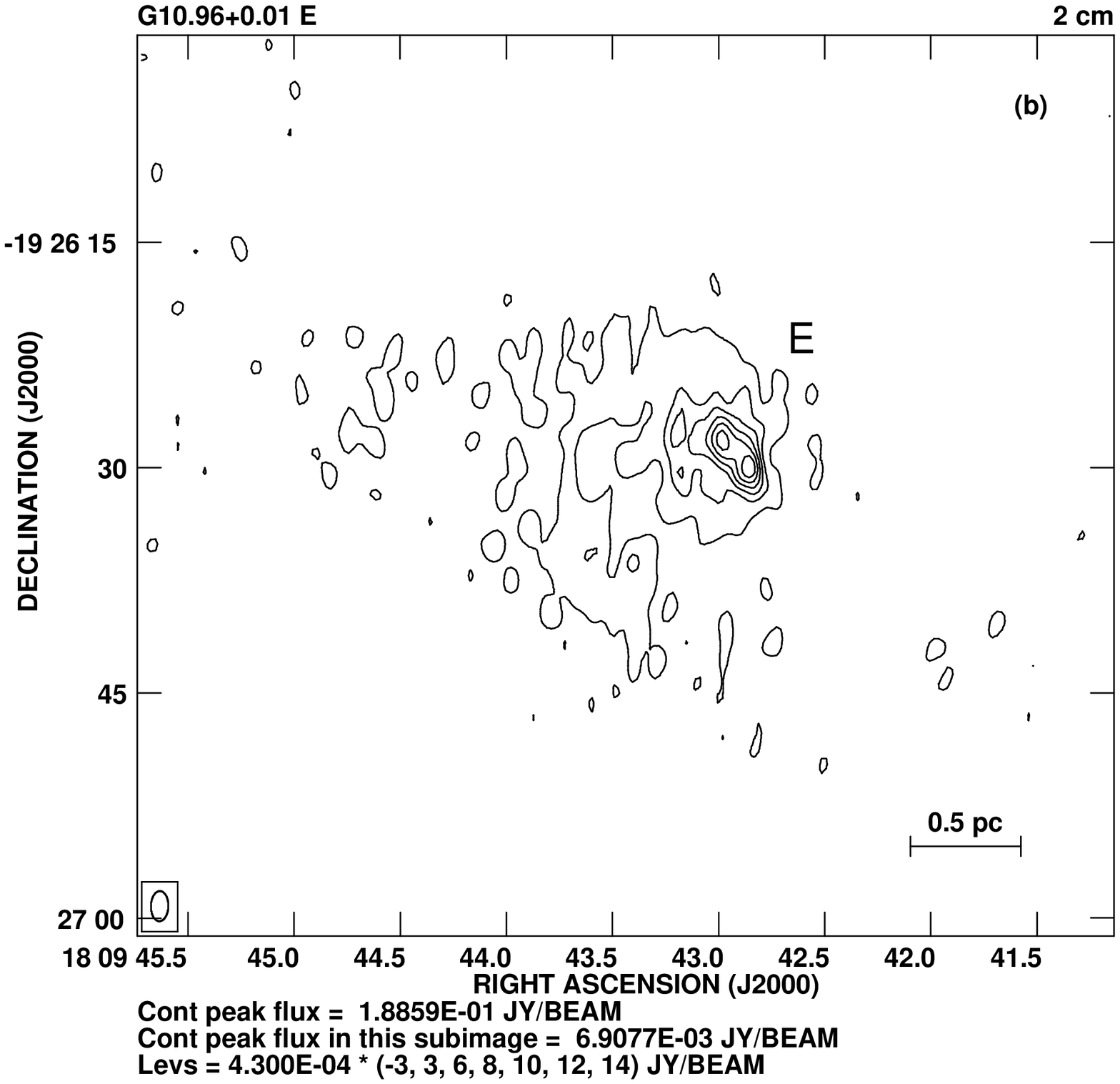}
\hfill
\hfill
\vspace*{0.6cm}
\hspace*{-0.4cm}
\includegraphics[width=0.495\textwidth]{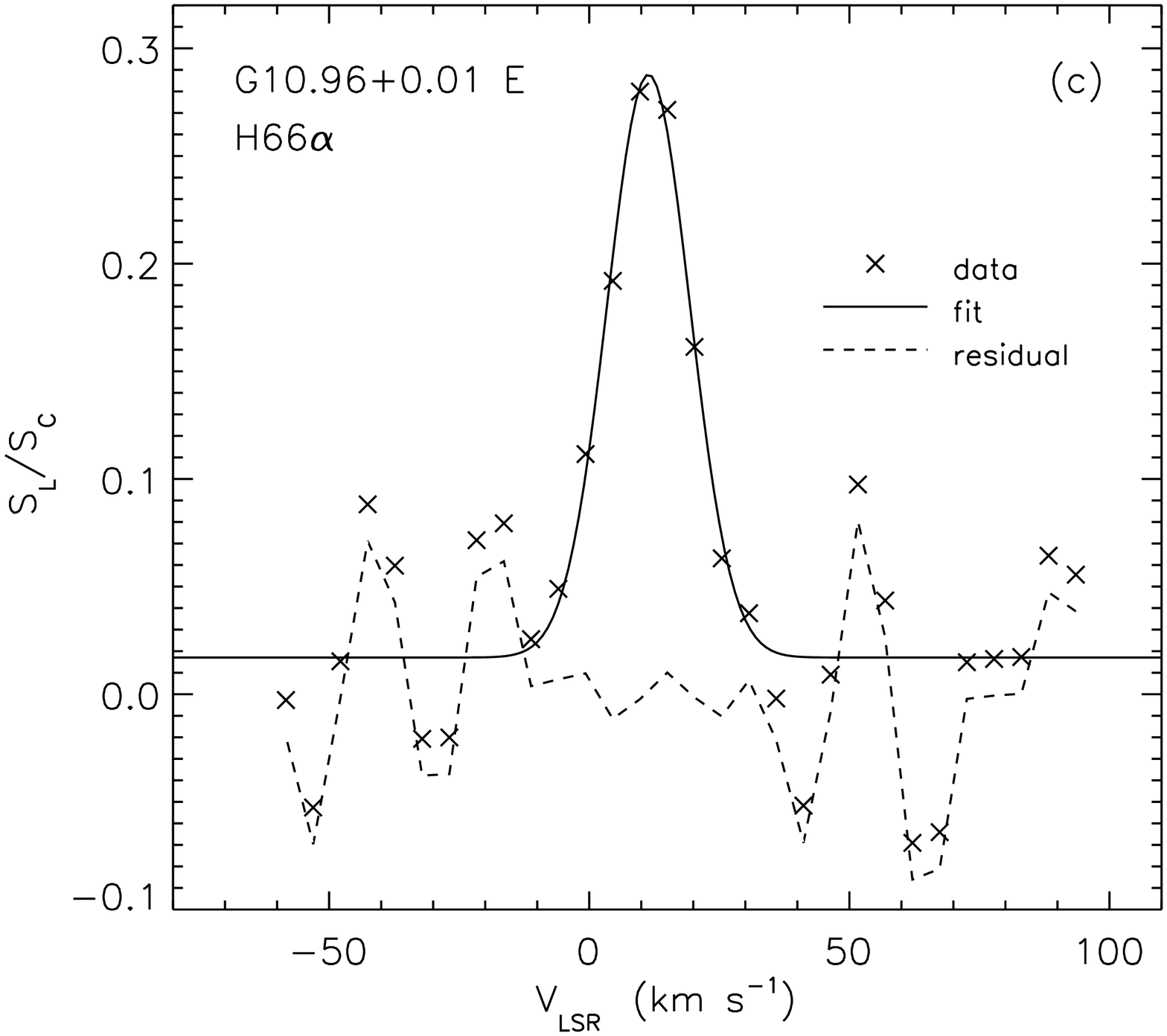}
\hfill
\includegraphics[width=0.495\textwidth]{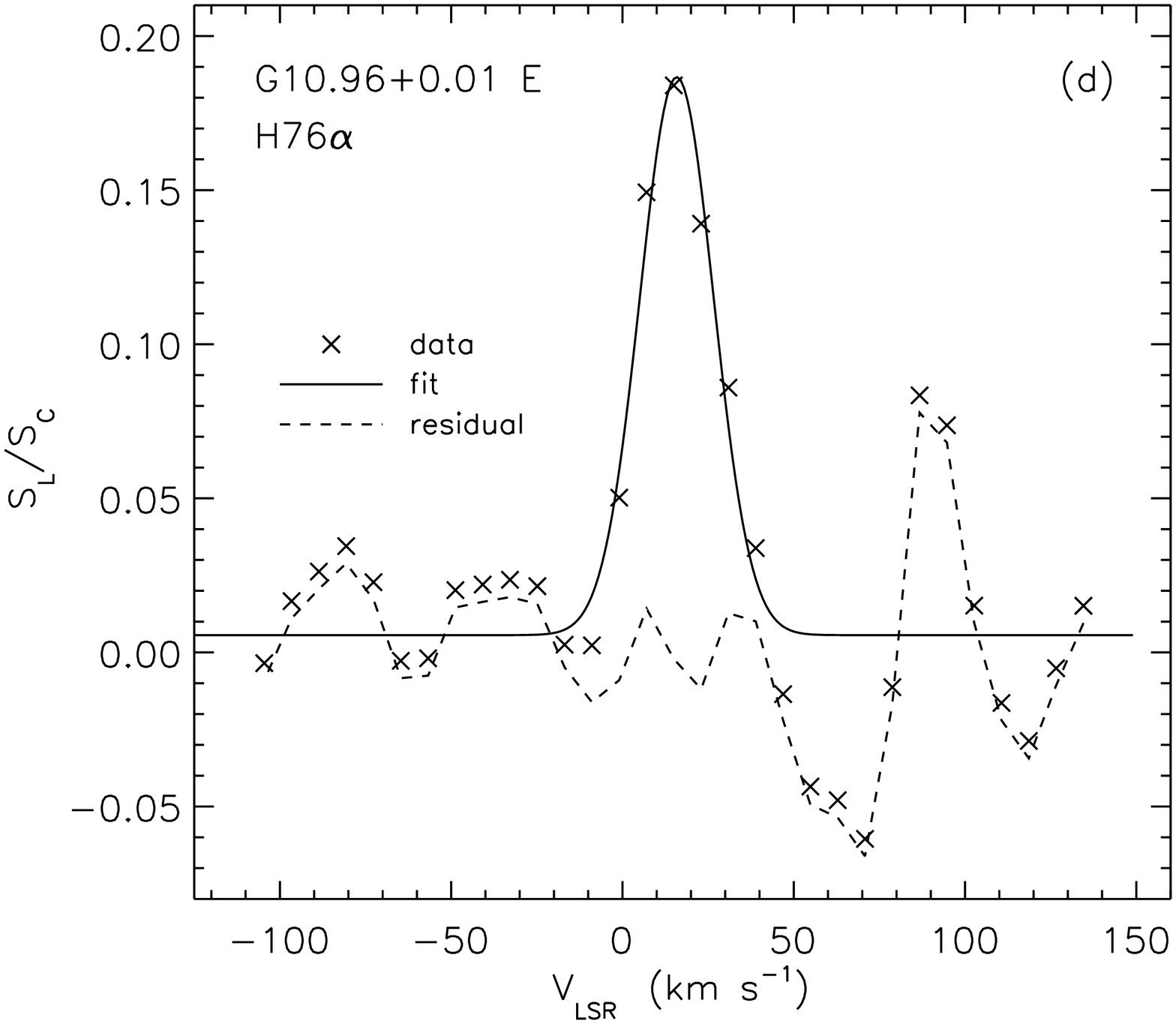}
\caption{The 1.3 (a) and 2 cm (b) VLA continuum images and the H66$\alpha$ (c) 
and H76$\alpha$ (d) line profiles from the E continuum component of the MSFR G10.96. 
The beam sizes shown in the lower left corners of the continuum images are 
$\sim$1$\rlap.{''}$1 and $\sim$1$\rlap.{''}$6 at 1.3 
and 2 cm, respectively (see Table \ref{tabinstr}). The 1.3 cm peak of maximum emission corresponds to one of 
the two peaks seen in the 2 cm image (E2 at south-west, see Table~\ref{tabpeak}).
\label{g10eradio}}
\end{figure}

The deconvolved linewidths of the H76$\alpha$ and H66$\alpha$
lines from G10.96 W reported in Table \ref{tabcorr} are 28.4 $\pm$ 1.5 
and 28.0 $\pm$ 0.6 km s$^{-1}$, respectively (see Figure~\ref{g10wradio}). These 
results indicate that the H92$\alpha$ line is strongly affected by 
pressure broadening (see the discussion in Section~\ref{discrrl}). 
For G10.96 E, both H76$\alpha$ and H66$\alpha$ lines are narrow 
with FWHMs of 19.4 $\pm$ 6.5 and 15.0 $\pm$ 4.9 km s$^{-1}$, respectively 
(see Figure~\ref{g10eradio}).
These lines are narrower than their H92$\alpha$ counterpart by over 10 km s$^{-1}$,
even if the 30\% uncertainty in the FWHMs of the higher frequency lines 
is accounted for. The difference in FWHMs over a range of
RRL transitions in a relatively low density gas indicates that  
other line broadening mechanisms may be at work in addition
to the thermal and turbulent broadening and a small contribution 
from pressure broadening (see Section~\ref{discrrl}).  Most of 
the diffuse 3.6 cm radio continuum emission observed near component E 
(Paper I; HPBW $\sim$ 2$\rlap.{''}$3) is resolved out 
in our 1.3 and 2 cm images (see Figure~\ref{g10eradio}).

Several maser species have been detected toward G10.96. Methanol
(CH$_{3}$OH) 6.67 GHz masers were first reported by \citet{s93} and
later confirmed by \citet{wh97}, both based on single-dish
observations. The high resolution interferometric survey of
\citet{w98} revealed two CH$_{3}$OH maser spots in the W component,
near the continuum emission peak (see Fig.  \ref{g10wradio}b).
Water masers and CS(2-1) emission are reported by \citet{co95} and
\citet{b96}, respectively, based on  single-dish observations. 

\begin{figure}
 \includegraphics[width=\textwidth]{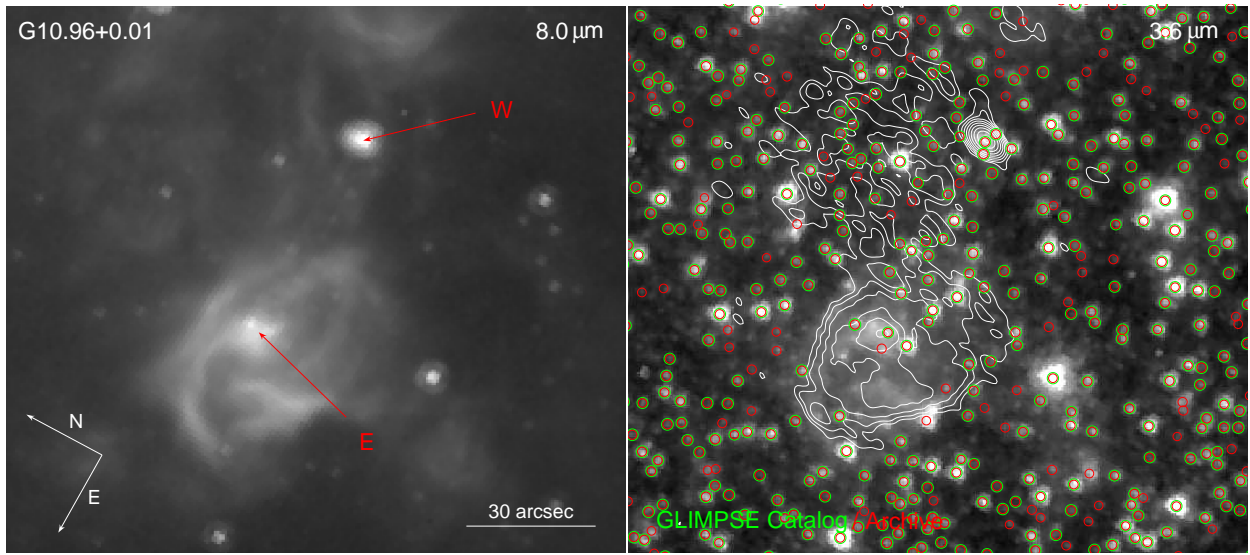}
\caption{\label{g10irac} The IRAC 8.0 ({\it left}) and 3.6 $\mu$m ({\it right}) 
images of G10.96. The positions of radio sources 
are indicated with red arrows in the 8.0 $\mu$m image, while the 3.6 cm radio 
contours are overlaid on the IRAC 3.6 $\mu$m image. The images are centered on 
the same coordinates and have the same size.  The sources from the GLIMPSE 
Catalog and Archive are marked in the 3.6 $\mu$m image with green and red circles, 
respectively. The intensity ranges from 0.5 mJy to 29.2 mJy for the 
8 $\mu$m image, and from $\sim$1 $\mu$Jy to 22.4 mJy for the 3.6 $\mu$m image.
Thirty arcesc are equal to $\sim$2 pc assuming a distance of 14 kpc 
(see Table \ref{trans}).}
\end{figure}

In Figure~\ref{g10irac} we present 3.6 and 8.0 $\mu$m images of the
G10.96 MSFR. These images indicate a close correspondence between the
distribution and morphology of dust and ionized gas. The diffuse dust
emission from G10.96 E is filamentary, with a size of
$\sim$60$''\times$40$''$ ($\sim$4.1$\times$2.7 pc at a distance of 
14 kpc; Table~\ref{trans}) in RA and Dec, respectively. This filamentary
structure is more prominent at 24 $\mu$m (MIPSGAL survey; \citealt{carey2009}), but is also
noticeable at shorter wavelengths. In the 3.6 $\mu$m image a possible 
central ionizing source (high-mass (proto)star) is prominent in the W component. 
The 3.6, 4.5, 5.8, and 8.0 $\mu$m flux densities of the
IRAC source (GLIMPSE Archive ID: SSTGLMA G010.9584+00.0219) are 
9.7 $\pm$ 0.9 mJy, 58.7 $\pm$ 7.5 mJy, 174.2 $\pm$ 12.1 mJy, and 
395.3 $\pm$ 30.7 mJy, respectively. 

The IRAC images also reveal a cluster of sources around the
radio emission peak in component E (see Figure~\ref{g10irac}). 
Some of these objects may be ionization sources; however, we can not 
rule out that some may be background objects.
The multiplicity of ionizing (proto)stars may give rise to 
the multiple-peak morphology of the UC H\,{\sc ii} region shown in 
Figure~\ref{g10eradio}.  The mid-IR source located near the 3.6 cm 
peak of G10.96 E in Figure~\ref{g10irac} corresponds to one of the 
two peaks (E2 to the south-west) of the 2 cm emission (Figure~\ref{g10eradio}b).
The GLIMPSE Archive provides the 3.6, 4.5, and 5.8 $\mu$m flux densities
for this source (SSTGLMA G010.9646+00.0098) of 8.2 $\pm$ 2.7 mJy, 
13.1 $\pm$ 4.0 mJy, and 58.9 $\pm$ 7.6 mJy, respectively.

\subsection{G28.20$-$0.04}
\label{g28_indiv}

The G28.20$-$0.04 MSFR hosts two radio continuum components: N and S.
Both components were detected in previous radio continuum (0.7 cm -- 6 cm, 
e.g. \citealt{k94, w98, s04, sew08}) and RRL (H92$\alpha$ and H53$\alpha$;
\citealt{s04,sew08}) observations. The N component has been identified as 
the HC H\,{\sc ii} region candidate by \citet{s04} based on the 3.6 cm
continuum and H92$\alpha$ RRL observations.  The H92$\alpha$ lines 
reported in Paper I have FWHMs of 74 $\pm$ 3 and 35 $\pm$ 1 km s$^{-1}$ 
for components N and S, respectively. Higher-frequency RRLs from 
G28.20 N (H53$\alpha$, \citealt{sew08}; H30$\alpha$, \citealt{k08}) are 
significantly narrower than the H92$\alpha$ line, with line widths of 
39.8 $\pm$ 1.7 km s$^{-1}$  and 20.9 $\pm$ 0.6 km s$^{-1}$, respectively.  

\begin{figure}
\includegraphics[width=0.44\textwidth]{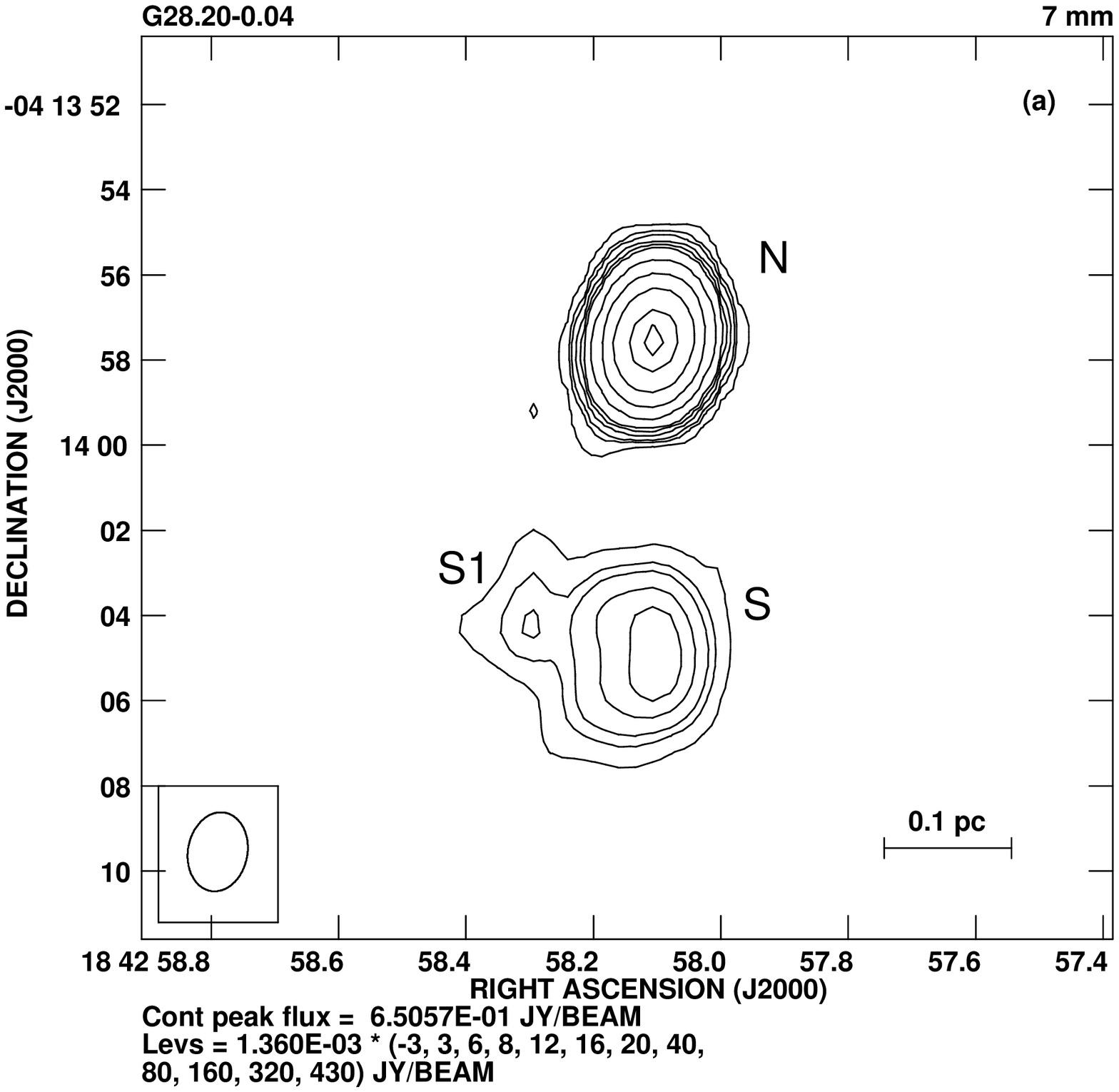}
\includegraphics[width=0.44\textwidth]{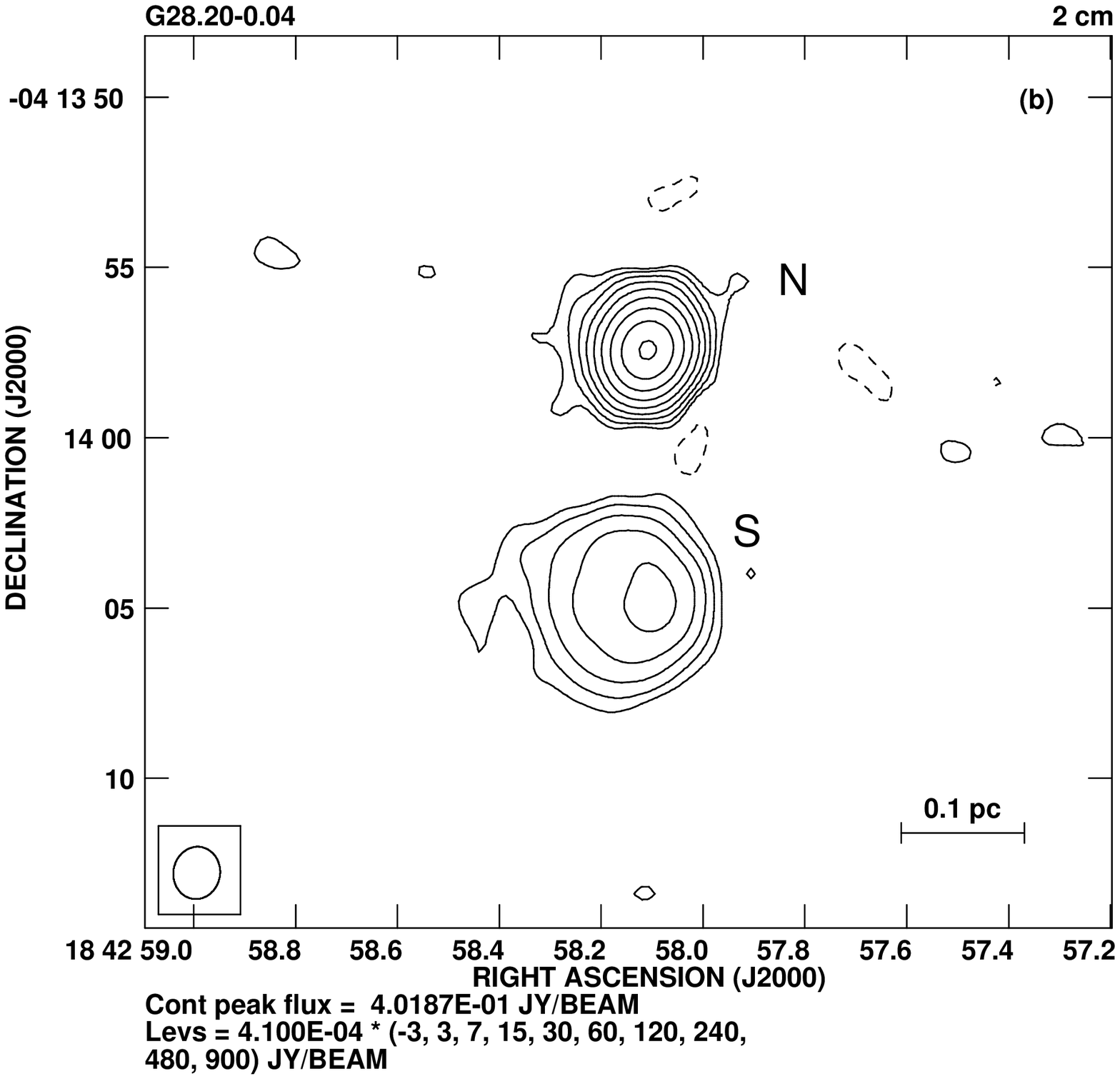}
\hfill
\hfill
\includegraphics[width=0.44\textwidth]{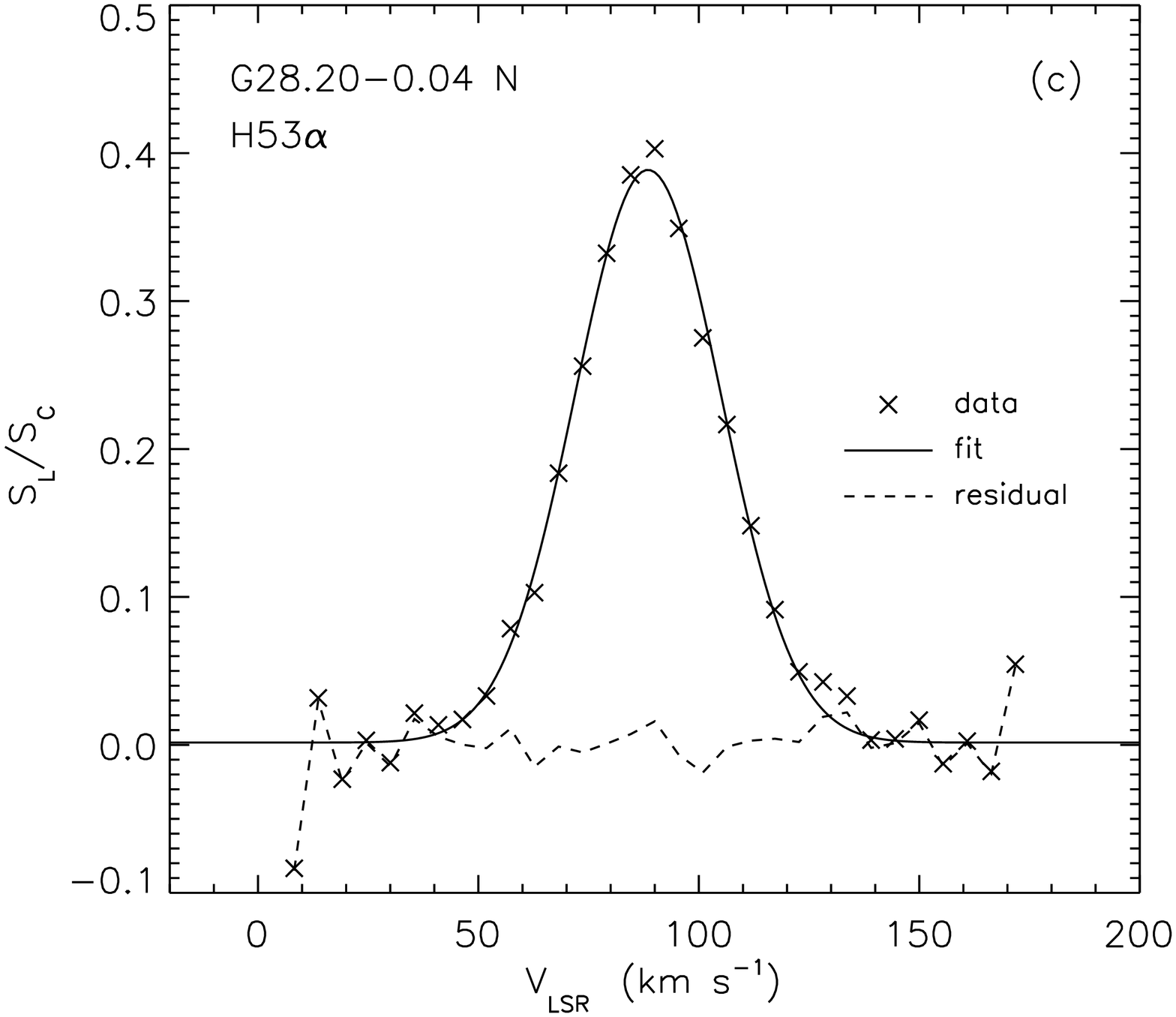}
\hfill
\includegraphics[width=0.44\textwidth]{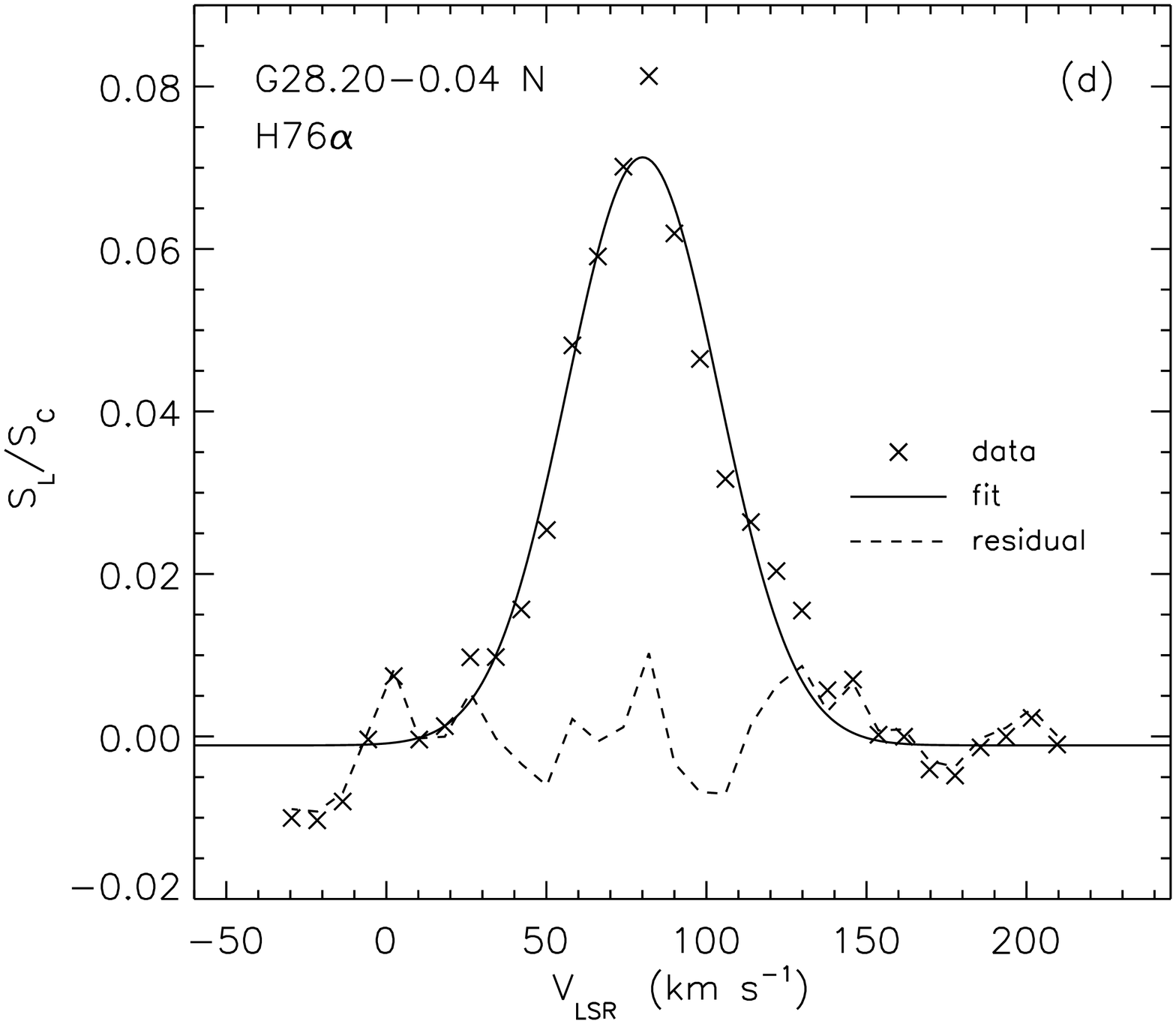}
\hfill
\hfill
\includegraphics[width=0.44\textwidth]{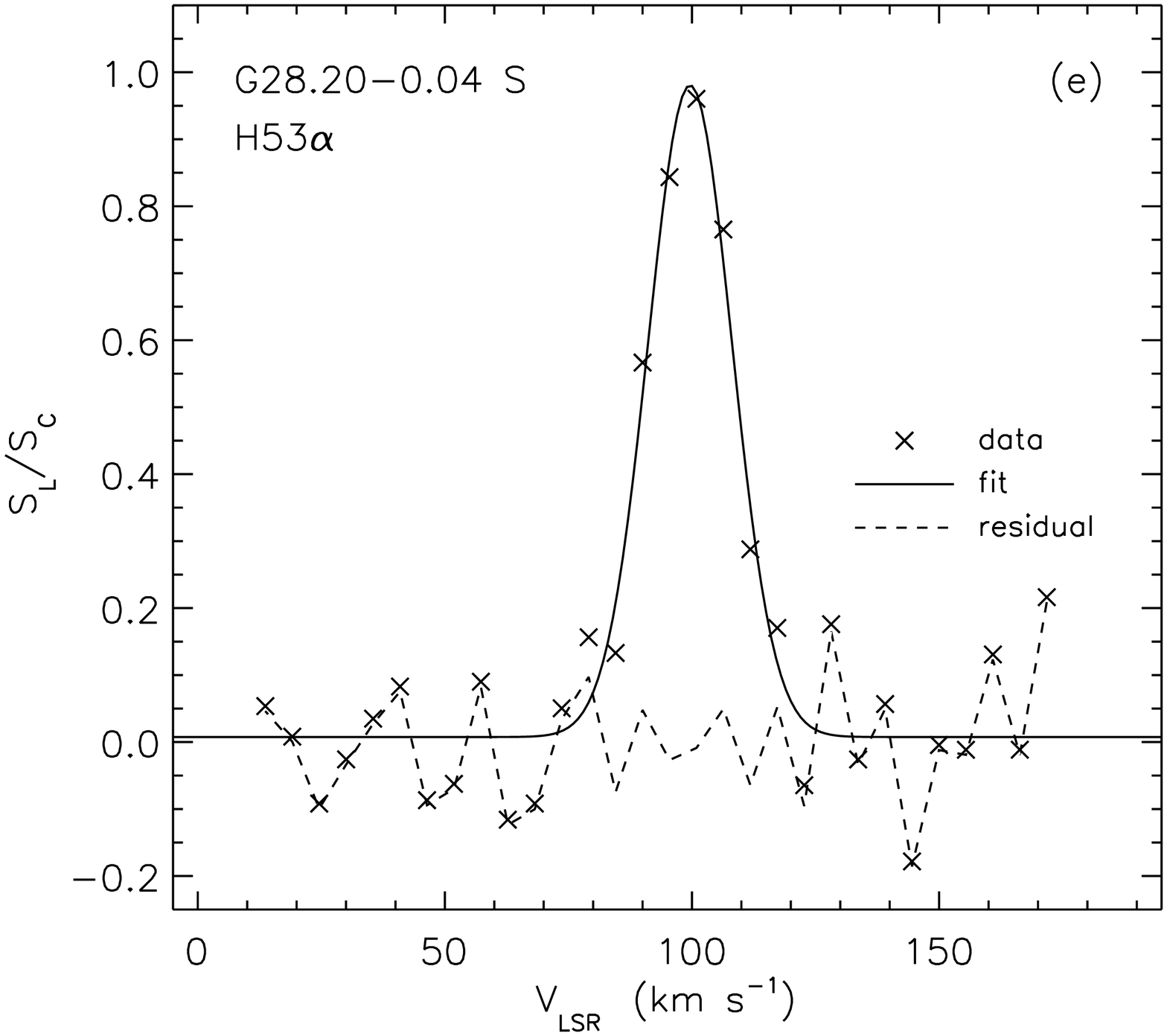}
\hfill
\includegraphics[width=0.44\textwidth]{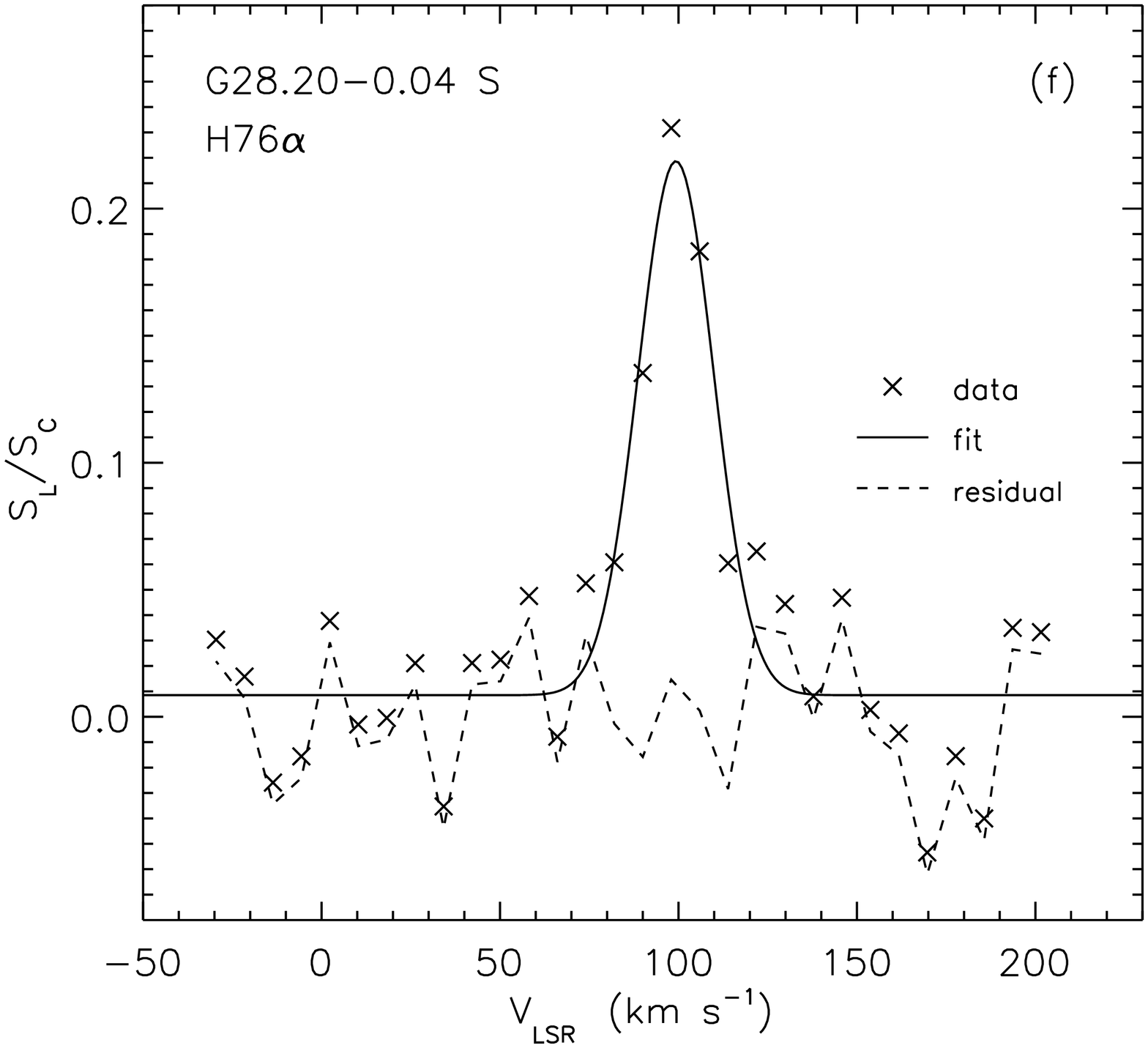}
\caption{(a) The 7 mm VLA continuum image and the integrated H53$\alpha$ line profiles 
from the N (c) and S (e) components of MSFR G28.20. (b) The 2 cm VLA continuum image 
and the integrated H76$\alpha$ line profiles from G28.20 N (d) and S (f). The 
beam sizes shown in the lower left corners in (a) and (b) are $\sim$1$\rlap.{''}$6 
and $\sim$1$\rlap.{''}$4, respectively (see Table \ref{tabinstr}). 
\label{g28radio}}
\end{figure}

\begin{figure}
 \includegraphics[width=\textwidth]{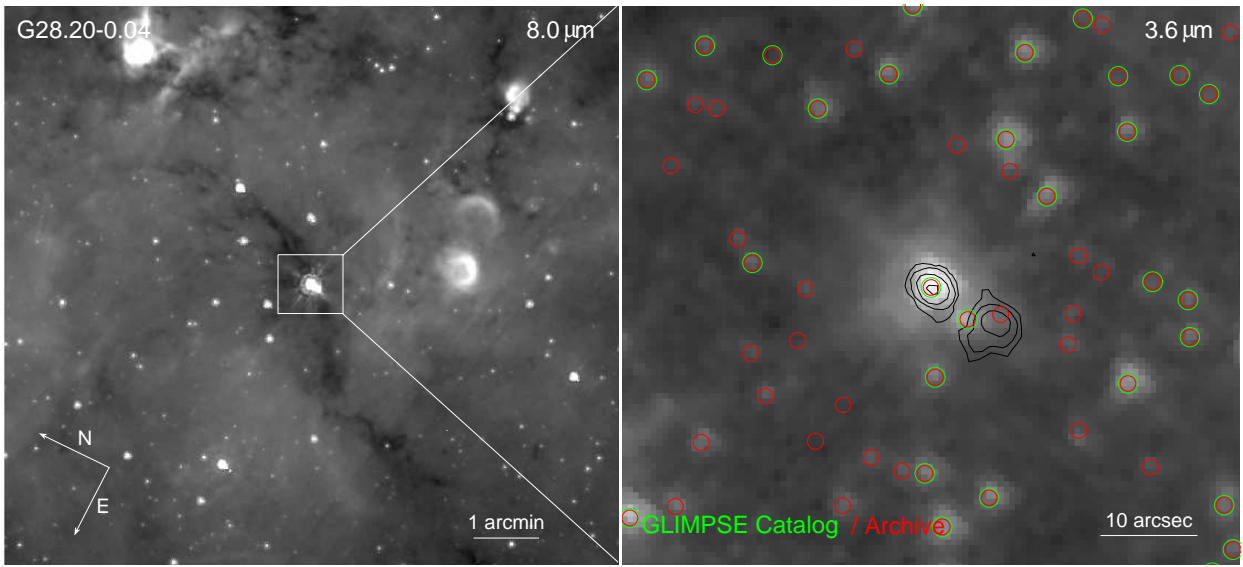}
\caption{\label{g28irac} The IRAC 8.0 $\mu$m ({\it left}) and 3.6 $\mu$m 
({\it right}) images of G28.20.   The 8.0 $\mu$m 
image shows that G28.20 is located toward an infrared dark cloud (see text). 
The 3.6 $\mu$m image is zoomed in on the source; 7 mm radio contours indicate 
the positions of radio sources (see Figure~\ref{g28radio}a). The point sources 
from GLIMPSE are marked. The source appearing to be blended with G28.20 N 
slightly to the south in the 8 \micron~ image is an image artifact produced 
by the ``electronic bandwidth effect'' (see Section~\ref{obsarch}). 
The intensity ranges from 0.5 mJy to 51.2 mJy for the 
8 $\mu$m image, and from $\sim$7 $\mu$Jy to 25.4 mJy for the 3.6 $\mu$m image.
One arcmin and 10 arcsec are equal to $\sim$1.7 pc and $\sim$0.3 pc, respectively, 
assuming a distance of 5.7 kpc (Table \ref{trans}).}
\end{figure}

In Fig.~\ref{g28radio}, we present a 2 cm and a new 7 mm continuum
image of G28.20, along with the corresponding H76$\alpha$ and
H53$\alpha$ line profiles.  Both RRLs are narrower than the
H92$\alpha$ line reported in Paper I for both the N
and S components. The H76$\alpha$ and H53$\alpha$ FWHMs are
55 $\pm$ 3  and  38 $\pm$ 1 km s$^{-1}$ for component N, and 24
$\pm$ 4 km and  19 $\pm$ 2 s$^{-1}$ for component S, respectively. 
The significant
increase in line width with increasing quantum number 
for G28.20 N indicates that pressure broadening is an
important (possibly dominant) line broadening mechanism for the
H92$\alpha$ line detected from this source.

Although not evident in Figure~\ref{g28radio}, we detect marginal
evidence for 7~mm continuum emission at the south-east border of
G28.20 N.  The emission is slightly more evident in the 6 cm
CORNISH image (C. Purcell, personal communication).  We do not
report parameters for this marginal detection, but note the possible
presence of a non-thermal source at this position.  In addition, our
continuum images --- at both 7~mm and 2~cm --- show evidence for a
new source, with a rising spectral index, just to the east of G28.20 S.
We refer to this source as G28.20 S1 (see Figure~\ref{g28radio}a), 
and report approximate values for its parameters in Table 3.

However, \citet{sew08} showed that gas dynamics also contributes to line
broadening in G28.20 N.  The morphology and velocity structure of G28.20 N 
was resolved in their high resolution ($\sim$0$\rlap.{''}$15) 7 mm continuum 
and H53$\alpha$ VLA observations. The source appears as a
shell-like structure with an inner radius of 1100 AU and an outer
radius of 2500 AU. They detected a velocity gradient of 10$^{3}$ km
s$^{-1}$ pc$^{-1}$ along the minor axis of the continuum source, from
which they inferred a rotating torus around an $\sim$30 M$_{\odot}$ 
central object. Based on line broadening and splitting along the major 
axis, they also suggested the presence of an outflow.

\subsubsection{Molecular Gas Kinematics}

G28.20 N has also been the target of 
high resolution molecular line observations by \citet{q08}, who
proposed the existence of a hot core in G28.20 N. They
used the SMA to image transitions in CH$_{3}$CN, CO, $^{13}$CO,
SO$_{2}$, OCS, and CH$_{3}$OH with $1''$ resolution. These authors derived a
CH$_{3}$CN rotation temperature of 308 $\pm$ 22 K, inferred an H$_{2}$
density $>$10$^{7}$ cm$^{-3}$, and an age of G28.20 N of 1.5 $\times$ 10$^{4}$
years.  These properties, together with the small size ($<$0.04 pc) fulfill
the standard hot core criteria (size $\lesssim$0.1 pc, density
$\gtrsim$10$^{7}$ cm$^{-3}$, and temperature $\gtrsim$100 K;
\citealt{ku00}). Spectra of SO$_{2}$, OCS, and CH$_{3}$OH have similar
radial velocities and line widths as CH$_{3}$CN; thus they likely
originate in the hot molecular core.  \citet{q08} argue that $^{13}$CO 
emission originates outside of the hot molecular core, since the excitation
temperature and H$_{2}$ density are lower than those derived from
the CH$_{3}$CN line.  The $^{13}$CO spectrum shows blueshifted
absorption and redshifted emission with the absorption
and emission peaks aligned with the source rotational axis (i.e., the SE-NW
direction; \citealt{sew08}). Based on the $^{13}$CO distribution and
line profile, \citet{q08} concluded that $^{13}$CO traces an outflow
from the molecular core over a linear scale of $\sim$0.1 pc.

The OCS position-velocity diagram for the NE-SW direction across the
continuum source G28.20 N shows two velocity peaks (95 and 97 km
s$^{-1}$) separated by 0$\rlap.{''}$2. This velocity difference
corresponds to a NE-SW velocity gradient of $\sim$400 km s$^{-1}$
pc$^{-1}$ \citep{q08}. A similar velocity pattern was observed in the
K=5 component of the \chcn (J=12-11) transition detected toward G28.20
N (S.-L. Qin, private communication).  The infall and outflow were
previously reported by \citet{so05}, based on VLA observations of
ammonia (HPBW from $\sim$0$\rlap.{''}$3 to $\sim$3$''$). 

Based on the same SMA data previously used by \citet{q08}, \citet{klaassen2009} 
reported a detection of the SO$_{2}$ and OCS velocity gradient of  $\sim$150 
km s$^{-1}$ pc$^{-1}$  in a roughly N-S direction (as opposed to the NE-SW direction
suggested by \citealt{q08}). \citet{klaassen2009} also suggested the presence of a 
CO outflow in the NE-SW direction. \citet{q08} found that the CO data are inconclusive due 
to the contamination from the infall; as mentioned above, they suggested an 
outflow in the SE-NW direction based on the $^{13}$CO data.  The detection of the 
H53$\alpha$ velocity gradient in the NE-SW direction by \citet{sew08} supports
the \citet{q08} interpretation.

\subsubsection{Maser Emission}

Numerous masers have been detected toward the G28.20 MSFR: 1665 and
1667 MHz OH \citep{a00}, 22 GHz H$_{2}$O (\citealt{h98};
\citealt{kh05}; this paper), and 6.67 GHz CH$_{3}$OH \citep{w98}. G28.20 was also
observed with arcminute resolution in the excited OH transitions at
6035 MHz (\citealt{cv95}; \citealt{c03}) and 6030 MHz \citep{c03}. 
The OH, H$_2$O, and CH$_3$OH maser emission is associated with component N
(see Fig.~\ref{g28masers}).

\subsubsection{Dust Emission}

Spitzer IRAC images of G28.20 at 8.0 and 3.6 $\mu$m are shown in Fig.
\ref{g28irac}.  The 8.0 $\mu$m image shows that G28.20 is located
toward an elongated region of high extinction (infrared dark cloud,
IRDC) stretching over at least 2$'$ ($\sim$3.3 pc at 5.7 kpc) in a
roughly NE-SW direction. This IRDC is probably the natal molecular
cloud from which the protostars in G28.20 are forming.
Figure  \ref{g28irac}b shows the 3.6 $\mu$m image of G28.20 MSFR. The
contours of the 2 cm radio emission indicate the position of the N
and S components. The mid-IR sources from the GLIMPSE Catalog/Archive
are indicated by circles. A bright source, detected in K$_{s}$-band and all 
four IRAC bands, coincides with G28.20 N. A much weaker source, detected
at 3.6 and 4.5 $\mu$m, coincides with G28.20 S.  Midway between the
two radio components, a third IR source is a near-IR (JHK$_{s}$) source
(also detected at 3.6 $\mu$m).  This source does not have a radio counterpart 
and may be a foreground object or a less deeply embedded object.

Both G28.20 N and G28.20 S have counterparts in the GLIMPSE Archive:
SSTGLMA G028.2003-00.0493 and SSTGLMAG028.1984-00.0501, respectively.
The mid-IR source corresponding to G28.20 N is significantly brighter
than the sources that matches G28.20 S. The 3.6 and 4.5 $\mu$m flux
densities are 358.4 $\pm$ 18.9 mJy and 2123.0 $\pm$ 99.3 mJy for
component N, and 2.1 $\pm$ 0.3 mJy and 8.9 $\pm$ 1.2 mJy for component S.
The 5.8 $\mu$m flux density of G28.20 N is 3892.0 $\pm$ 98.6 mJy.

\subsection{G34.26$+$0.15}
\label{g34_indiv}

Four radio continuum components (A--D) and two infrared sources
\citep[E and F,][]{ce00} have been identified in the G34.26
MSFR. Component C is a prototypical cometary UC H\,{\sc ii} region
with a compact head toward the east and a diffuse tail to the west
\citep[e.g.,][]{rh85, g86, s04}. Components A and B, located to the
east of component C, are HC H\,{\sc ii} region candidates.  A more
evolved component (D), located south-east of components A--C, is thought
to be an H\,{\sc ii} shell with a diameter $\sim$1$'$, expanding
within an ambient density gradient \citep{fg94}.


We detected H53$\alpha$ emission toward G34.26 components A, B, and C.
The line parameters are presented in Tables \ref{tabfits} and \ref{tabcorr};
line profiles and the 7 mm continuum image are shown in Figure \ref{g34radio}.

The H76$\alpha$ (Paper I) and H53$\alpha$ (this paper) lines from
G34.26 A are narrow, with FWHMs of $\sim$22 km s$^{-1}$, consistent
with a combination of thermal and turbulent broadening.  The
H53$\alpha$ linewidth of component C is $50.2\pm6.4$ km~s$^{-1}$, in
good agreement with the H76$\alpha$ linewidth of $55.5\pm1.9$
km~s$^{-1}$ reported by \citet{g86}.  These authors demonstrate that
this relatively broad linewidth probably results from microturbulent
motions having a velocity of $\sim$30 km~s$^{-1}$.  Component B has a 
broad H76$\alpha$
line of FWHM 48.4$\pm$4.4 km s$^{-1}$ (Paper I); however the
H53$\alpha$ line is $\sim$15 km s$^{-1}$ narrower (see Table
\ref{tabcorr}) indicating that the H76$\alpha$ line has a substantial
contribution from pressure broadening.  Also notable is the higher
H53$\alpha$ line center velocity for component B; the velocity is more than 15
km~s$^{-1}$ higher than for either component A or C.  It is unclear if
this velocity difference is important to understand the nature of
component B.  \citet{h01} considered component B as a strong candidate
for the driving source of the SiO outflow; if this is
the case, there may be a dynamic explanation for the higher H53$\alpha$
velocity.

If we assume no contribution to the component A and B line widths
from large scale motions and a non-thermal (turbulent) contribution
of 5 km s$^{-1}$ (FWHM; \citealt{c91}, \citealt{p97},
\citealt{sew08}), then we can determine an upper limit for pressure
broadening from the H76$\alpha$ line.  The electron temperatures for
the A and B components are 5700 K and 5300 K, respectively (Paper
I). Thus, the thermal plus turbulent contribution to the line width is
$\sim$17 and $\sim$16 \kms for components A and B, respectively. The resulting
upper limits for the pressure broadening contribution to the
H76$\alpha$ linewidth (calculated as described in \citealt{sew08}) are
$\sim$10 \kms and $\sim$43 \kms for the A and B components,
respectively.

Components A and B meet the criteria for HC H\,{\sc ii} regions
(Paper I; \citealt{ava06}; this paper). The diameters, electron
densities and emission measures derived in Paper I based on high-resolution
($\sim0\rlap.{''}5$) 2 cm observations are 0.008 pc and 0.006 pc (assuming a
distance of 3.7 kpc); 1.4 $\times$ 10$^5$ cm$^{-3}$ and 2.2 $\times$
10$^5$ cm$^{-3}$; 2.2 $\times$ 10$^8$ pc cm$^{-6}$ and 4.3
$\times$ 10$^8$ pc cm$^{-6}$, respectively.  However, both components 
A and B have a relatively low turnover frequency of $\sim$ 10 GHz \citep{ava06}.
The physical parameters derived in this paper (see Table~\ref{tabphys}) for 
component A agree with the Paper I results.  No physical parameters 
for component B can be derived because this source is partially blended with 
component C at the resolution of 1\farcs5.

\citet{ava09} resolved G34.26 A and B with $0\rlap.{''}05$ resolution
VLA observations and detected possible limb brightening in both
sources. They used spherical shell models with power-law
density gradients ($n_e \propto r^{-\alpha}$) to explain their
observations. They showed that the 7 mm intensity profiles, radio
continuum spectra, and angular sizes of both G34.26 A and B can be
reproduced by a shell of inner and outer radii $\sim$400 AU and
$\sim$1000 AU, respectively, and a density gradient $\alpha
\sim$ 0.3--1.0. This physical picture is quite similar to what
has been proposed for G24.78$+$0.08 A1 \citep{bel07} and G28.20-0.04 N
\citep{sew08}.

\subsubsection{Maser Emission}
\label{g34_indiv_masers}

 Maser emission is detected in G34.26 including H$_2$O
 \citep{hc96}, OH at 1665 and 1667 MHz (\citealt{g85}; \citealt{g87},
 \citealt{g02}), OH at 6 GHz (\citealt{cv95}), and CH$_3$OH at 6.67
 GHz (\citealt{m91}) and 44 GHz \citep{khv04}.  The H$_2$O maser spots
 are distributed ahead of and projected onto the cometary arc of
 component C, while most of the OH masers are more tightly confined to
 a parabolic arc along the eastern edge of component C
 \cite[e.g.,][]{g02}. A cluster of OH 1665 MHz masers is associated
 with component B. No OH, H$_{2}$O, or CH$_{3}$OH maser emission is 
detected toward component A.   Several of the authors cited above provide 
detailed discussions of the relation between the masers and other
physical phenomena within the MSFR.  

 High angular resolution 
molecular line observations  \citep{k92,g90,hlb89,hc96,moo07} indicate 
that the molecular gas lies roughly between  components A, B, and C (hot core).  
In particular, there is no evidence for localized molecular peaks coincident 
with components A or B.

\subsubsection{Dust Emission}

\begin{figure}
 \includegraphics[width=\textwidth]{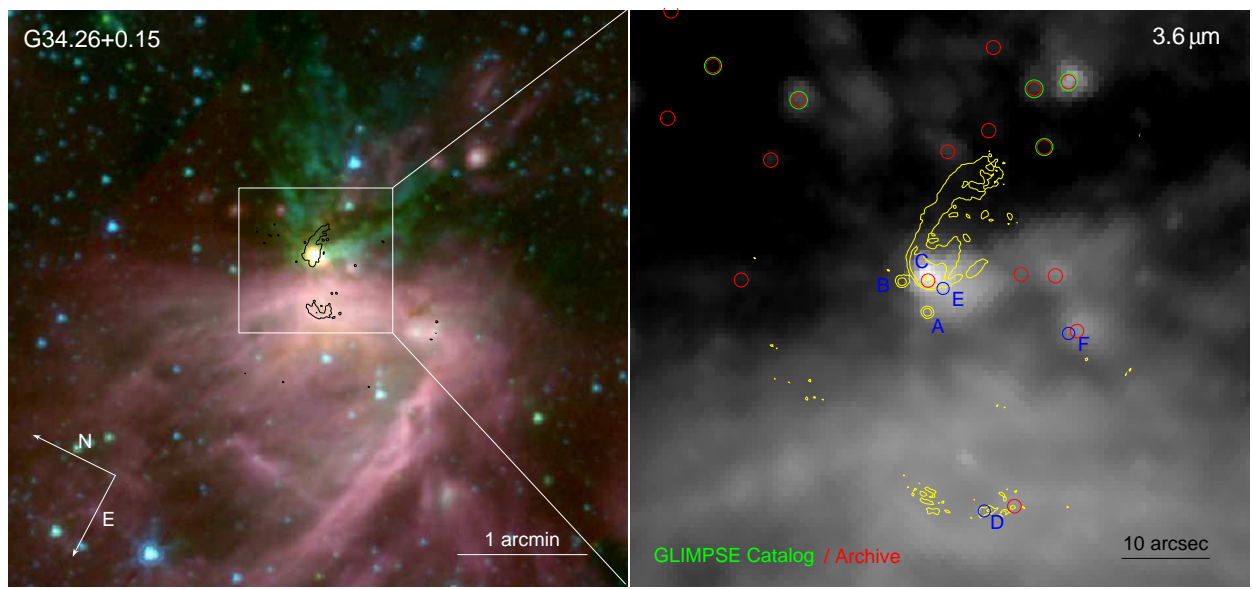} 
\caption{{\it Left:} The three color composite image of G34.26.  Red, green, 
and blue correspond to IRAC 8.0 $\mu$m, 4.5 $\mu$m, and 3.6 
$\mu$m, respectively. The 7 mm 3$\sigma$ radio  contour (see Figure  \ref{g34radio}) 
is overlaid to indicate the radio positions. A striking feature of the image is 
a massive outflow detected at 4.5 $\mu$m  (see Section 4.3.2). An incomplete ring 
of emission dominating the central and lower parts of the image corresponds to 
radio component D. {\it Right:} The IRAC 3.6 $\mu$m image of G34.26 with the 
high-resolution 2 cm radio contours \citep{s04} overlaid. Mid-IR components 
E and F from \citet{ce00} and the peak of radio D component \citep{fg94} 
are marked with blue circles and labeled.  The GLIMPSE point sources are marked.
The 8.0 $\mu$m image suffers from the ``electronic bandwidth effect'' 
(see Section~\ref{obsarch}). The intensity ranges are:  ($\sim$1 $\mu$Jy, 20.7 mJy),
($\sim$1 $\mu$Jy, 23.4 mJy), and (0.3 mJy, 55.9 mJy) for 3.6, 4.5, and 8.0 $\mu$m image,
respectively. One arcmin and 10 arcsec are equal to $\sim$1.1 pc 
and $\sim$0.2 pc, respectively, assuming a distance of 3.7 kpc (see Table \ref{trans}).  
\label{g34irac}}
\end{figure}

Mid-IR emission from components A, C, and D of G34.26 was
detected at wavelengths from 8 to 20.6 $\mu$m with resolutions $<2''$
by \citet{k92} and \citet{ce00}. Their observations failed to detect
component B, suggesting that this source is more deeply embedded. They also
showed that component C does not have a cometary morphology in the
mid-IR; the general discrepancy between mid-IR and radio morphologies
is discussed by \citet{h07}.  \citet{ce00} detected two additional
sources, which they designated E and F. Component F is located about
17$''$ south of component C in the 20.6 $\mu$m image; component E is
located 2$''$ south of and is blended with component C \citep{ce00}.
Mid-IR sources C and E coincide with the arc of the cometary H {\small
  II} region (see Fig.  \ref{g34irac}) and were resolved into four
individual sources in the sub-arcsecond resolution 10.5 $\mu$m and
18.1 $\mu$m images of \citet{db03}.

The IRAC images of G34.26 illustrate how active and dynamic this MSFR
is. Fig.  \ref{g34irac} shows the 4.5 $\mu$m and 3.6 $\mu$m images. The
8.0 and 5.8 $\mu$m emission is similar to that shown in
Fig. \ref{g34irac} except that the outflow is only detected at 4.5
$\mu$m (see below).  The IR counterparts of radio components A and C
cannot be distinguished from one another, owing to very bright,
diffuse emission in the IRAC bands. Also, IR components E and C are
blended. IR component F was detected in all
bands. The large ring of emission south-east of components A-C (see
left panel of Fig. \ref{g34irac}) corresponds to the D radio
component of the G34.26 complex. Comparison of the 2 cm image of
\citet{fg94} with the IRAC images shows that this IR ring has roughly
the same angular extent ($\sim$1$'$ radius) and brightness
distribution (75\% complete ring of emission with an opening to the
NE) as at radio wavelengths.

The IRAC images reveal multiple IR sources that were not detected by
earlier infrared observations. Nevertheless, component B remains
undetected, even in the IRAC images. The point sources listed in the GLIMPSE
Catalog and Archive are shown in Fig.  \ref{g34irac}. The
high-resolution 2 cm radio contours (Paper I) are overlaid on the 3.6
$\mu$m image to indicate the positions of the A--C radio
components; IR sources E and F are also marked. Only
G34.26 C has a counterpart in the GLIMPSE Archive (SSTGLMA
G034.2572+00.1533) with flux densities of 4857 $\pm$ 278 mJy and
4481 $\pm$ 155 mJy at 4.5 and 5.8 $\mu$m, respectively.

One of the most prominent features in the color composite image  of
G34.26 (Fig. \ref{g34irac}) is an outflow detected as an Extended Green Object (EGO,
Cyganowski et al. 2009).  Molecular outflows are particularly strong
in the IRAC 4.5 $\mu$m band because of shocked H$_2$ and/or CO line
emission (Cyganowski et al. 2008).  The source of the outflow is
unknown.  Moreover, other outflows appear in
SiO (Hatchell et al. 2001) and $^{13}$CO (Matthews et al. 1987).
The latter two outflows are not well-aligned with the EGO, suggesting
that they are different outflows arising from different driving sources.

\subsection{G45.07$+$0.13}
\label{g45_indiv}

\begin{figure}
\includegraphics[width=0.49\textwidth]{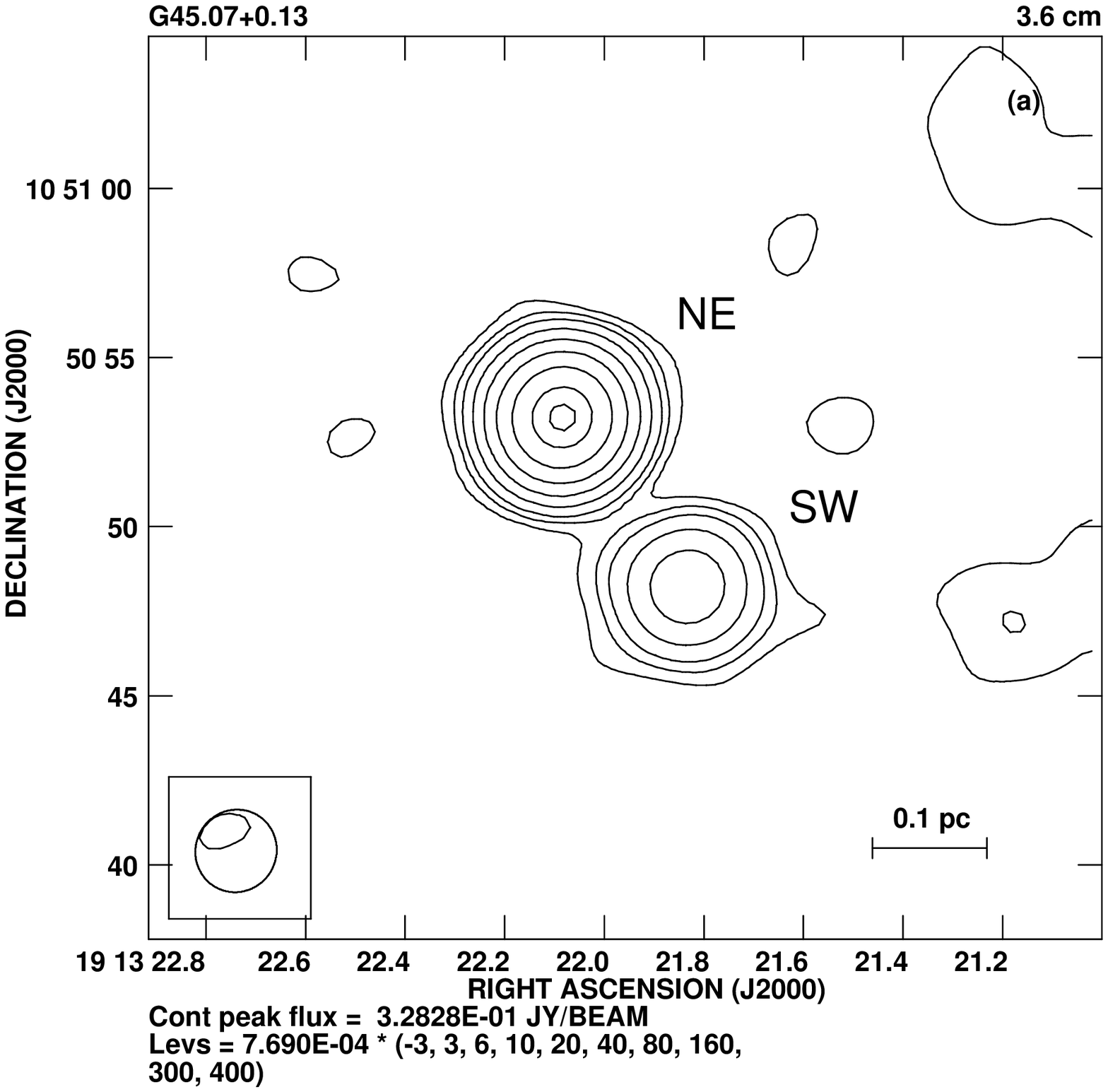}
\hfill
\includegraphics[width=0.46\textwidth]{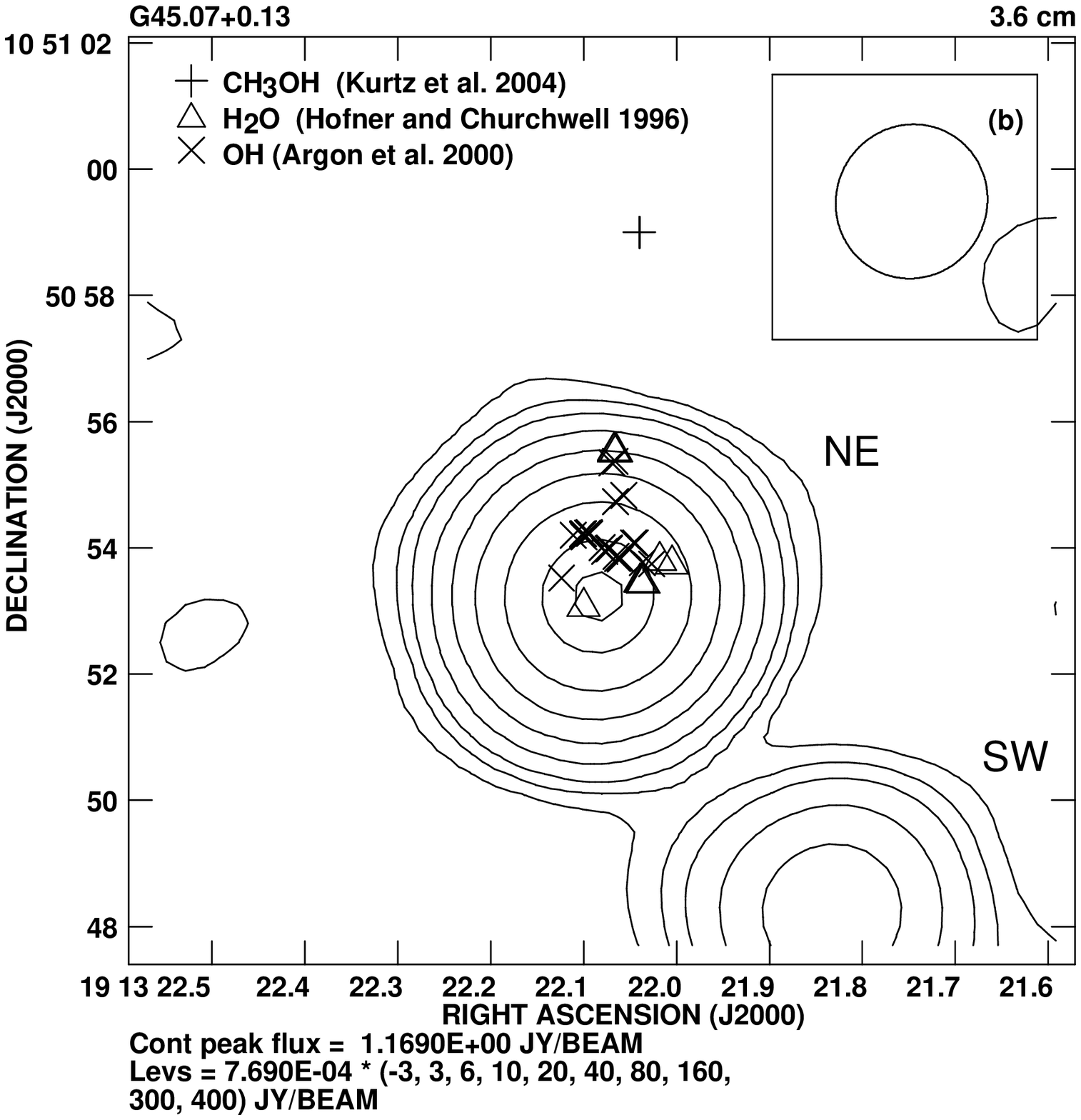}
\caption{(a) The 3.6 cm VLA continuum image of the NE and SW components of 
the MSFR G45.07.  (b) The magnification of (a) centered on G45.07 NE. 
Locations of 22 GHz H$_{2}$O, 44 GHz CH$_{3}$OH, 1665 MHz and 1667 MHz 
OH masers in the vicinity of G45.07 NE are indicated. 
The VLA beam  is shown in both images (HPBW$\sim$2$\rlap.{''}$5, 
see Table~\ref{tabinstr}). \label{g45radio}}
\end{figure}

G45.07 consists of a bright NE component (S$_{2 cm}$ = 620 mJy) and a
much fainter SW component (S$_{2 cm}$ = 40 mJy), separated by
$\sim$6$''$ (see Fig.  \ref{g45radio}). The SW component was first
detected by \citet{g86} at 2 cm and designated G45.07 A. Both NE and
SW components are unresolved in our radio continuum maps. High
resolution (HPBW $\sim$0$\rlap.{''}$1) observations by \citet{tm84} of
the NE component revealed a clumpy ring with a gap to the west which
they interpreted as a shell with an inner cavity produced by a stellar
wind from a hot central star. However, \citet{g86} suggested that the
ring-like structure may be the ionized inner wall of an accretion
disk.  They further suggest that the ring is expanding at $\sim
10$~km~s$^{-1}$ (owing to a hot stellar wind) and is inclined
by $\sim$75$^{\circ}$ to the line of sight. The H76$\alpha$ RRL
integrated over the whole NE component is broad with a FWHM of 48.1
$\pm$ 0.9 km s$^{-1}$ \citep{g86}. \citet{g85} also detected broad
H66$\alpha$ emission with a FWHM of 42.3 $\pm$ 2.3 km
s$^{-1}$. \citet{k08} reported a detection of the H30$\alpha$ line
toward the NE component with FWHM of 33.2 $\pm$ 4.2 km s$^{-1}$.

We detected H92$\alpha$ emission toward both the NE and SW components
of G45.07; however, we were not able to determine reliable line
parameters and physical properties of the sources based on these
data. The H92\al line observed toward the SW component is very narrow
relative to the channel width (16.9 km~s$^{-1}$) preventing a reliable
Gaussian analysis. We estimate that the peak
line-to-continuum ratio of this line is $\ga$ 0.07. For the
NE component, the H92\al line has a very poor signal-to-noise ratio, likely
the result of the high optical depth which attenuates the
line. The peak line-to-continuum ratio for this line is $\sim$0.03 and
the deconvolved FWHM of 40 $\pm$ 9 \kms.

\citet{h97} reported a molecular outflow centered on G45.07 NE based
on observations at millimeter and submillimeter wavelengths and in the
CO(6-5) and CS(2-1) molecular lines. Infall was also detected by
redshifted CS(2-1) absorption, confirming the young age of this
source. The outflow axis is aligned spatially and kinematically with
the H$_{2}$O maser spots in G45.07 NE \citep{hc96}, suggesting that
the masers form in warm and dense shocked gas at the inner
edges of the outflow lobes \citep{h97}.

Several masers have been detected in the NE component of G45.07, while
no maser emission is associated with the SW component (see
Fig. \ref{g45radio}b). \citet{hc96} detected four clusters of water
masers in the area, three of which are closely associated with the
cm continuum emission from the NE component. The fourth group is
located about 2$''$ North of the NE component and is not coincident
(in high resolution images)
with any radio continuum emission. The H$_{2}$O masers are intermixed
with 1665 and 1667 MHz OH masers at these two locations
(\citealt{g85}; \citealt{a00}).  A 44 GHz CH$_{3}$OH maser was
detected $\sim$6$''$ north of the NE component by \citet{khv04}.
Single-dish observations by \citet{m91} and \citet{ce95} revealed
methanol 6.67 GHz masers.

\subsubsection{Mid-Infrared Emission}

High-resolution observations (from 1$\rlap.{''}$1 to 1$\rlap.{''}$7;
\citeauthor{db03} \citeyear{db03}, \citeyear{db05}; \citeauthor{k03}
\citeyear{k03}) at mid-IR wavelengths from 10.5 $\mu$m to 20.6 $\mu$m,
revealed three mid-IR components associated with the G45.07 MSFR,
designated KJK 1, 2, and 3 by \citet{k03}. The brightest component,
KJK 1, is coincident with G45.07 NE and contributes roughly half of
the total (KJK 1--3) flux at 12.5 $\mu$m and 20.6 $\mu$m
\citep{k03}. The second brightest component, KJK 2, lies 2$''$ north
of KJK 1, coincident with the water maser group 2$''$ north of the NE
component \citep[see discussion above]{hc96}; this source is 
a good candidate for a hot molecular core (HMC). Based on
the mid-IR luminosities, \citet{db03} estimated that the putative HMC 
in G45.07 would contain an O-type star if it is heated by
a single star. KJK 2 may drive the outflow reported by \citet{h97}.
The third mid-IR component, KJK 3, corresponds to the SW component of
G45.07.

\begin{figure}
 \includegraphics[width=\textwidth]{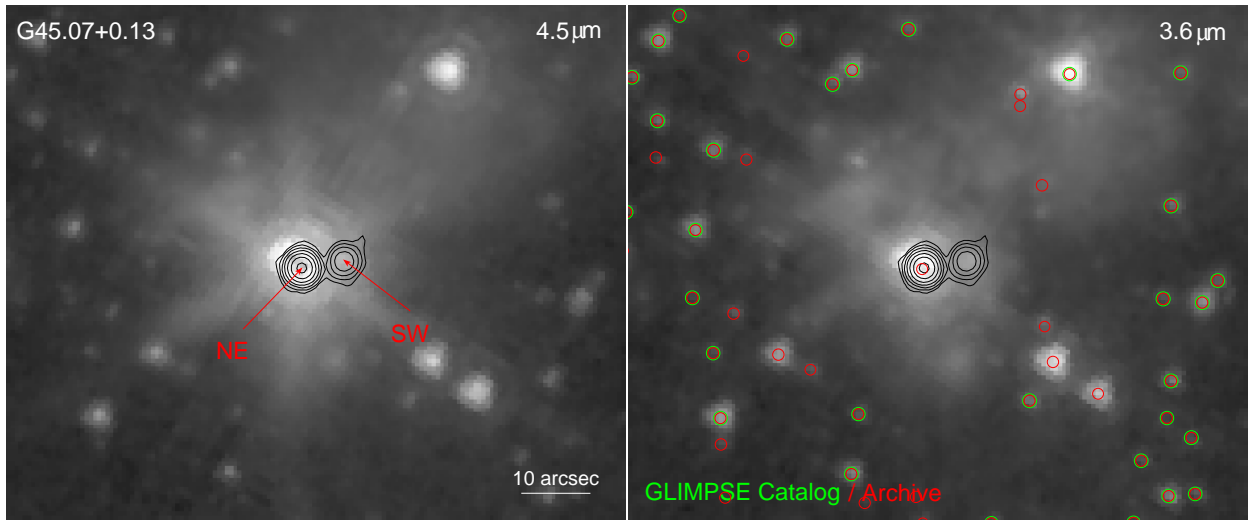}
\caption{The IRAC 4.5 $\mu$m ({\it left}) and 3.6 $\mu$m ({\it right}) images of G45.07. 
The images have the same size. The 3.6 cm radio 
contours are overlaid on both images and the NE/SW components are indicated with red 
arrows in the 4.5 $\mu$m image.  Sources from the GLIMPSE Catalog and Archive are marked 
in the 3.6 $\mu$m image with green and red circles, respectively. 
The intensity ranges from $\sim$1 $\mu$Jy to 28.8 mJy for the 
4.5 $\mu$m image, and from $\sim$1 $\mu$Jy to 21.6 mJy for the 3.6 $\mu$m image.
Ten arcsec are equal to 0.3 pc at a distance of 6 kpc (see Table \ref{trans}). 
\label{g45irac}}
\end{figure}

GLIMPSE 3.6 and 4.5 $\mu$m IRAC images of the G45.07 MSFR are
presented in Fig.  \ref{g45irac}. At IRAC bands, G45.07 is a 
bright source.  The
KJK 1 and 2 sources are not well-resolved in these images but are
clearly apparent in Fig.  \ref{g45irac}. KJK 3 is very faint at
shorter IRAC bands, but is bright at 8 $\mu$m where PAH emission 
dominates.

The GLIMPSE Catalog and Archive sources shown in Fig.  \ref{g45irac}
reveal the positions of (proto)stars that may be responsible for
exciting PAH emission.  In general there is a good correspondence
between the distribution of dust and gas in G45.07; however the dust
emission extends beyond the region of ionized gas emission at 3.6 cm
because the PAH contribution to this band traces the PDR that surrounds
the nebula. The correspondence in projected emission does not necessarily
imply that ionized gas and dust are well mixed. Additionally, the VLA 
filters out extended emission and much of the extended IR emission is 
instrumental (sidelobes of the {\it Spitzer} point spread function),
making the interpretation of the data more difficult.

\section{Discussion}

\subsection{Radio Recombination Lines}
\label{discrrl}

\begin{figure}
\includegraphics[width=0.49\textwidth]{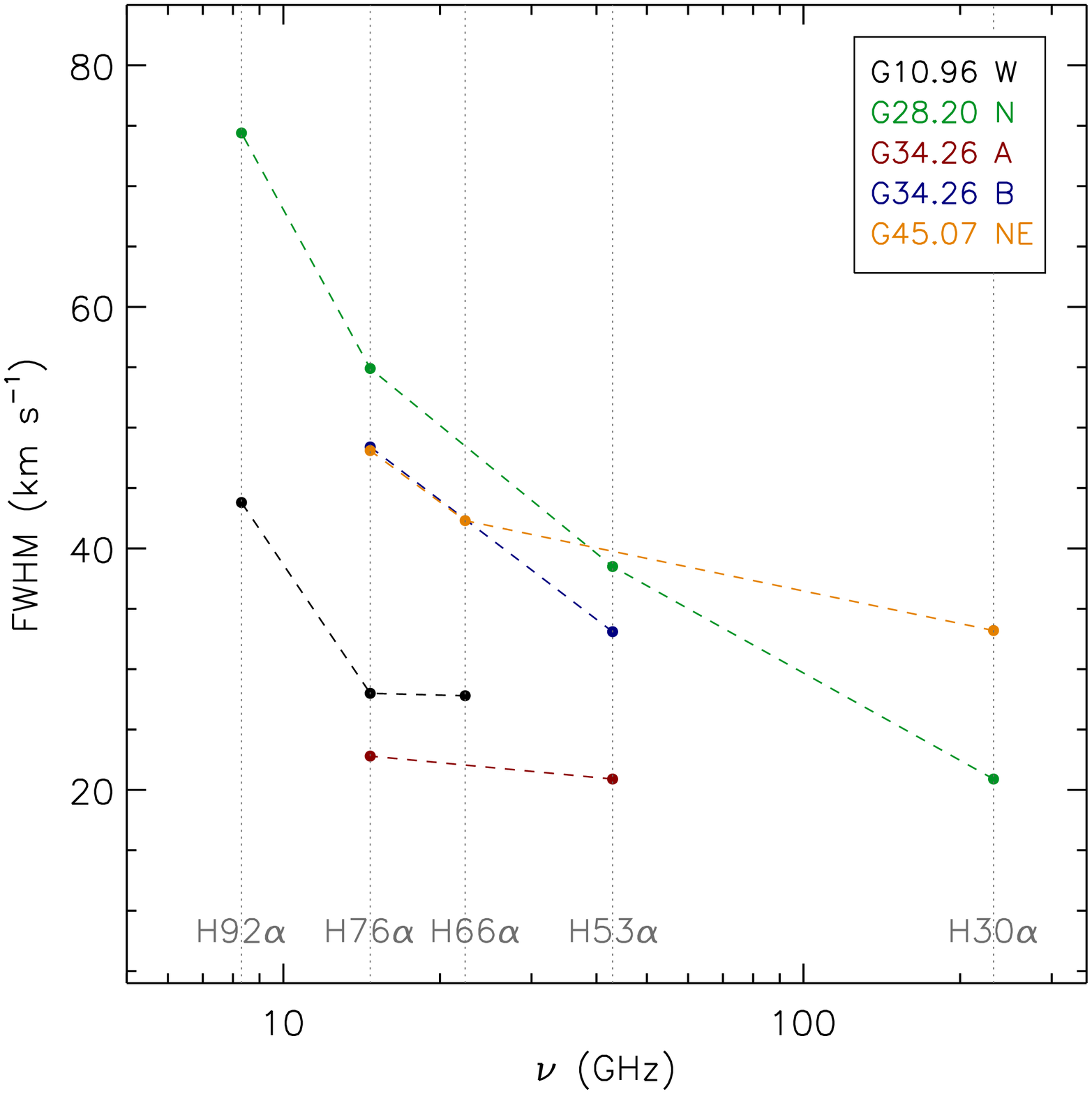}
\includegraphics[width=0.49\textwidth]{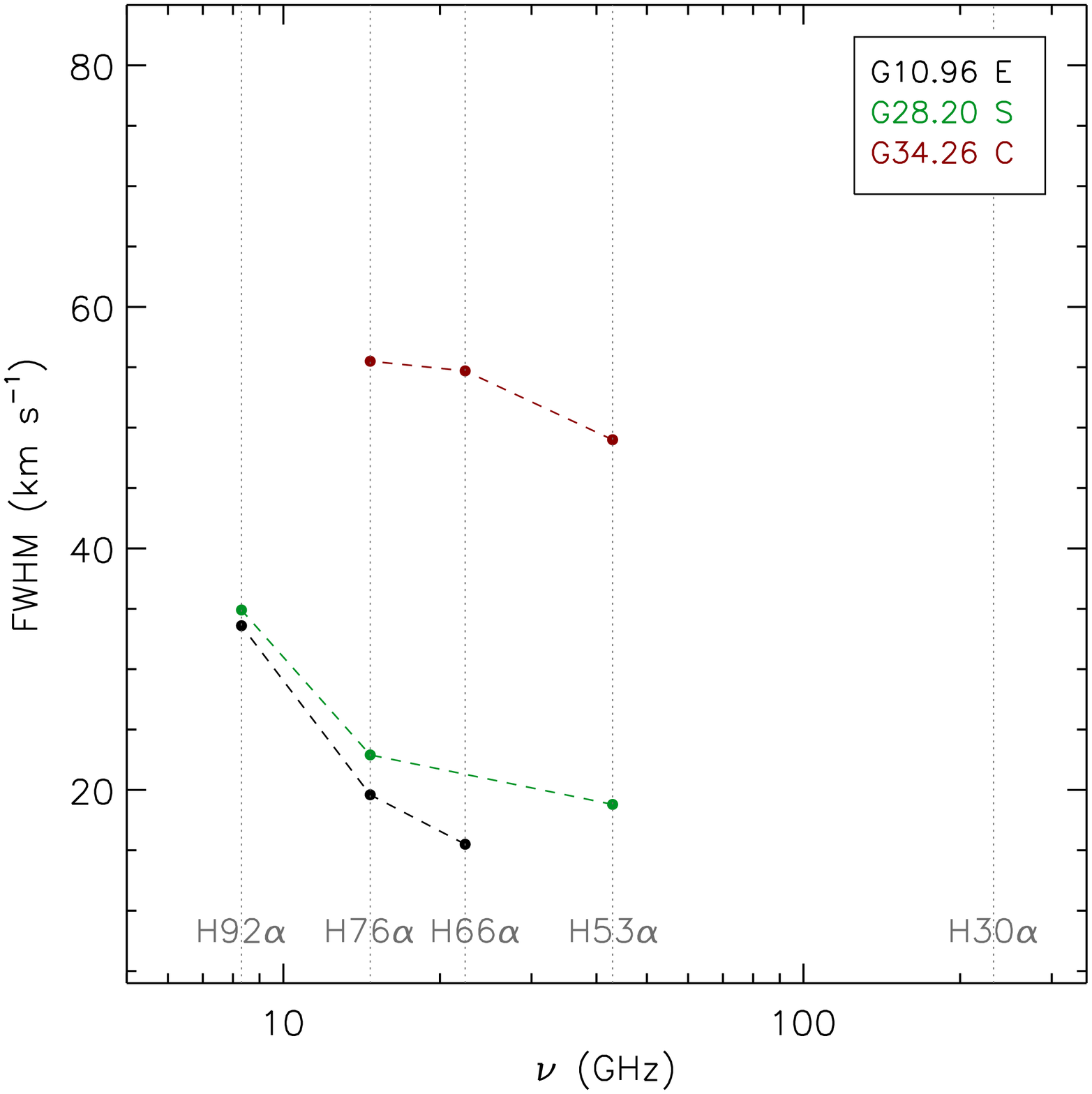}
\caption{Plots showing the relation between the half-power widths of the RRLs (FWHMs) and 
frequency for HC H\,{\sc ii} region candidates ({\it left}) and UC H\,{\sc ii} 
regions ({\it right}) from our sample (Paper I and this paper). The plots include only 
the sources observed in more than one RRL. The FWHMs with the uncertainties, as well as 
references can be found in Section~\ref{indiv}. The values measured in this paper are 
also listed in Table~\ref{tabcorr}.  See the discussion of the plot in text 
(Section~\ref{discrrl}).
\label{nfwhm}}
\end{figure}

We detected RRLs from all four sources in our
sample (see Table~\ref{trans} for details on the observed transitions).
Figure~\ref{nfwhm} shows the relation between the widths (FWHMs) of the 
RRLs and frequency for these sources. Regions classified as HC H\,{\sc ii} 
candidates and UC H\,{\sc ii} regions are plotted separately in 
Figure~\ref{nfwhm}a and \ref{nfwhm}b, respectively. The dashed vertical 
lines indicate the rest frequencies of the observed transitions.    

The change of observed Gaussian line widths with frequency (or principal
quantum number) presented in Figure~\ref{nfwhm} clearly illustrates the effects of
pressure broadening based on our data.  The limited sensitivity that is
typical of very high spatial resolution RRL observations rarely permits
the Voigt profile to be seen \citep{rg92}.  Very broad
bandwidths and good signal-to-noise ratios in the weak RRL wings are
a prerequiste to fit Voigt profiles; our data are not adequate for
such a fitting procedure.

Figure~\ref{nfwhm}a shows that the lines from HC H\,{\sc ii} regions 
(except G34.26 A), particularly 
the lower-frequency transitions, are too broad to be the result of a 
combination of thermal and turbulent broadening only. If an electron 
temperature of 10,000 K and turbulent velocity of 5 km s$^{-1}$ are 
assumed, the contribution from thermal and turbulent motions to the 
line widths would be only $\sim$22 km s$^{-1}$.  The change in line width 
with transition apparent in Figure~\ref{nfwhm}a  strongly suggests that 
pressure broadening is a dominant broadening mechanism. Owing to the
rapid change of pressure broadening with principal quantum number $n$
(FWHM of pressure broadening $\propto n^{7.4}$; \citealt{bs72}), the lower-frequency
lines are affected much more strongly by pressure broadening than the 
higher-frequency lines. At high frequencies, pressure broadening 
becomes negligible, thus line widths should be dominated by thermal, 
turbulent, and bulk motions. Even though pressure broadening depends 
so strongly on $n$, accurate adherence to the $n^{7.4}$ dependence 
is not expected because of other dependences on electron density and 
electron temperature, which vary within H\,{\sc ii} regions (e.g. \citealt{rg92}).
 
Pressure broadening is less strongly dependent on electron density 
than on the principal quantum number (FWHM $\propto$ $n_{e,true}$, where 
$n_{e,true}$ is the local true electron density; \citealt{bs72}). The true electron density
traced by line widths is higher than the rms electron density measured 
from the continuum data. The difference depends on the filling 
factor $f$ of the H\,{\sc ii} region ($f = n_{e,rms}^{2} / n_{e,true}^{2}$). 
The filling factor is the fraction of the total volume that is occupied by 
dense gas; thus it is a measure of clumpiness in the H\,{\sc ii} region.  
The higher electron density results in more pressure broadening, thus 
this effect may explain the observed increase in line width for the 
relatively low rms electron density UC H\,{\sc ii} regions G10.96 E 
and G28.20 S (see Figure~\ref{nfwhm}b).  

Sources G28.20~N, G28.20~S, G34.26~B and G34.26~C have been observed
in high frequency RRLs: H30$\alpha$ for G28.20~N and H53$\alpha$ for
the remaining three sources.  By comparing these high frequency lines
with lower frequency (and hence pressure-broadened) lines, we can
determine $n_{e,true}$.  Comparing these densities with the {\it rms}
values from the continuum observations then provides an estimate of
the filling factor.  Sources showing the steepest slopes in the
FWHM versus frequency plot of Figure~\ref{nfwhm} will have $n_{e,true}$ much
greater than $n_{e,rms}$, and hence have the smallest filling factors.

The largest change of linewidth with frequency of the RRL is obtained
for G28.20~N.  Sewilo et al. (2008) obtained an $n_{e,true}$ more than
an order of magnitude higher than $n_{e,rms}$ and a filling factor of
$6 \times 10^{-4}$ for this source.   G28.20~S, on the other
hand, shows only a 4 km~s$^{-1}$ change in linewidth between the
H76$\alpha$ and H53$\alpha$ lines.  We determine an $n_{e,true}$ of
$(1.2 \pm 0.7) \times 10^5$ cm$^{-3}$ compared to the continuum value
of $n_{e,rms} = 8 \times 10^3$ cm$^{-3}$, thus yielding a filling factor
of $5 \times 10^{-3}$.

For G34.26 components B and C we calculate true electron densities of
$(4.2 \pm 0.9) \times 10^5$ cm$^{-3}$ and $(4 \pm 2) \times 10^5$
  cm$^{-3}$, respectively, resulting in filling factors of 0.3 and 0.06.

We note that the true electron densities and hence the filling factors
are calculated under the assumption that the H53$\alpha$ linewidth
has {\it no} contribution from pressure broadening.  If this
assumption is not true, then the values we report should be
considered as limits; the true electron densities will be higher
and the true filling factors will be smaller.

The broad linewidths (FWHM $\geq$ 40 km s$^{-1}$) detected toward the
HC H{\sc ii} candidates (Figure~\ref{nfwhm}a) probably have a
contribution from bulk motions: G28.28 N has been related to rotation
and possible outflow; G34.26 B to an outflow; and G45.07 NE to
expansion (see Section~\ref{g45_indiv}). All these phenomena may
contribute to line broadening. For the UC H{\sc ii} region G34.26 C
(Figure~\ref{nfwhm}b), \citet{g86} proposed that the relatively broad
H76$\alpha$ line results from microturbulent motions of $\sim$30 km
s$^{-1}$ (see Section~\ref{g34_indiv}).

\subsection{Continuum Emission}

\begin{figure}
\centering
\includegraphics[width=0.8\textwidth]{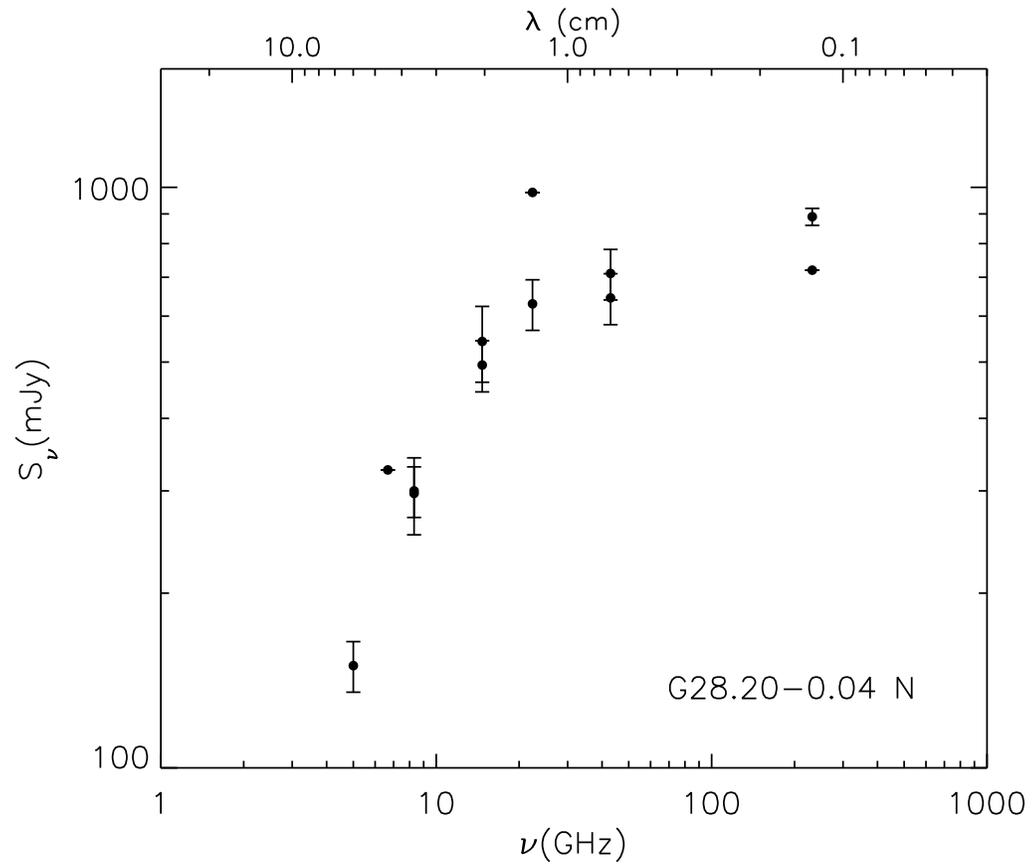}
\caption{Spectral energy distribution of G28.20 N from 5 GHz to 231.9 GHz. Flux 
densities and beam sizes, along with references, are listed in Table \ref{g28sed_tab}. 
The uncertainties are plotted when available. \label{g28sed}}
\end{figure}

The continuum observations of MSFRs G10.96 and G28.20 at two wavelengths
(Table~\ref{tabcont}) allow a determination of the spectral index 
for the individual sources in these regions, provided that the reliable
flux density measurements can be made (G10.96 W, G28.20 N). For consistency, 
the two continuum images for each region (1.3 cm and 2 cm for G10.96; 7 mm and 2 cm 
for G28.20) were convolved to a common resolution before
measuring the flux densities. For G10.96 W, the 1.3 cm and 2 cm 
flux densities of 296$\pm$15 mJy and 288$\pm$15 mJy, respectively, give
a spectral index of +(0.07$\pm$0.17), indicating that this source is in the
optically thin regime. The 7 mm and 2 cm flux densities of G28.20 N are 
723$\pm$37 mJy and 490$\pm$25 mJy, respectively, indicating a spectral index 
of $+$(0.4$\pm$0.1). This value suggests that the H{\sc ii} region has a
turnover wavelength in the 2 cm -- 7 mm range. The flux density 
distribution for G28.20 N, based on high-resolution interferometric 
observations covering the wavelength range from 6 cm to 1.3 mm, is shown in 
Fig. \ref{g28sed}. The plot shows the spectrum becoming optically thin in 
the 20 -- 50 GHz range.  

We found that G28.20 N is very similar to the well-studied HC H\,{\sc
ii} region G24.78$+$0.08 A1 \citep{bel07} in terms of physical properties, kinematics
(rotation, infall, outflow), morphology (shell-like), and association
with masers. The flux density distributions of these sources are 
similar in terms of the shape and the turn-over frequency (\citealt{bel07}; 
Fig.~\ref{g28sed}). Neither of 
these HC H\,{\sc ii} regions has a rising spectrum to millimeter wavelengths 
(as do G9.62+0.19 E and G75.78+0.34 (H$_2$O), see \citealt{franco00}). 
The shell-like morphology appears to be a common feature of the HC H\,{\sc ii}
regions. Besides G28.20 N and G24.78$+$0.08 A1, the A and B components 
in G34.26, and also G45.07 NE, show a shell morphology in high resolution 
images reported by \citet{ava09} and \citet{tm84}, respectively.

\subsection{Distribution of Masers}

We investigated the distribution of maser emission in MSFRs
from our sample.  We found that maser emission is only detected toward 
sources with broad RRLs (typically the most compact and dense sources 
in the UC/HC H\,{\sc ii} region complexes): G10.96 W, G28.20 N, G34.26 
B and C, and G45.07 NE.

\begin{figure}
\centering
\includegraphics[width=0.48\textwidth]{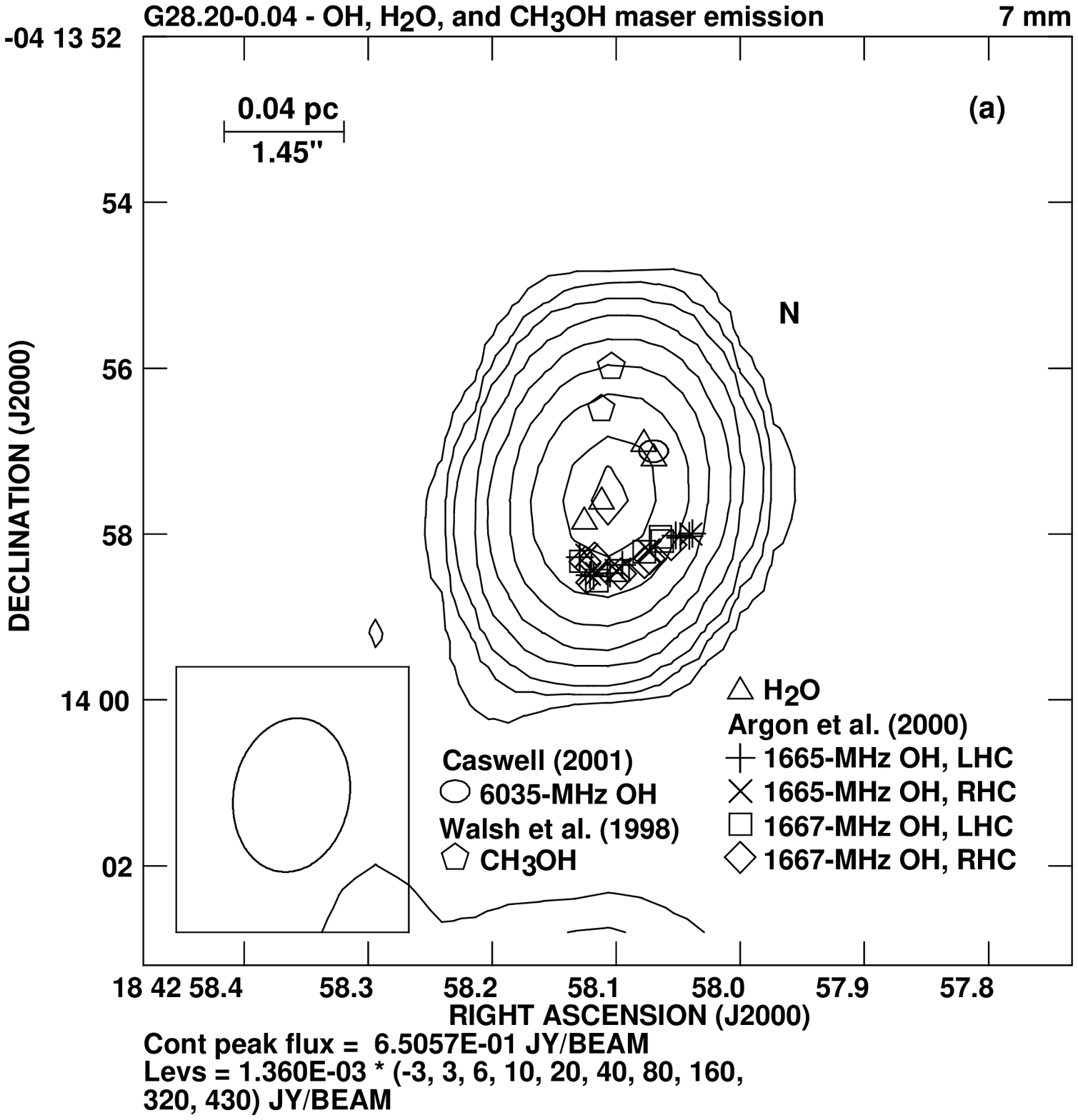}
\hfill
\includegraphics[width=0.49\textwidth]{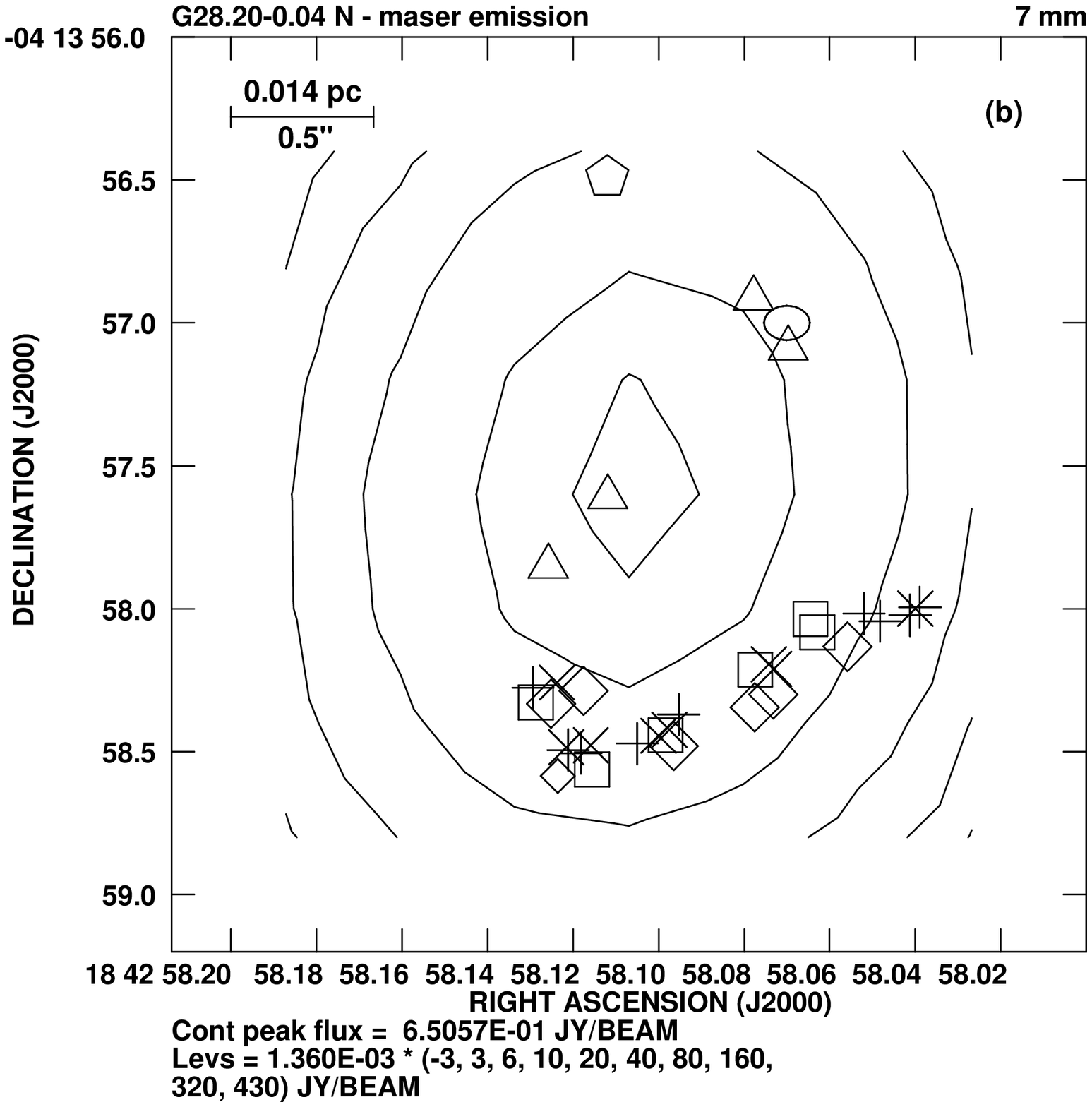}
\vspace*{0.2cm}
\hfill 
\includegraphics[width=0.49\textwidth]{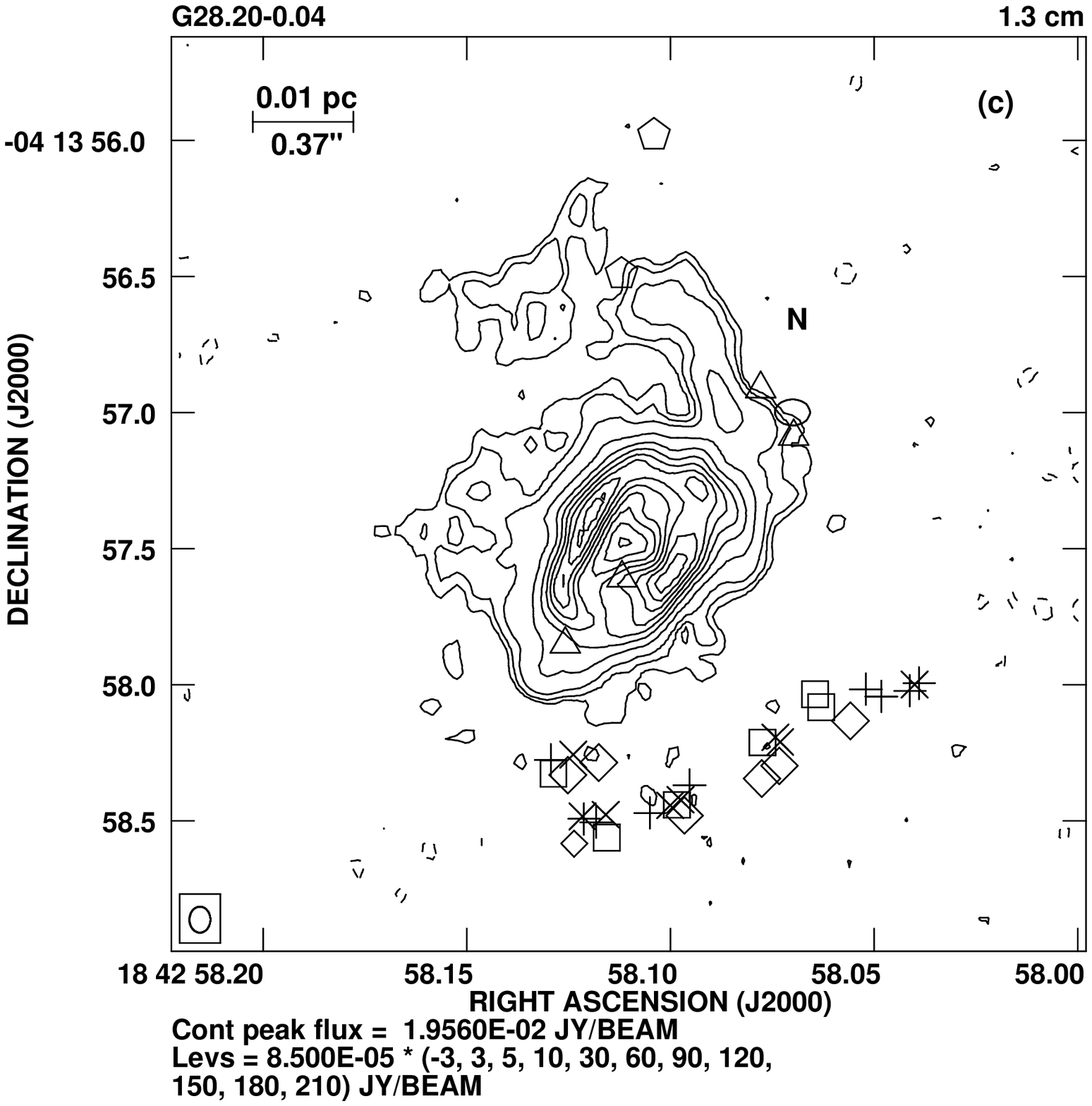} 
\hfill \hfill
\includegraphics[width=0.49\textwidth]{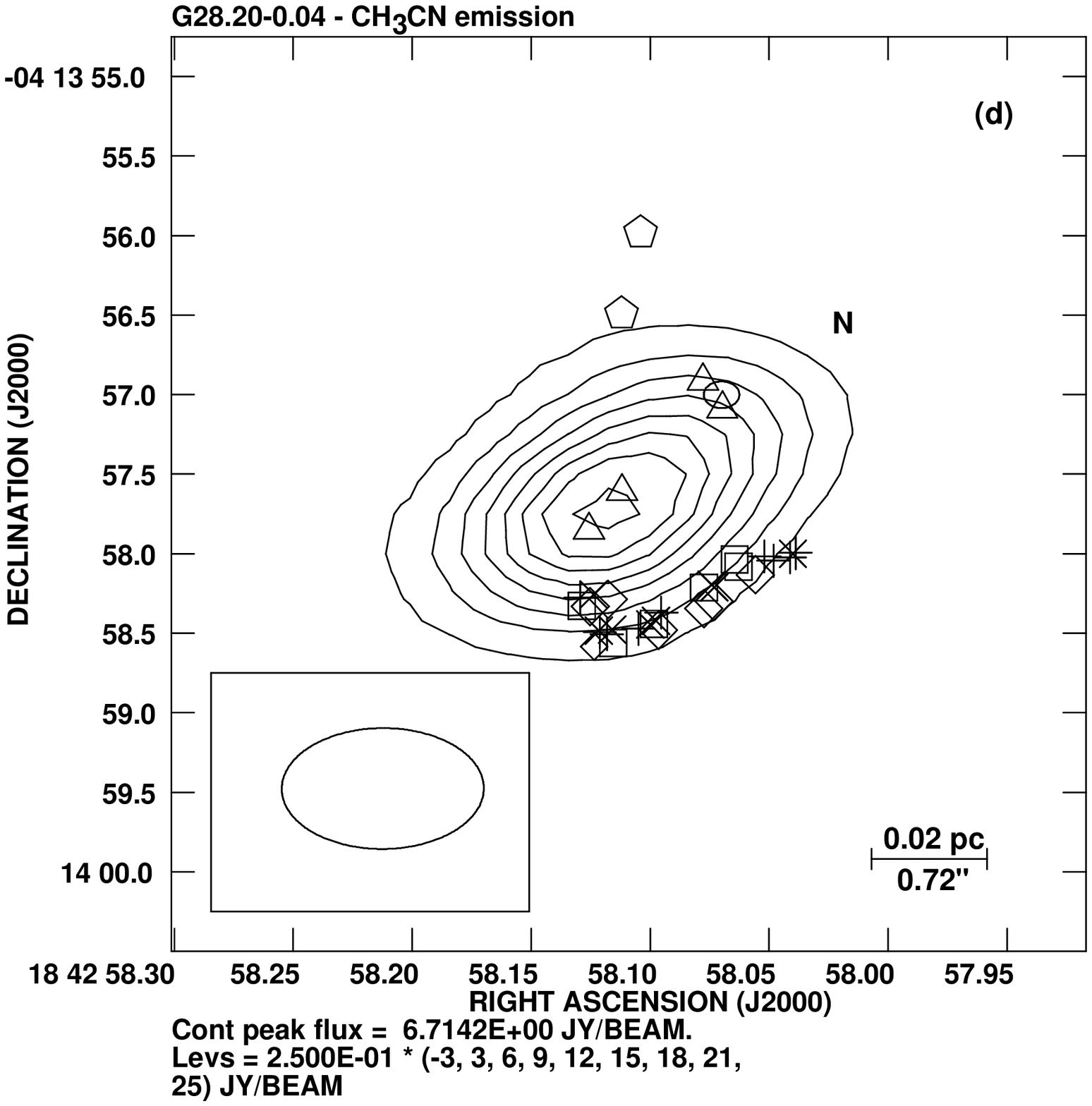} 
\caption{\small The distribution of OH and CH$_{3}$OH masers in G28.20~N 
superimposed on: the 7 mm VLA image (a and b: b is a magnification of the 
central portion of the image shown in a; this paper), the 1.3 cm VLA image 
(c; VLA archival data, project AZ168), and the CH$_3$CN SMA image 
(d; \citealt{q08}). The synthesized beams of the VLA shown in the lower 
left insets in (a) and (c) are $\sim$1$\rlap.{''}$6 and  0$\rlap.{''}$09,
respectively (see Table \ref{tabinstr}).  The synthesized beam of the SMA  
is $\sim$1$\rlap.{''}$5 (see Section \ref{g28_indiv}) and is shown in the 
lower left corner in (d). \label{g28masers}}
\end{figure}

In Figure \ref{g28masers} we show the maser positions for
G28.20 N overlaid on the continuum and thermal molecular line
emission. The OH masers are distributed linearly over a region $\sim$1$\rlap.{''}$4 
(or $\sim$0.04 pc; Figure~\ref{g28masers}a) with a velocity gradient of
$\sim$150 km s$^{-1}$ pc$^{-1}$.  The velocities increase
from $\sim$94 km s$^{-1}$ at the north-western end of the linear feature to
$\sim$100 km s$^{-1}$ at the south-eastern end.  The small cluster of maser 
spots north of the south-eastern end has a velocity of $\sim$95 km s$^{-1}$.  
 The 1.3 cm image (Figure \ref{g28masers}c), with 
20$\times$ higher angular resolution than the 7 mm image (Figure~\ref{g28masers}a), 
shows the OH masers situated along the south-western boundary of 
the continuum source, at the edge of the CH$_{3}$CN emission 
(Figure~\ref{g28masers}d). The location of the OH masers suggests that 
they arise within a molecular shell surrounding the ionized gas.

The OH maser velocities in G28.20 N roughly agree with the CH$_3$CN
velocity of $\sim$95-96 km s$^{-1}$; however, they are redshifted
relative to the H\,{\sc ii} region velocity of $\sim$93 km s$^{-1}$
estimated by the systemic velocity of the highest frequency RRL
observed toward G28.20 N (H30$\alpha$ at 1.3 mm, \citealt{k08};
\citealt{wm87}). A similar redshift is seen for CH$_{3}$OH masers,
detected over the velocity range 94 -- 105 km s$^{-1}$.  These
redshifts for OH and CH$_{3}$OH masers, together with the fact that
they appear projected against the optically thick continuum, suggest
that they arise on the near side of G28.20 N, thus supporting the
scenario that the masers are located in a collapsing molecular
envelope surrounding the H\,{\sc ii} region. A similar scenario was
proposed for W3(OH) \citep{rei80} and several other UC H\,{\sc ii}
regions (including G34.26 C, and G45.07 NE; \citealt{g85}) based on
the maser and radio continuum emission in those regions.  This
interpretation can also apply to G10.96 W from our sample, 
where the two CH$_{3}$OH maser spots are near the continuum emission peak, 
and have velocities ($\sim$25 km s$^{-1}$; \citealt{w98}) redshifted by 
$\sim$14 km s$^{-1}$ relative to the H66$\alpha$ velocity (see
Table~\ref{tabfits}).

The OH maser velocities in G28.20 N appear to be inconsistent with the
velocity gradient reported for the H53$\alpha$ line by \citet{sew08}.
In particular, the H53$\alpha$ velocity gradient, when extrapolated to
the position of the arc of maser emission, gives velocities
of 115 -- 120 km s$^{-1}$, significantly higher than the OH maser
velocities of 93 -- 101 km s$^{-1}$.  The differing direction and
magnitude of the H53$\alpha$ and the OH velocity gradients suggests
that dynamics of the ionized and molecular gas trace different
phenomena.

The orientation of the H$_{2}$O masers (SE-NW) in G28.20 N
is consistent with the outflow suggested by \citet{sew08} and \citet{q08}.
However, the velocity structure of the masers is ambiguous and
does not provide strong evidence for the outflow. 

In G45.07 NE, the H$_{2}$O maser spots are aligned spatially and 
kinematically with the outflow axis detected in CO(6-5) and 
CS(2-1) molecular lines (\citealt{hc96}; see Section~\ref{g45_indiv}), 
suggesting that the masers form in warm and dense shocked gas at the 
inner edges of the outflow lobes \citep{h97}.

\subsection{{\it Spitzer}/IRAC Mid-IR Emission}

We find that all the radio components have point source counterparts in the 
GLIMPSE Archive except G34.26 A and B, and G45.07 SW.  A faint 
mid-IR source was detected at the position of G45.07 SW; however, 
it was not included in the GLIMPSE catalog due to the close 
proximity of a much brighter source (G45.07 NE) that prevented 
an accurate photometric measurement. G34.26 A does not have a 
clear counterpart in the IRAC image; this source was detected 
in the mid-IR by earlier studies, unlike component B which is 
undetected in the mid-IR. 

In Table~\ref{glimpse}, we list GLIMPSE counterparts to the 
radio sources (the closest neighbors). The distance between
the mid-IR and radio positions ranges from 0$\rlap.{''}$2 
for G45.07 NE to 2$\rlap.{''}$4 for G28.20 S (see Table~\ref{glimpse}).
The proximity of the mid-IR sources to the radio peaks of
HC H\,{\sc ii} regions suggests that these sources probably 
are the ionizing (proto)stars.

The relatively low density of stars nearby to G10.96 E 
and G28.20 relative to more distant parts of these fields, suggests
extinction of background stars in these regions. The majority of 
IR sources shown in Figures~\ref{g10irac}-\ref{g45irac} 
are probably background stars; however, we expect to find a 
number of young stellar objects (YSOs) in these MSFRs.  The YSOs
can be identified by their mid-IR colors and by fitting their 
spectral energy distributions with  YSO models. This would be 
particularly interesting for G28.20, which is located 
in an IRDC (see Figure~\ref{g28irac}). 

The {\it Spitzer} images of the sources from our sample suggest that
generally the large-scale spatial distribution of gas and dust in these
regions are similar; G34.26 C being the principal exception in our sample.
Higher resolution infrared observations would be worthwhile, however, to
separate diffuse emission from the mid-IR emission from outflows and disks 
(e.g., NGC7538 IRS 1; De Buizer \& Minier 2005).

\section{Summary}

We present high spatial resolution (0$\rlap.{''}$9 to 2$\rlap.{''}$3),
multi-transition (H53\al to H92$\alpha$) Very Large Array observations
of four massive star formation regions hosting HC H\,{\sc ii} region
candidates identified in earlier studies. G10.96, G28.20, and G34.26
contain HC H\,{\sc ii} region candidates with broad H92$\alpha$
(G10.96 W, G28.20 N) or H76$\alpha$ (G34.26 B) lines \citep{s04}.  An
additional HC H\,{\sc ii} region candidate (G45.07 NE), known to have
broad H66$\alpha$ and H76$\alpha$ lines, small size, high electron
density and emission measure, was also included in this survey. We
observed up to two additional transitions per source to investigate
the change of line width with frequency.  We found a strong
correlation of line width with principal quantum number, i.e., higher
frequency lines of sources with broad H92\al or H76\al lines have
narrower linewidths ($<$40 km s$^{-1}$) indicating that pressure
broadening is the main line broadening mechanism at lower
frequencies. This result is not surprising because of the high electron
density in HC H\,{\sc ii} regions.  However, large scale dynamics is
also an important contributor to line broadening \citep{sew08}. We
determined physical properties of the H\,{\sc ii} regions
based on both the continuum and line data. Multi-frequency RRLs
reported in this paper are well-separated in principal quantum number
and will provide good constraints for non-LTE models of these regions.

We also present a comparison of the radio
continuum and mid-IR emission in HC H\,{\sc ii} regions from our
sample.  Based on the {\it Spitzer} GLIMPSE survey, covering a
wavelength range from 3.6 to 8.0 $\mu$m, we find  a generally good 
correspondence between the radio continuum and dust
distributions for the sources in our sample.  The mid-IR images, however,
show more complex features and reveal multiple sources that do not 
correspond so well to the radio continuum.

We report the distribution of H$_{2}$O, OH, and CH$_{3}$OH masers with
respect to the radio continuum.  We also present 
high-resolution VLA archival data for the water masers in G28.20 N for
the first time.  We compile information from the literature on maser
emission in G10.96, G28.20, and G45.07. A detailed
discussion of the maser distribution in G34.26 can be found elsewhere
(e.g., \citealt{hc96}, \citealt{g02}). As expected, all of the massive
star formation regions studied in this paper are associated with maser
emission; however, the masers are not coincident with all the radio
components in these regions. In G10.96, G28.20, and G45.07 the
masers are associated only with the HC H\,{\sc ii} regions (G10.96 W,
G28.20 N, G45.07 NE).  In G34.26 the masers are coincident with both
the HC H\,{\sc ii} region G34.26 B and also with the cometary UC H\,{\sc
ii} region G34.26 C. All of the sources associated with masers have
broad RRLs. 

The question of the origin of BRRLs is not fully answered, although
our observations show that pressure broadening is an important
mechanism. To distinguish the effects of pressure broadening from
large scale motions of the gas and at the same time resolve the
structures of HC H\,{\sc ii} regions will require high-resolution
observations of millimeter RRLs such as H42\al and H30\al.
With its high spatial resolution and sensitivity, ALMA will be the 
instrument to address this issue.

\section{Acknowledgements}
MS acknowledges the support from the NASA ADP grant 
NNX10AD43G. EC acknowledges partial support from NSF grant AST-080811.
PH acknowledges partial support from NSF grant AST-0908901. 
We thank Thushara Pillai for useful discussion on the water masers in
G28.20$-$0.04. We also thank an anonymous referee for comments 
and suggestions that improved the manuscript.


\begin{deluxetable}{p{6.0cm}ll}
\tabletypesize{\footnotesize}
\tablecaption{Instrumental Parameters for the VLA Observations \label{tabinstr}}
\tablewidth{0pt}

\tablehead{
\colhead{Parameter} &
\colhead{2003 Mar 20} &
\colhead{2004 Mar 7, 8\tablenotemark{a}}    
}

\startdata
Program  \dotfill & AS753 & AS797 \\
Transitions observed \dotfill & H53$\alpha$&  a) H66$\alpha$\\
&\nodata& b) H76$\alpha$ \\
&\nodata&c) H92$\alpha$ \\
Observed sources \dotfill & G28.20$-$0.04&G10.96$+$0.01 (ab) \\
                          &G34.26$+$0.15 & G28.20$-$0.04 (b)  \\
                          & \nodata        & G45.07$+$0.13 (c) \\
Total observing time \dotfill & 10 hr &  a) 3 hr \\  
&\nodata& b) 9 hr \\
&\nodata& c) 1 hr \\
Configuration\tablenotemark{b} \dotfill & D &  C\\
Synthesized beam/FWHM ($''$), PA($\arcdeg$): & & \\
\hspace{0.5cm} G10.96$+$0.01 \dotfill & \nodata & 1.4 $\times$ 0.8, $-$8 (a)\\
 \hspace{2.6cm} \dotfill & \nodata & 2.0 $\times$ 1.2, $-$2 (b)\\
\hspace{0.5cm} G28.20$-$0.04  \dotfill & 1.9 $\times$ 1.4, $-$10 & 1.5 $\times$ 1.4, $-$10 (b) \\
\hspace{0.5cm} G34.26$+$0.15  \dotfill & 1.6 $\times$ 1.5, $-$4 & \nodata \\
\hspace{0.5cm} G45.07$+$0.13  \dotfill & \nodata & 2.5 $\times$ 2.4, $-$31 (c)\\
Rest frequency of the H line\dotfill & 42,951.97 MHz &  a) 22,364.17 MHz\\
&\nodata& b) 14,689.99 MHz\\
&\nodata&c) 8,309.38 MHz\\
Observing mode \dotfill & 1A &  2AD\\
Bandwidth \dotfill & 25 MHz&  12.5 MHz\\
Number of channels \dotfill & 31 &  31 \\
Channel separation  \dotfill  & 781.25 kHz&  390.625 kHz\\
Velocity resolution\tablenotemark{c} \dotfill & 6.5 km s$^{-1}$ &  a) 6.2 km s$^{-1}$\\
&\nodata& b) 9.6 km s$^{-1}$  \\ 
&\nodata& c) 16.9 km s$^{-1}$ \\ 
Flux density calibrator (Jy): & & \\ 
\hspace{0.5cm}\underline{3C 286} for G10.96$+$0.01 \dotfill & \nodata &  2.53 (a), 3.50 (b) \\
\hspace{2.1cm}           G28.20$-$0.04 \dotfill & 1.47    &  3.50 (b) \\
\hspace{2.1cm}           G34.26$+$0.15 \dotfill & 1.47    &  \nodata \\
\hspace{2.1cm}           G45.07$+$0.13 \dotfill & \nodata   & 5.27 (c) \\

Phase calibrators (Jy): & &  \\
\hspace{0.5cm}\underline{1820$-$254} for G10.96$+$0.01 \dotfill & \nodata  & 0.610 $\pm$ 0.004 (a) \\
\hspace{2.52cm}                            \dotfill & \nodata & 0.692 $\pm$ 0.003 (b) \\
\hspace{0.5cm}\underline{1851$+$005} for G28.20$-$0.04 \dotfill & 0.75 $\pm$ 0.02 &  0.917 $\pm$ 0.003 (b)\\
\hspace{2.52cm}              G34.26$+$0.15 \dotfill & 0.72 $\pm$ 0.02 &  \nodata \\
\hspace{0.5cm}\underline{1925$+$211} for G45.07$+$0.13 \dotfill &  \nodata & 2.57 $\pm$ 0.01 (c) \\
\enddata
\tablenotetext{a}{a, b, and c indicate values for the H66$\alpha$, H76$\alpha$, and H92$\alpha$ lines, respectively.}
\tablenotetext{a}{Primary beams are: 1$'$ for AS753; 2$'$, 3$'$, and 5$\rlap.{'}$4 for AS797 (a), (b), and (c), respectively.}
\tablenotetext{c}{The observations were done with uniform spectral weighting, thus the velocity resolution 
is 1.2 times the channel separation.}
\end{deluxetable}


\begin{deluxetable}{p{4cm}cccc|lc}
\tablecaption{Transitions Observed toward Individual Sources and Adopted
  Distances \label{trans}}
\tablewidth{0pt}
\tablehead{
\colhead{Source} &
\colhead{H53$\alpha$} &
\colhead{H66$\alpha$} &
\colhead{H76$\alpha$} &
\colhead{H92$\alpha$} &
\colhead{D (kpc)} &
\colhead{Ref.} }
\startdata
G10.96+0.01  \dotfill &  \nodata & $+$ & $+$ & $+$\tablenotemark{a} & 14.0 $^{+1.3}_{-0.9}$ & 1 \\
G28.20-0.04  \dotfill & $+$ & \nodata & $+$ & $+$\tablenotemark{a} & 5.7 $^{+0.5}_{-0.8}$ & 2 \\
G34.26+0.15  \dotfill &  $+$ & \nodata & $+$\tablenotemark{a} & \nodata & 3.7 & 3 \\
G45.07+0.13  \dotfill & \nodata & \nodata & \nodata & $+$ & 6.0 & 3  \\
\enddata
\vspace*{-0.5cm}
\tablecomments{The plus signs ($+$) indicate the observed transitions.}
\tablenotetext{a}{Transitions reported in Sewi{\l}o et al. (2004)}
\tablerefs{(1) \citeauthor{s04} \citeyear{s04}; (2) \citeauthor{f03} \citeyear{f03}; (3) \citeauthor{a02} \citeyear{a02}}
\end{deluxetable}


\begin{deluxetable}{p{2.2cm}ccccccccc} 
\tabletypesize{\scriptsize}
\rotate
\tablecaption{Continuum Parameters\tablenotemark{a} \label{tabcont}}
\tablewidth{0pt}
\tablehead{

\colhead{} &
\colhead{} &
\colhead{} &
\colhead{} &
\colhead{} &
\colhead{Time} &
\colhead{Source} &
\colhead{} &
\colhead{Integrated} &
\colhead{} \\

\colhead{} &
\colhead{} &
\colhead{} &
\multicolumn{2}{c}{Position} &
\colhead{on} &
\colhead{Diameter} &
\colhead{Source} &
\colhead{Flux} &
\colhead{RMS} \\

\cline{4-5}

\colhead{Source} &
\colhead{$\lambda$} &
\colhead{Comp.} &
\colhead{$\alpha (J2000)$} &
\colhead{$\delta (J2000)$} &
\colhead{Source} &
\colhead{$\theta_{maj} \times \theta_{min}$} & 
\colhead{PA\tablenotemark{a}} &
\colhead{Density} &
\colhead{in Image} \\

\colhead{} &
\colhead{} &
\colhead{} &
\colhead{(h~m~s)}  & 
\colhead{($\arcdeg$  $'$  $''$) }   & 
\colhead{(h)} &
\colhead{($''~\times~''$)} &
\colhead{($\arcdeg$)} &
\colhead{(mJy)}& 
\colhead{(mJy beam$^{-1}$)} 
}

\startdata

&&&&&&&& \\ 
G10.96$+$0.01 \dotfill & 1.3 cm & E\tablenotemark{b} & 18 09 42.85 & $-$19 26 29.6   & 0.5 & \nodata & \nodata &  \nodata & 0.4   \\ 
& \nodata & W & 18 09 39.35 & $-$19 26 28.1 & \nodata & 1.0 $\times$ 0.9 & 177 & 291 \\ 

\dotfill & 2 cm & E/E1\tablenotemark{b} & 18 09 42.98  & $-$19 26 28.2 & 1.4 & \nodata & \nodata & \nodata & 0.4 \\ 
\dotfill  & \nodata & E/E2\tablenotemark{b} & 18 09 42.85 & $-$19 26 30.0 & \nodata & \nodata & \nodata & \nodata & \nodata  \\
& \nodata & W & 18 09 39.35 & $-$19 26 28.1  &  \nodata & 1.1 $\times$ 0.9 & 53 & 289 & \nodata \\
&&&&&&&& \\ 
G28.20$-$0.04 \dotfill & 0.7 cm   & N & 18 42 58.11 & $-$04 13 57.5 & 0.6 & 0.7 $\times$ 0.5 & 159 & 710 & 1.4 \\ 
                       & \nodata  & S & 18 42 58.12 & $-$04 14 04.9 & \nodata & 2.9 $\times$ 2.4 & 177 & $<$108\tablenotemark{c} & \nodata\\ 
                       & \nodata  & S1& 18 42 58.29 & $-$04 14 04.4 & \nodata & \nodata & \nodata & \tablenotemark{c} & \nodata \\ 
\dotfill               & 2 cm     & N & 18 42 58.11 & $-$04 13 57.4 & 2.6 & 0.8 $\times$ 0.6 & 150 & 494 & 0.4  \\
   & \nodata  & S & 18 42 58.13 & $-$04 14 04.7 & \nodata & 2.9 $\times$ 2.7 & 28 & 132 & \nodata\\ 
&&&&&&&& \\ 
G34.26$+$0.15 \dotfill & 0.7 cm  & A & 18 53 18.79 & $+$01 14 56.3 & 1.0 & 0.3 $\times$ 0.2 & 54 & $>$86\tablenotemark{d} & 2.8   \\
& \nodata & B\tablenotemark{e} & 18 53 18.66 & $+$01 15 00.5 & \nodata & \nodata & \nodata & \nodata& \nodata\\  
& \nodata & C\tablenotemark{f} & 18 53 18.56 & $+$01 14 58.2 & \nodata & \nodata & \nodata & 6000 \tablenotemark{g}& \nodata\\ 

&&&&&&&& \\ 
G45.07$+$0.13 \dotfill & 3.6 cm  & NE& 19 13 22.08 & $+$10 50 53.2 & 0.6 & 0.6 $\times$ 0.6 & 87& 350 & 0.8  \\ 
& \nodata & SW& 19 13 21.83 & $+$10 50 48.2 & \nodata & 1.4 $\times$ 1.1 & 82 & 60 & \nodata\\ 
\enddata

\tablenotetext{a}{Source positions, deconvolved angular sizes, position angles, and integrated flux densities were determined by elliptical Gaussian fitting using the AIPS routine JMFIT.  We estimate the positional uncertainty to be $\sim$0$\rlap.{''}$1.  Uncertainties in source sizes depend on both the brightness and the intrinsic source extent; we estimate them to be accurate to within $\sim$10\%.  The flux density uncertainties are 10\% - this value takes into account the error of the fit (JMFIT) and systematic effects (e.g. calibration errors, mismatches between the Gaussian model and the actual source shape, etc.).}
\tablenotetext{b}{The E component of the MSFR G10.96 is overresolved at both 1.3 and 2 cm. Most of the flux is filtered out, thus we do not give either the total flux density or source size at these wavelengths. The 1.3 cm coordinates correspond to the maximum pixel. The peak flux density at 1.3 cm is 3.9 mJy beam$^{-1}$. The 2 cm flux density from the region that encloses two compact components (9 $\sigma$ level; see Fig.  \ref{g10eradio}b) is 38 $\pm$ 3 mJy beam$^{-1}$.}
\tablenotetext{c}{The faint S1 component emerging at 7 mm is partially blended with component S; the reported  
flux density is the total integrated flux density from both components. The flux density of the S1 component 
constitutes $\sim$15\% of the total flux. The peak 7 mm flux density of S1 is 26.6 mJy beam$^{-1}$.}
\tablenotetext{d}{The A component is partially blended with component C; the integrated flux density should be considered the lower limit.}
\tablenotetext{e}{The B component is not fully resolved from the cometary component C, thus we cannot give either its angular diameter or flux density. The position given in the table was determined from the 3.6 cm continuum map in Sewi{\l}o et al. (2004).}
\tablenotetext{f}{The coordinates correspond to the maximum pixel in the cometary component C. The peak flux density is 2.78 Jy beam$^{-1}$.}
\tablenotetext{g}{The integrated flux density was calculated using the AIPS task IMEAN from the area of $\sim$22$''$ by $\sim$12$''$ in R.A. and decl., respectively, that encloses all three components of the G34.26  H\,{\sc ii} region complex. The total flux density of the A component is $\sim$92 mJy, the remaining $\sim$5908 Jy arise from components B and C. Only a small fraction of this flux density ($<$ 3\%) comes from the B component (see text).}
\end{deluxetable}


\begin{deluxetable}{ccccr}
\tablecaption{Positions and Observed Parameters of Water Masers in G28.20$-$0.04 N \label{vlawater}}
\tablewidth{0pt}
\tablehead{
\colhead{water} &
\colhead{$\alpha (J2000)$} &
\colhead{$\delta (J2000)$} &
\colhead{$v_{LSR}$\tablenotemark{a}} &
\colhead{$S_{peak}$\tablenotemark{b}}\\
\colhead{maser} &
\colhead{(h~m~s)} &
\colhead{($\arcdeg$  $'$  $''$)} &
\colhead{(km s$^{-1}$)} &
\colhead{(Jy)}}
\startdata
1 &  18 42 58.07 & -04 13 56.9 &  97.7 &   1.2 \\
2 &  18 42 58.12 & -04 13 57.8 &  95.6 &  10.7 \\
3 &  18 42 58.11 & -04 13 57.6 &  93.1 &   1.6 \\
4 &  18 42 58.06 & -04 13 57.0 &  91.0 &  22.2 \\
\enddata
\tablenotetext{a}{The velocity of the peak channel.}
\tablenotetext{b}{The uncertainties of peak flux densities are dominated by systematic errors. 
We adopt a standard 10\% uncertainty for the VLA at K-band.}
\end{deluxetable}


\begin{deluxetable}{lcccccccccc}
\tablecaption{Source Averaged Physical Parameters based on Continuum Observations\tablenotemark{a} \label{tabphys}}
\rotate
\tablewidth{0pt}
\tablehead{
\colhead{} &
\colhead{} &
\colhead{} &
\colhead{diameter} &
\colhead{} &
\colhead{} &
\colhead{} &
\colhead{} &
\colhead{} &
\colhead{} &
\colhead{} \\
\colhead{Source} &
\colhead{} &
\colhead{$\lambda$} &
\colhead{of sphere} &
\colhead{$\rm \tau_{\rm c}$} &
\colhead{n$_{\rm e,rms}$/10$^{4}$} &
\colhead{EM/10$^{7}$}&
\colhead{U}&
\colhead{M$_{\rm H{II}}$}  &
\colhead{log N$_{\rm c}^{'}$} &
\colhead{Spectral\tablenotemark{b}} \\
\colhead{} &
\colhead{} &
\colhead{(cm)} &
\colhead{(mpc)} &
\colhead{} &
\colhead{(cm$^{-3}$)} &
\colhead{(pc cm$^{-6}$)} &
\colhead{(pc cm$^{-2}$)}&
\colhead{(M$_{\odot}$)}&
\colhead{(s$^{-1}$)} &
\colhead{Type}}
\startdata
G10.96$+$0.01\tablenotemark{c}   & W  & 1.3 & 121 & 0.02 & 2.9 & 5.3 & 57.6 & 0.6 & 48.8 & O7.5V\\
\dotfill               & \nodata &        2 & 124  & 0.06 & 2.7 & 5.3 & 56.4 & 0.6 & 48.8 & \nodata \\
G28.20$-$0.04 \dotfill & N       & 0.7 & 28 & 0.04 & 17 & 36 & 43.9 & 0.05 & 48.4 & O8V\\
\dotfill               & \nodata &   2 & 35 & 0.2 & 9.8  & 19 & 37.1 & 0.05 & 48.2 & \nodata \\
\dotfill               & S       & 0.7 & 131 & 0.0003 & 0.7 & 0.3 & 23.4 & 0.2 & 47.6 & O9.5V \\
\dotfill               & \nodata &   2 & 120 & 0.01 & 0.8 & 0.4 & 23.9 & 0.2 & 47.6 & \nodata \\
G34.26$+$0.15\tablenotemark{d} \dotfill & A  & 0.7 & 8 & 0.03 & 27 & 24 & 16.3 & 0.002 & 47.1 & B0.5V\\
G45.07$+$0.13 \dotfill & NE      & 3.6 & 32 & 0.7 & 9.4 & 19 & 33.4 & 0.04 & 48.1 & O8.5V \\
\dotfill               & SW      & 3.6 & 65 & 0.03 & 1.4 & 0.8 & 18.7 & 0.05 & 47.3 & B0V\\
\enddata
\vspace*{-0.5cm}
\tablecomments{The physical parameters are: 
the linear diameter of a spherical H\,{\sc ii} region (assumed model), 
the continuum optical depth ($\rm \tau_{\rm c}$), the rms electron density (n$_{e,rms}$), 
the emission measure (EM), the excitation parameter (U), the mass of ionized gas (M$_{\rm H {II}}$),  
and the Lyman continuum photon flux required to maintain ionization of the nebula (N$_{c}^{'}$).}
\tablenotetext{a}{We estimate an uncertainty of the derived physical parameters to be up to 20\%.  These parameters
depend on the different combinations of the integrated flux density, observed source size, and assumed distance. All 
of these quantities have uncertainties of $\sim$10\%.}
\tablenotetext{b}{$\rm{L}$ower limits to the spectral types of the
  exciting stars are derived from N$_{\rm c}^{'}$ using the model stellar
  atmosphere results of \citet{sm02}. See text for details.}
\tablenotetext{c}{We do not report the physical parameters for component E of G10.96$+$0.01, because most of the emission from this source if filtered out. See footnotes to Table \ref{tabcont}.}
\tablenotetext{d}{We do not report the physical parameters for
  components B and C of G34.26$+$0.15. They are blended at
  our spatial resolution, hence we cannot determine their angular sizes or
  flux densities.}
\end{deluxetable}


\begin{deluxetable}{lcccccccc}
\tablecaption{Peak Values of the Physical Parameters\tablenotemark{a} \label{tabpeak}}
\rotate
\tablewidth{0pt}
\tablehead{

\colhead{Source} &
\colhead{} &
\colhead{$\nu$} &
\colhead{S$_{\nu, peak}$} &
\colhead{$\Delta s$} &
\colhead{T$_{B}$} &
\colhead{$\tau_{c}$} &
\colhead{EM/10$^{7}$}&
\colhead{N$_{\rm e,rms}$/10$^{4}$} \\

\colhead{} &
\colhead{} &
\colhead{(GHz)} &
\colhead{(mJy/beam)} &
\colhead{(pc)} &
\colhead{(K)} &
\colhead{} &
\colhead{(pc cm$^{-6}$)} &
\colhead{(cm$^{-3}$)} }

\startdata
G10.96$+$0.01 \dotfill & W & 22.4 & 147.2 $\pm$ 0.4 & 0.07 & 325 & 0.03 & 7.1 & 3.3 \\
\dotfill               & \nodata  & 14.7  & 188.6 $\pm$ 0.4 & 0.07 & 460 & 0.05 & 4.2 & 2.4 \\
\dotfill               & E & 22.4 &  3.9 $\pm$ 0.4 & \nodata & 9 & $>$0.001 & 0.2 & \nodata \\
\dotfill               & E1\tablenotemark{b}  & 14.7 & 6.5 $\pm$ 0.4 & \nodata & 16 & 0.002 & 0.1 & \nodata\\
\dotfill               & E2\tablenotemark{b}  & 14.7 & 6.9 $\pm$ 0.4 & \nodata & 17 & 0.002 & 0.1 & \nodata \\
G28.20$-$0.04 \dotfill & N & 42.9 & 651 $\pm$ 1 & 0.015 & 165 & 0.017 & 14.4 & 9.8 \\
\dotfill               & \nodata  & 14.7 & 401.9 $\pm$ 0.4 & 0.02 & 1079 & 0.1 & 10.1 & 7.3 \\
\dotfill               & S & 42.9 & 27 $\pm$ 1 & 0.07 & 7 & $>$0.001 & 0.6 & 0.9 \\
\dotfill               & \nodata  & 14.7 & 30.7 $\pm$ 0.4& 0.08 & 82 & 0.01 & 0.7 & 1.0 \\
G34.26$+$0.15 \dotfill & A & 42.9 & 86 $\pm$ 3& 0.004 & 24 & 0.002 & 2.1 & 7.1 \\
\dotfill               & B & 42.9 & 120\tablenotemark{c} $\pm$  3& \nodata & 33 & 0.003 & 2.9 & \nodata \\
\dotfill               & C & 42.9 & 2777 $\pm$ 3& \nodata & 773 & 0.1 & 69.7 & \nodata \\
G45.07$+$0.13 \dotfill & NE & 8.3 & 328.3 $\pm$ 0.7 & 0.02 & 988 & 0.1 & 2.7 & 4.0 \\
\dotfill               & SW & 8.3 & 49.4 $\pm$ 0.7 & 0.04 & 149 & 0.02 & 0.4 & 1.1 \\
\enddata
\tablenotetext{a}{We estimate an uncertainty of the derived physical parameters to be up to 20\%.  These parameters
depend on combinations of the integrated flux density, observed source size, and assumed distance. All 
of these quantities have uncertainties of order 10\%.}
\tablenotetext{b}{Two maxima/subcomponents can be distinguished in the 2 cm image of component E of G10.96$+$0.01 (see Fig.  \ref{g10eradio}). E1 and E2 correspond to the NE and SW maxima, respectively.}
\tablenotetext{c}{The peak flux density of G34.26$+$0.15 component B is an upper limit because of blending with component C.}
\end{deluxetable}


\begin{deluxetable}{p{2.7cm}cccccccc}
\tablecaption{Gaussian Fits to the H53$\alpha$, H66$\alpha$, H76$\alpha$, and
  H92$\alpha$ Radio Recombination Lines  \label{tabfits}}
\rotate
\tablewidth{0pt}
\tablehead{

\colhead{Source} &
\colhead{} &
\colhead{RRL} &
\colhead{V$_{\rm LSR}$} &
\colhead{FWHM$^{\tablenotemark{a}}$}&
\colhead{S$_{L}$}&
\colhead{S$_{C}^{\tablenotemark{b}}$}&
\colhead{${\frac{S_{L}}{S_{C}}}$}  &
\colhead{$\int \frac{S_{L}}{S_{C}}\,\mathrm{d}v $} \\

\colhead{} &
\colhead{} &
\colhead{} &
\colhead{(km~s$^{-1}$)} &
\colhead{(km~s$^{-1}$)}&
\colhead{(mJy)} &
\colhead{(mJy)} &
\colhead{(peak)}  &
\colhead{(km~s$^{-1}$)} }

\startdata



G10.96$+$0.01 \dotfill & 
E               & 
H66$\alpha$\tablenotemark{c}    & 
11.4 $\pm$ 0.7 & 
18.7 $\pm$ 1.8 & 
12.6 $\pm$ 1.0 & 
46.3 $\pm$ 2.3 & 
0.27 $\pm$ 0.03 &  
 5.4 $\pm$ 0.7 \\

\dotfill & 
\nodata         & 
H76$\alpha$\tablenotemark{c}     & 
15.8 $\pm$ 1.0 & 
25.3 $\pm$ 2.5 & 
9.7  $\pm$ 0.8 & 
53.8 $\pm$ 2.7 & 
0.18 $\pm$ 0.02 &  
4.9  $\pm$ 0.7 \\

\dotfill & 
W               & 
H66$\alpha$     & 
13.6  $\pm$ 0.4 & 
28.5  $\pm$ 1.0 & 
35.9  $\pm$ 1.1 & 
147 $\pm$ 7 & 
0.24  $\pm$ 0.01 &  
 7.4  $\pm$ 0.5 \\

\dotfill & 
\nodata         & 
H76$\alpha$     & 
13.3  $\pm$ 0.5 & 
29.6  $\pm$ 1.3 & 
19.9  $\pm$ 0.7 & 
131 $\pm$ 7 & 
0.15  $\pm$ 0.01 &  
4.8   $\pm$ 0.4 \\

&&&&&&&&\\

G28.20$-$0.04 \dotfill & 
N               & 
H53$\alpha$     & 
88.5   $\pm$ 0.5 & 
39.0   $\pm$ 1.4 & 
144  $\pm$ 4 & 
372 $\pm$ 19 & 
0.39   $\pm$ 0.02  &  
16.1   $\pm$ 1.1 \\

\dotfill & 
\nodata         & 
H76$\alpha$     & 
80.1  $\pm$ 0.8 & 
55.7  $\pm$ 2.2 & 
17.0  $\pm$ 0.5 & 
235 $\pm$ 12 & 
0.07  $\pm$ 0.01 &  
4.3   $\pm$ 0.3  \\

\dotfill & 
S               & 
H53$\alpha$     & 
99.6 $\pm$ 0.4 & 
19.9 $\pm$ 0.9 & 
63.6 $\pm$ 2.4 & 
65.3 $\pm$ 3.3 & 
0.97 $\pm$ 0.06 &  
20.7 $\pm$ 1.6 \\

\dotfill & 
\nodata         & 
H76$\alpha$     & 
99.3 $\pm$ 0.5 & 
24.8 $\pm$ 1.3 & 
7.3  $\pm$ 0.3 & 
34.9 $\pm$ 1.7 & 
0.21 $\pm$ 0.01 &  
5.6  $\pm$ 0.5 \\

&&&&&&&&\\

G34.26$+$0.15 \dotfill & 
A               & 
H53$\alpha$     & 
50.2 $\pm$ 0.4 & 
21.9 $\pm$ 1.0 & 
46.3 $\pm$ 1.7 & 
62.9 $\pm$ 3.1 & 
0.74 $\pm$ 0.05 &  
17.2 $\pm$ 1.3  \\

\dotfill & 
B         & 
H53$\alpha$\tablenotemark{c}     & 
64.9 $\pm$ 1.0 & 
34.8 $\pm$ 2.6 & 
20.5 $\pm$ 1.2 & 
58.6 $\pm$ 2.9 & 
0.35 $\pm$ 0.03 &  
12.9 $\pm$ 1.4 \\

\dotfill & 
C               & 
H53$\alpha$     & 
47.9   $\pm$ 1.2 & 
49.4   $\pm$ 3.7 & 
589  $\pm$ 32 & 
2587 $\pm$ 129 & 
0.23   $\pm$ 0.02 &  
12.0   $\pm$ 1.2 \\
\enddata
\tablenotetext{a}{Full width at half-maximum.}
\tablenotetext{b}{S$_{C}$ is a continuum flux density of the same area of the source over which a RRL was integrated. It was determined using AIPS task IMEAN.}
\tablenotetext{c}{The spectrum was Hanning smoothed.}
\end{deluxetable}


\begin{deluxetable}{p{2.7cm}ccccc}
\tablecaption{Corrected H53$\alpha$, H66$\alpha$, H76$\alpha$, and H92$\alpha$ Radio Recombination Line Parameters
and the LTE Electron Temperatures  \label{tabcorr}}
\tablewidth{0pt}
\tablehead{

\colhead{Source} &
\colhead{} &
\colhead{RRL} &
\colhead{FWHM$_{\rm deconv}^{\tablenotemark{a}}$}&
\colhead{$({\frac{S_{L}}{S_{C}}})_{\rm corr}^{\tablenotemark{b}}$}  &
\colhead{T$_{e}^{\star}$ \tablenotemark{c}} \\

\colhead{} &
\colhead{} &
\colhead{} &
\colhead{(km~s$^{-1}$)} &
\colhead{(peak)}  &
\colhead{(K)}  }

\startdata
G10.96$+$0.01 \dotfill & E       & H66$\alpha$ & 15.5 $\pm$ 2.1 & \nodata & 11400 $\pm$ 1600 \\
\dotfill               & \nodata & H76$\alpha$ & 19.6 $\pm$ 3.2 & \nodata & 8800 $\pm$ 1400 \\
\dotfill               & W       & H66$\alpha$ & 27.8 $\pm$ 1.1 & 0.25 $\pm$ 0.01 & 7500 $\pm$ 500 \\
\dotfill               & \nodata & H76$\alpha$ & 28.0 $\pm$ 1.4 & 0.16 $\pm$ 0.01 & 7400 $\pm$ 500 \\
&&&&& \\
G28.20$-$0.04 \dotfill & N       & H53$\alpha$ & 38.5\tablenotemark{d} $\pm$ 1.5 & 0.39 $\pm$ 0.02 & 7000 $\pm$ 400 \\
\dotfill               & \nodata & H76$\alpha$ & 54.9\tablenotemark{d} $\pm$ 2.3 & 0.08 $\pm$ 0.01 & 7300 $\pm$ 500 \\
\dotfill               & S       & H53$\alpha$ & 18.8 $\pm$ 0.9 &\nodata  & 5900 $\pm$ 400 \\
\dotfill               & \nodata & H76$\alpha$ & 22.9 $\pm$ 1.4 & \nodata & 6800 $\pm$ 500 \\
&&&&& \\
G34.26$+$0.15 \dotfill & A       & H53$\alpha$ & 20.9 $\pm$ 1.0 & 0.75 $\pm$ 0.05 & 6800 $\pm$ 500 \\
\dotfill               & B       & H53$\alpha$ & 33.1 $\pm$ 2.8 & \nodata  & 8800 $\pm$ 900 \\
\dotfill               & C       & H53$\alpha$ & 49.0 $\pm$ 3.7 & 0.24\tablenotemark{e} $\pm$ 0.02 & 8700 $\pm$ 800 \\
\enddata
\tablenotetext{a}{Full width at half-maximum (FWHM) of the spectral lines corrected for broadening 
introduced by spectrometer and smoothing. The convolving function in the spectral frequency domain 
was assumed to be a Gaussian function with FWHM of 1.2 times a channel width 
(or twice a channel width if a spectrum is Hanning smoothed)
and deconvolved from the observed line widths. Velocity resolutions for all frequencies are given in Table \ref{tabinstr}.}
\tablenotetext{b}{We adopt the same uncertainties as for the fitted line-to-continuum ratios given in Table \ref{tabfits}.}
\tablenotetext{c}{For optically thin sources, the values of $(\frac{S_{L}}{S_{C}})_{\rm peak}$ from 
Table \ref{tabfits} were used to calculate T$_{e}^{\star}$.}
\tablenotetext{d}{For G28.20 N, the H53$\alpha$ and H76$\alpha$ line widths are slightly smaller that those 
reported in \citet{sew08} based on the same data: 39.7 $\pm$ 1.3 km s$^{-1}$ and 57.6 $\pm$ 2.2 km s$^{-1}$ for 
H53$\alpha$ and H76$\alpha$ lines, respectively. The difference in linewidths is due to different integration 
areas used in both papers. While \citet{sew08} integrated the lines over the entire source, the lines reported in 
this paper were integrated over the area enclosed by the 50\% contour.}
\tablenotetext{e}{The line-to-continuum ratios corrected for opacity for this source was derived from the peak optical depths listed in Table \ref{tabpeak}.}
\end{deluxetable}


\begin{deluxetable}{rcccc}
\tablecaption{Flux densities of G28.20-0.04 N \label{g28sed_tab}}
\tablewidth{0pt}
\tablehead{
\colhead{$\nu$} &
\colhead{Instr.} &
\colhead{Synth. Beam} &
\colhead{S$_{\nu}$} &
\colhead{Ref.}\\

\colhead{(GHz)} &
\colhead{} &
\colhead{($''~\times ~''$)} &
\colhead{(mJy)} &
\colhead{}
}
\startdata
5.0   &  VLA  & 2.0$\times$1.6     & 150$\pm$15  &  1  \\
6.7   &  ATCA &  $\sim$1.9         & 326\tablenotemark{a}   &  2  \\
8.3   &  VLA  & 0.9$\times$0.8     & 297$\pm$45   &  3  \\ 
8.3   &  VLA  & 4.9$\times$2.7     & 300$\pm$30   &  4  \\
14.7  &  VLA  & 1.5$\times$1.4     & 494$\pm$50 &  9  \\
14.7  &  VLA  & 0.6$\times$0.5     & 543$\pm$81  &  3  \\
22.4  &  VLA  & 0.3$\times$0.2     & 980\tablenotemark{a,b}   &  5  \\
22.4  &  VLA  & 0.1$\times$0.1     & 630$\pm$63    &  9  \\
43.0  &  VLA  & 1.9$\times$1.4     & 710$\pm$70  &  9  \\
43.0  &  VLA  & 0.17$\times$0.13   & 645$\pm$65   &  6  \\     
231.9 &  SMA  &       1            & 720\tablenotemark{a,c}    &  7  \\
231.9 &  SMA  & 0.6$\times$0.2     & 890$\pm$30\tablenotemark{c}   &  8  \\     
 \enddata
\tablenotetext{a}{Uncertainties are not provided in the referenced papers.}
\tablenotetext{b}{The 1.3 cm flux density given by \citet{so05} seems to be anomously high relative to other measurements (see Fig.  \ref{g28sed}). }
\tablenotetext{c}{Both 231.9 GHz flux densities are from the same SMA observations. Any difference in flux density 
represents a difference in the data reduction.}
\tablerefs{(1) \citealt{pur08}; (2) \citealt{w98} ; (3) \citealt{k94}; (4) \citealt{s04}; (5) \citealt{so05}; (6) \citealt{sew08}; (7) \citealt{k08}; (8) \citealt{q08}; (9) this paper}
\end{deluxetable}

\clearpage


\begin{deluxetable}{lcccp{1mm}ccccc}
\tablecaption{GLIMPSE Counterparts to the Radio Continuum Sources \label{glimpse}}
\tablewidth{0pt}
\tiny
\rotate
\tablehead{
\colhead{Source} &
\colhead{GLIMPSE ID\tablenotemark{a}} &
\multicolumn{2}{c}{d\tablenotemark{b}} &
\colhead{} &
\multicolumn{5}{c}{Flux densities\tablenotemark{c} (mJy)} \\

\cline{3-4}
\cline{6-10}

\colhead{} &
\colhead{} &
\colhead{($''$)} &
\colhead{(mpc)} &
\colhead{} &
\colhead{$K_{s}$} &
\colhead{$3.6 \mu m$} &
\colhead{$4.5 \mu m$} &
\colhead{$5.8 \mu m$} &
\colhead{$8.0 \mu m$} 
}
\startdata
G10.96 W & SSTGLMA G010.9584+00.0219 & 2 & 136 && \nodata & 9.7 $\pm$ 0.9 & 59 $\pm$ 8 &  174 $\pm$ 12 &  395 $\pm$ 31 \\
G10.96 E2 & SSTGLMA G010.9646+00.0098 & 0.3 & 20 && \nodata & 8 $\pm$ 3 & 13 $\pm$ 4 & 59  $\pm$ 8 & \nodata \\
G28.20 N & SSTGLMA G028.2003-00.0493 & 2.2 & 61 && 9.0 $\pm$ 0.4 & 358 $\pm$ 19 & 2123 $\pm$ 99 & 3892 $\pm$ 99 & \nodata \\
G28.20 S &  SSTGLMAG028.1984-00.0501 & 2.4 & 66 && \nodata & 2.1 $\pm$ 0.3 & 9 $\pm$ 1 & \nodata & \nodata\\
G34.26 C & SSTGLMA G034.2572+00.1533 & 0.7 & 13 && \nodata & \nodata & 4857 $\pm$ 278 & 4481 $\pm$ 155 & \nodata \\
G45.07 NE & SSTGLMA G045.0712+00.1321 & 0.2& 6 && 40 $\pm$ 1 & 2240 $\pm$ 184 & \nodata &\nodata & \nodata\\
\enddata 
\tablenotetext{a}{The Galactic coordinates of the {\it Spitzer} sources are part of the GLIMPSE IDs.}
\tablenotetext{b}{The distance between the radio continuum source and the nearest mid-IR GLIMPSE point source; 
                  kinematic distances to MSFRs listed in Table~\ref{trans} were used to calculate distances in mpc.}
\tablenotetext{c}{None of the sources has J and H fluxes available in the 2MASS Point Source Catalog.}
\end{deluxetable}

\normalsize

\clearpage


\clearpage


\clearpage


\clearpage


\clearpage


\clearpage


\clearpage


\clearpage


\clearpage


\clearpage


\clearpage


\clearpage


\clearpage

\begin{thebibliography}{}
\bibitem[\protect\citeauthoryear{{Allamandola}~et~al.}{{Allamandola}~et~al.}{1989}]{all89} Allamandola L. J, Tielens A. G. G. M.,  \&
Barker, J. R. 1989. Ap. J. Suppl. 71, 733.
\bibitem[\protect\citeauthoryear{{Altenhoff} et~al.}{{Altenhoff} et~al.}{1960}]{a60} Altenhoff W. J., Mezger P. G., Wendker H., Westerhout G. 1960, Ver\"{o}ff Sternw\"{a}rte Bonn, No. 59, 48
\bibitem[\protect\citeauthoryear{{Araya}~et~al.}{{Araya}~et~al.}{2002}]{a02} Araya, E., Hofner, P., Churchwell, E. \& Kurtz, S. 2002, \apjs, 138, 63 
\bibitem[\protect\citeauthoryear{{Argon}~et~al.}{{Argon}~et~al.}{2000}]{a00} Argon, A. L., Reid, M. J., \& Menten, K. M. 2000, \apjs, 129, 159 
\bibitem[\protect\citeauthoryear{{Avalos}~et~al.}{{Avalos}~et~al.}{2006}]{ava06} Avalos, M., Lizano, S., Rodr\'\i guez, L. F., Franco-Hern\'andez, R., Moran, J. M. 2006, \apj, 641, 406
\bibitem[\protect\citeauthoryear{{Avalos}~et~al.}{{Avalos}~et~al.}{2009}]{ava09} Avalos, M., Lizano, S., Franco-Hern\'andez, R., Rodr\'\i guez, L. F., \& Moran, J. M. 2009, \apj, 690, 1084
\bibitem[\protect\citeauthoryear{{Becker}~et~al.}{{Becker}~et~al.}{1994}]{becker94} Becker, R. H., White, R. L., Helfand, D. J.,\& Zoonematkermani, S. 1994, \apjs, 91, 347
\bibitem[\protect\citeauthoryear{{Beltr\'{a}n}~et~al.}{{Beltr\'{a}n}~et~al.}{2004}]{bel04} Beltr\'{a}n, M. T., et al. 2004, \apj, 601, L187
\bibitem[\protect\citeauthoryear{{Beltr\'{a}n}~et~al.}{{Beltr\'{a}n}~et~al.}{2005}]{bel05} Beltr\'{a}n, M. T., Cesaroni, R., Neri, R., Codella, C., Furuya, R. S., Testi, L., \& Olmi, L. 2005, \aap, 435, 901
\bibitem[\protect\citeauthoryear{{Beltr\'{a}n}~et~al.}{{Beltr\'{a}n}~et~al.}{2006}]{bel06} Beltr\'{a}n, M. T., Cesaroni, R., Codella, C., Testi, L., Furuya, R.
S., \& Olmi, L. 2006, Nature, 443, 427
\bibitem[\protect\citeauthoryear{{Beltr\'{a}n}~et~al.}{{Beltr\'{a}n}~et~al.}{2007}]{bel07} Beltr\'{a}n, M., Cesaroni, R., Moscadelli, L., \& Codella, C. 2007, \aap, 471, L13 
\bibitem[\protect\citeauthoryear{{Benjamin}~et~al.}{{Benjamin}~et~al.}{2003}]{b03} Benjamin, R. A., et al. 2003, \pasp, 115, 953
\bibitem[\protect\citeauthoryear{Brocklehurst \& Seaton}{Brocklehurst \& Seaton}{1972}]{bs72} Brocklehurst, M., \& Seaton, M. J. 1972, MNRAS, 157, 179 
\bibitem[\protect\citeauthoryear{{Bronfman}~et~al.}{{Bronfman}~et~al.}{1996}]{b96} Bronfman, L., Nyman L.-A., May, J. 1996, \aaps, 115, 81
\bibitem[\protect\citeauthoryear{{Campbell}~et~al.}{{Campbell}~et~al.}{2000}]{ce00} Campbell, M. F., Garland, C. A., Deutsch, L. K., Hora, J. L., Fazio, G. G., Dayal, A., \& Hoffmann, W. F. 2000, \apj, 536, 816
\bibitem[\protect\citeauthoryear{{Carey}~et~al.}{{Carey}~et~al.}{2009}]{carey2009} Carey, S.~J., et al.  2009, \pasp, 121, 76
\bibitem[\protect\citeauthoryear{{Caswell}~et~al.}{{Caswell}~et~al.}{1995a}]{ce95} Caswell, J. L., Vaile, R. A., Ellingsen, S. P., Whiteoak, J. B., \& Norris, R. P. 1995, \mnras, 272, 96
\bibitem[\protect\citeauthoryear{Caswell \& Vaile}{Caswell \& Vaile}{1995b}]{cv95} Caswell, J. L., \& Vaile, R. A. 1995, \mnras, 273, 328
\bibitem[\protect\citeauthoryear{Caswell}{Caswell}{2001}]{c01} Caswell, J. L. 2001, \mnras, 326, 805
\bibitem[\protect\citeauthoryear{Caswell}{Caswell}{2003}]{c03} Caswell, J. L. 2003, \mnras, 341, 551
\bibitem[\protect\citeauthoryear{{Cesaroni}~et~al.}{{Cesaroni}~et~al.}{1991}]{c91} Cesaroni, R., Walmsley, C. M., Kompe, C., \& Churchwell, E. B. 1991, \aap, 252, 278
\bibitem[\protect\citeauthoryear{{Churchwell}~et~al.}{{Churchwell}~et~al.}{2009}]{churchwell09} Churchwell, E., Babler, B. L., Meade, M. R., Whitney, B. A., Benjamin, R., Indebetouw, R., Cyganowski, C., Robitaille, T. P., Povich, M., Watson, C., Bracker, S. 2009, PASP, 121, 213
\bibitem[\protect\citeauthoryear{{Codella}~et~al.}{{Codella}~et~al.}{1995}]{co95} Codella, C., Palumbo, G. G. C., Pareschi, G., Scappini, F., Caselli, P., \& Attolini, M. R. 1995, \mnras, 276, 57
\bibitem[\protect\citeauthoryear{{Codella}~et~al.}{{Codella}~et~al.}{1997}]{cod97} Codella, C., Testi, L., \& Cesaroni, R. 1997, \aap, 325, 282
\bibitem[\protect\citeauthoryear{Crowther \& Conti}{Crowther \& Conti}{2003}] {cc03} Crowther, P. A., \& Conti, P. S. 2003, \mnras, 343, 143
\bibitem[\protect\citeauthoryear{{Cyganowski}~et~al.}{{Cyganowski}~et~al.}{2008}]{cyg08} Cyganowski, C. J., et al. 2008, \aj, 136, 2391
\bibitem[\protect\citeauthoryear{{Cyganowski}~et~al.}{{Cyganowski}~et~al.}{2009}]{cyg09} Cyganowski, C. J., Brogan, C. L., Hunter, T. R., Churchwell, E. 2009, \apj, 702, 1615
\bibitem[\protect\citeauthoryear{{De Buizer}~et~al.}{{De Buizer}~et~al.}{2003}]{db03} De Buizer, J. M., Radomski, J. T., Telesco, C. M., \& Pina, R. K. 2003, \apj, 598, 1127
\bibitem[\protect\citeauthoryear{{De Buizer}~et~al.}{{De Buizer}~et~al.}{2005}]{db05} De Buizer, J. M., Radomski, J. T., Telesco, C. M., \& Pina, R. K. 2005, \apjs, 156, 179
\bibitem[\protect\citeauthoryear{{De Pree}~et~al.}{{De Pree}~et~al.}{1996}]{depree96} De Pree, C. G., Gaume, R. A., Goss, W. M., \& Claussen, M. J. 1996, \apj, 464, 788
\bibitem[\protect\citeauthoryear{{De Pree}~et~al.}{{De Pree}~et~al.}{1997}]{depree97} De Pree, C. G., Mehringer, D. M., \& Goss, W. M. 1997, \apj, 482, 307
\bibitem[\protect\citeauthoryear{Elitzur \& de Jong}{Elitzur \& de Jong}{1987}]{elitzur78} Elitzur, M., de Jong, T. 1978, A\&A, 67, 323
\bibitem[\protect\citeauthoryear{{Fazio}~et~al.}{{Fazio}~et~al.}{2004}]{f04} Fazio, G. G., et al. 2004, \apjs, 154, 10
\bibitem[\protect\citeauthoryear{{Fey}~et~al.}{{Fey}~et~al.}{1994}]{fg94} Fey, A. L., Gaume, R. A., Nedoluha, G. E., \& Claussen, M. J. 1994, \apj, 435, 738
\bibitem[\protect\citeauthoryear{{Fish}~et~al.}{{Fish}~et~al.}{2003}]{f03} Fish, V. L., Reid, M. J., Wilner, D. J., \& Churchwell, E. 2003, \apj, 587, 701
\bibitem[\protect\citeauthoryear{Forster \& Caswell}{Forster \& Caswell}{1989}]{for89} Forster, J. R., \& Caswell, J. L. 1989, \aap, 213, 339
\bibitem[\protect\citeauthoryear{{Franco}~et~al.}{{Franco}~et~al.}{2000}]{franco00} Franco, J., Kurtz, S., Hofner, P., Testi, L., Garc\'\i a-Segura, G., \& Martos, M. 2000, \apj, 542, L143
\bibitem[\protect\citeauthoryear{{Furuya}~et~al.}{{Furuya}~et~al.}{2002}]{fur02} Furuya, R. S., et al. 2002, \aap, 390, L1
\bibitem[\protect\citeauthoryear{{Garay}~et~al.}{{Garay}~et~al.}{1985}]{g85} Garay, G., Reid, M. J., \& Moran, J. M. 1985, \apj, 289, 681
\bibitem[\protect\citeauthoryear{Garay \& Rodr\'\i guez}{Garay \& Rodr\'\i guez}{1990}]{g90} Garay, G., \& Rodr\'\i guez, L. F., 1990 \apj, 362, 191
\bibitem[\protect\citeauthoryear{{Garay}~et~al.}{{Garay}~et~al.}{1986}]{g86} Garay, G., Rodr\'\i guez, L. F. \& van Gorkom, J. H. 1986, \apj, 309, 553
\bibitem[\protect\citeauthoryear{{Gasiprong}~et~al.}{{Gasiprong}~et~al.}{2002}]{g02} Gasiprong, N., Cohen, R. J., \& Hutawarakorn, B. 2002, \mnras, 336, 47 
\bibitem[\protect\citeauthoryear{Gaume \& Mutel}{Gaume \& Mutel}{1987}]{g87} Gaume, R. A. \& Mutel, R. L. 1987 \apjs, 65, 193 
\bibitem[\protect\citeauthoryear{{Han}~et~al.}{{Han}~et~al.}{1998}]{h98} Han, F., Mao, R. Q., Lu, J., Wu, Y. F., Sun, J., Wang, J. S., Pei, C. C., Fan, Y., Tang, G. S., Ji, H. R. 1998, \aaps, 127, 181
\bibitem[\protect\citeauthoryear{{Hatchell}~et~al.}{{Hatchell}~et~al.}{2001}]{h01} Hatchell, J., Fuller, G. A., \& Millar, T. J. 2001, \aap, 372, 281
\bibitem[\protect\citeauthoryear{{Heaton}~et~al.}{{Heaton}~et~al.}{1989}]{hlb89} Heaton, B. D., Little, L. T., \& Bishop, L. S., \aap, 213, 148
\bibitem[\protect\citeauthoryear{{Hoare}~et~al.}{{Hoare}~et~al.}{2007}]{h07} Hoare, M. G., Kurtz, S. E., Lizano, S., Keto, E., \& Hofner, P., 2007 in Protostars \& Planets V, ed. B. Reipurth, D. Jewitt, \& K. Keil, (Tucson: University of Arizona Press), 181
\bibitem[\protect\citeauthoryear{Hofner \& Churchwell}{Hofner \& Churchwell}{1996}]{hc96} Hofner, P., \& Churchwell, E. 1996, \aaps, 120, 283
\bibitem[\protect\citeauthoryear{{Hollenbach}~et~al.}{{Hollenbach}~et~al.}{1994}]{hj94} Hollenbach, D., Johnston, D., Lizano,
S., \& Shu, F. 1994, \apj, 428, 654
\bibitem[\protect\citeauthoryear{{Hunter}~et~al.}{{Hunter}~et~al.}{1997}]{h97} Hunter, T. R., Phillips, T. G., Menten,
  K. M. 1997, \apj, 478, 283
\bibitem[\protect\citeauthoryear{Ignace \& Churchwell}{Ignace \& Churchwell }{2004}]{ic04} Ignace, R., \& Churchwell,
E. 2004, \apj, 610, 351
\bibitem[\protect\citeauthoryear{{Keto}~et~al.}{{Keto}~et~al.}{1992}]{k92} Keto, E., Proctor, D., Ball, R., Arens, J.,
  \& Jernigan, G. 1992, \apj, 401, L113
\bibitem[\protect\citeauthoryear{Keto}{Keto}{2003}]{ke03} Keto, E. 2003, \apj, 599, 1196
\bibitem[\protect\citeauthoryear{{Keto}~et~al.}{{Keto}~et~al.}{2008}]{k08} Keto, E., Zhang, Q., \& Kurtz, S. 2008, \apj, 672, 423
\bibitem[\protect\citeauthoryear{{Klaassen}~et~al.}{{Klaassen}~et~al.}{2009}]{klaassen2009} Klaassen, P. D., Wilson, C. D., Keto, E. R., \& Zhang, Q. 2009, \apj, 703, 1308
\bibitem[\protect\citeauthoryear{{Kraemer}~et~al.}{{Kraemer}~et~al.}{2003}]{k03} Kraemer, K. E., et al. 2003, \apj, 588, 918
\bibitem[\protect\citeauthoryear{{Kurtz}~et~al.}{{Kurtz}~et~al.}{1994}]{k94} Kurtz, S., Churchwell, E., \& Wood, D. O. S. 1994, \apjs, 91, 659
\bibitem[\protect\citeauthoryear{{Kurtz}~et~al.}{{Kurtz}~et~al.}{2000}]{ku00} Kurtz, S., Cesaroni, R., Churchwell, E., Hofner, P., \& Walmsley, C. M. 2000, in ``Protostars and Planets IV'', ed. V. Mannings, A. Boss, \& S. Russell (Tucson: University of Arizona Press), 299
\bibitem[\protect\citeauthoryear{{Kurtz}~et~al.}{{Kurtz}~et~al.}{2004}]{khv04} Kurtz, S., Hofner, P., \& Vargas Alvares,
C. 2004, \apjs, 155, 149
\bibitem[\protect\citeauthoryear{{Kurtz}}{{Kurtz}}{2005}]{kurtz05} Kurtz, S. 2005, "Massive Star Birth: A Crossroads of Astrophysics", IAU Symposium 227, eds.  R.~Cesaroni, M.~Felli, E.~Churchwell, \& M.~Walmsley, p. 111
\bibitem[\protect\citeauthoryear{Kurtz \& Hofner}{Kurtz \& Hofner}{2005}]{kh05} Kurtz, S., \& Hofner, P. 2005, \apj,
130, 711
\bibitem[Lizano et al.(1996)]{lizano96} Lizano, S., Canto, J., 
Garay, G., \& Hollenbach, D.\ 1996, \apj, 468, 739 
\bibitem[Lugo et al.(2004)]{lugo04} Lugo, J., Lizano, S., \& Garay, G.\ 2004, \apj, 614, 807
\bibitem[\protect\citeauthoryear{{Matthews}~et~al.}{{Matthews}~et~al.}{1987}]{m87} Matthews, N., Little, L. T., MacDonald, G. H., Andersson, M., Davies, S. R., Riley, P. W., Dent, W. R. F., \& Vizard, D. 1987, \aap, 184, 284
\bibitem[\protect\citeauthoryear{{Meade}~et~al.}{{Meade}~et~al.}{2009}]{meade2009} Meade, M., Whitney, B. A., Babler, B.,
Indebetouw, R., Bracker, S., Cohen, M., Robitaille, T., Benjamin, R., \& Churchwell, E. 2009, GLIMPSE documentation
available at the {\it Spitzer} Science Center (http://ssc.spitzer.caltech.edu/spitzermission/observingprograms/legacy/glimpse/) 
\bibitem[\protect\citeauthoryear{Menten}{Menten}{1991}]{m91} Menten, K. M. 1991, \apj, 380, L75
\bibitem[\protect\citeauthoryear{Mezger \& Henderson}{Mezger \& Henderson}{1967}]{mh67} Mezger, P. G., \& Henderson,
  A. P. 1967, ApJ, 147, 471
\bibitem[\protect\citeauthoryear{{Mookerjea}~et~al.}{{Mookerjea}~et~al.}{2007}]{moo07} Mookerjea, B., Casper, E., Mundy, L. G., \& Looney, L. W. 2007, \apj, 659, 447
\bibitem[\protect\citeauthoryear{{Moscadelli}~et~al.}{{Moscadelli}~et~al.}{2007}]{mos07} Moscadelli, L., Goddi, C., Cesaroni, R., Beltr\'{a}n, M. T., \& Furuya, R. S. 2007, \aap, 472, 867
\bibitem[\protect\citeauthoryear{Oster}{Oster}{1961}]{o61} Oster, L. 1961, Reviews of Modern Physics, 33, 525
\bibitem[\protect\citeauthoryear{Panagia \& Walmsley}{Panagia \& Walmsley}{1978}]{pw78} Panagia, N., \& Walmsley,
C. M. 1978, \aap, 70, 411
\bibitem[\protect\citeauthoryear{{Plume}~et~al.}{{Plume}~et~al.}{1997}]{p97} Plume, R., Jaffe, D. T., Evans, N. J., II, Mart\'\i n-Pintado, J., \& G\'omez-Gonz\'alez, J.  1997, \apj, 476, 730
\bibitem[\protect\citeauthoryear{{Povich}~et~al.}{{Povich}~et~al.}{2007}]{pov07}  Povich, M. S., Stone, J. M., Churchwell, E., Zweibel, E. G., Wolfire, M. G., Babler, B. L., Indebetouw, R., Meade, M. R., \& Whitney, B. A. 2007, \apj, 660, 346
\bibitem[\protect\citeauthoryear{{Price}~et~al.}{{Price}~et~al.}{2001}]{pri01} Price, S. D., Egan, M. P., Carey, S. J., Mizuno, D., Kuchar, T. 2001, \aap, 121, 2819
\bibitem[\protect\citeauthoryear{{Purcell}~et~al.}{{Purcell}~et~al.}{2008}]{pur08} Purcell, C. R., Hoare, M. G., \& Diamond, P.
2008, ``Massive Star Formation: Observations Confront Theory'', eds. H. Beuther, H. Linz, \& T. Henning, ASP Conference Series, Vol. 387, p. 389
\bibitem[\protect\citeauthoryear{{Qin}~et~al.}{{Qin}~et~al.}{2008}]{q08}  Qin, S.-L., Huang, M., Wu, Y., Xue, R., \& Chen, S. 2008, \apj, 686, 21
\bibitem[\protect\citeauthoryear{{Reid}~et~al.}{{Reid}~et~al.}{1980}]{rei80} Reid, M. J., Haschick, A. D., Burke, B. F., Moran, J. M,; Johnston, K. J., \& Swenson, G. W., Jr. 1980, \apj, 239, 89
\bibitem[\protect\citeauthoryear{Reid \& Ho}{Reid \& Ho}{1985}]{rh85} Reid, M. J., \& Ho, P. T. P. 1985, \apj,
  288, L17
\bibitem[\protect\citeauthoryear{Rohlfs \& Wilson}{Rohlfs \& Wilson}{1986}]{rw86} Rohlfs, K., \& Wilson, T. L. 1986, ``Tools of Radio Astronomy'', 2nd edition, Springer-Verlag Berlin Heidelberg
\bibitem[\protect\citeauthoryear{Roelfsema \& Goss}{Roelfsema \& Goss}{1992}]{rg92} Roelfsema, P. R., \& Goss, W. M. 1992, A\&ARv, 4, 161 
\bibitem[\protect\citeauthoryear{{Schutte}~et~al.}{{Schutte}~et~al.}{1993}]{s93} Schutte, A. J., van der Walt, D. J.,
Gaylard, M. J., \& MacLeod, G. C. 1993, \mnras, 261, 783
\bibitem[\protect\citeauthoryear{{Sellgren}}{{Sellgren}}{1984}]{sel84} Sellgren K. 1984, \apj, 277, 623
\bibitem[\protect\citeauthoryear{{Sewi{\l}o}~et~al.}{{Sewi{\l}o}~et~al.}{2004a}]{s04} Sewi{\l}o, M., Churchwell, E., Kurtz, S.,
Goss, W. M., \& Hofner, P. 2004a, \apj, 605, 285 (Paper I)
\bibitem[\protect\citeauthoryear{{Sewi{\l}o}~et~al.}{{Sewi{\l}o}~et~al.}{2004b}]{swa04} Sewi{\l}o, M., Watson, C., Araya, E.,
Churchwell, E., Hofner, P., \& Kurtz, S. 2004b, \apjs, 154, 553
\bibitem[\protect\citeauthoryear{{Sewi{\l}o}~et~al.}{{Sewi{\l}o}~et~al.}{2008}]{sew08} Sewi{\l}o, M., Churchwell, E., Kurtz, S., Goss, W. M., \& Hofner, P. 2008, \apj, 681, 350
\bibitem[\protect\citeauthoryear{{Smith}~et~al.}{{Smith}~et~al.}{2002}]{sm02} Smith, L. J., Norris, R. P. F., \& Crowther, P. A. 2002, \mnras, 337. 1309
\bibitem[\protect\citeauthoryear{{Sollins}~et~al.}{{Sollins}~et~al.}{2005}]{so05} Sollins, P. K., Zhang, Q., Keto, E., \& Ho, P. T. P. 2005, \apj, 631, 399
\bibitem[\protect\citeauthoryear{Tan \& McKee}{Tan \& McKee}{2003}]{tm03} Tan, J. C., \& McKee, C. F. 2003, {\it ASP Conference Series 221}, also astro-ph/0309139
\bibitem[\protect\citeauthoryear{{Tielens}~et~al.}{{Tielens}~et~al.}{1993}]{tie93} Tielens A. G. G. M., et al. 1993, Science 262, 86
\bibitem[\protect\citeauthoryear{Turner \& Matthews}{Turner \& Matthews}{1984}]{tm84} Turner, B. E., \& Matthews, H. E. 1984, \apj, 277, 164
\bibitem[\protect\citeauthoryear{{Voit}}{{Voit}}{1992}]{voi92} Voit, G. M 1992, MNRAS, 258, 841
\bibitem[\protect\citeauthoryear{{Walsh}~et~al.}{{Walsh}~et~al.}{1997}]{wh97} Walsh, A. J., Hyland, A. R., Robinson, G.,
\& Burton, M. G. 1997, \mnras, 291, 261
\bibitem[\protect\citeauthoryear{{Walsh}~et~al.}{{Walsh}~et~al.}{1998}]{w98} Walsh, A. J., Burton, M. G., Hyland, A. R.,
\& Robinson, G. 1998, \mnras, 301. 640
\bibitem[\protect\citeauthoryear{Welch \& Marr}{Welch \& Marr}{1987}]{wm87} Welch, W. J., \& Marr, J. 1987, \apj, 317, L21
\bibitem[\protect\citeauthoryear{{Wilcots}~et~al.}{{Wilcots}~et~al.}{2003}]{wilcots2003} Wilcots, E. M., 
 Brinks, E., \& Higdon, J. 2003, {\it ``A Guide for VLA Spectral Line Observers''}, http://www.vla.nrao.edu/astro/guides/sline/current/
\bibitem[\protect\citeauthoryear{Wood \& Churchwell}{Wood \& Churchwell}{1989}]{wc89} Wood, D. O. S., \& Churchwell, E. 1989, \apjs, 69, 831
\end{thebibliography}
\end{document}